\documentclass[manuscript]{aastex}

\shorttitle{MANOS: photometric results}
\shortauthors{Thirouin et al.}

\begin{document}

\title{The Mission Accessible Near-Earth Objects Survey (MANOS): \\
first photometric results }

\author{A. Thirouin\altaffilmark{1}}
\affil{Lowell Observatory, 1400 W Mars Hill Rd, Flagstaff, AZ 86001, United States of America.}
\email{thirouin@lowell.edu}

\author{N. Moskovitz\altaffilmark{1}}
\affil{Lowell Observatory, 1400 W Mars Hill Rd, Flagstaff, AZ 86001, United States of America.}

\author{R.P. Binzel\altaffilmark{2}}
\affil{Massachusetts Institute of Technology (MIT), 77 Massachusetts Ave, Cambridge, MA 02139, United States of America. }

\author{E. Christensen\altaffilmark{3}}
\affil{University of Arizona, Tucson, AZ, United States of America. }

\author{F.E. DeMeo\altaffilmark{2}}
\affil{Massachusetts Institute of Technology (MIT), 77 Massachusetts Ave, Cambridge, MA 02139, United States of America. }

\author{M.J. Person\altaffilmark{2}}
\affil{Massachusetts Institute of Technology (MIT), 77 Massachusetts Ave, Cambridge, MA 02139, United States of America. }

\author{D. Polishook\altaffilmark{5}}
\affil{Department of Earth and Planetary Science, Weizmann Institute, Herzl St 234, Rehovot, 7610001, Israel. }

\author{C.A. Thomas\altaffilmark{6, 7}}
\affil{Planetary Science Institute (PSI), 1700 E Fort Lowell Rd 106, Tucson, AZ 85719, United States of America. }
\affil{USRA/NASA Goddard Space Flight Center, 8800 Greenbelt Rd., Greenbelt, MD 20771, United States of America. }

\author{D. Trilling\altaffilmark{8}}
\affil{Department of Physics and Astronomy, PO Box 6010, Northern Arizona University, Flagstaff AZ 86001, United States of America. }

\author{M. Willman\altaffilmark{9}}
\affil{University of Hawaii, Pukalani, HI 96788, United States of America.}

\author{M. Hinkle\altaffilmark{8}}
\affil{Department of Physics and Astronomy, PO Box 6010, Northern Arizona University, Flagstaff AZ 86001, United States of America. }

\author{B. Burt\altaffilmark{1}}
\affil{Lowell Observatory, 1400 W Mars Hill Rd, Flagstaff, AZ 86001, United States of America.}

\author{D. Avner\altaffilmark{8}}
\affil{Department of Physics and Astronomy, PO Box 6010, Northern Arizona University, Flagstaff AZ 86001, United States of America. }

\and

\author{F.J. Aceituno\altaffilmark{10}}
\affil{Instituto de Astrof\'{i}sica de Andaluc\'{i}a (IAA-CSIC), Glorieta de la Astronom\'{i}a, S/N, Granada, 18008, Spain. }

\begin{abstract}
The Mission Accessible Near-Earth Objects Survey (MANOS) aims to physically characterize sub-km Near-Earth Objects (NEOs). We report first photometric results from the survey which began in August, 2013. Photometric observations were performed using 1~m to 4~m class telescopes around the world. We present rotational periods and lightcurve amplitudes for 86 sub-km NEOs, though in some cases, only lower limits are provided. Our main goal is to obtain lightcurves for small NEOs (typically, sub-km objects) and estimate their rotational periods, lightcurve amplitudes, and shapes. These properties are used for statistical study to constrain overall properties of the NEO population. A weak correlation seems to indicate that smaller objects are more spherical than the larger ones. We also report 7 NEOs that are fully characterized (lightcurve and visible spectra) as the most suitable candidates for a future human or robotic mission. Viable mission targets are objects fully characterized, with a $\Delta v$$^{NHATS}$ $\leq$12~km~s$^{-1}$, and a rotational period P$>$1~h. Assuming a similar rate of object characterization as reported in this paper, approximately 1,230 NEOs need to be characterized in order to find 100 viable mission targets.   

\end{abstract}


%%%%%%%%%%%%%%%%%%%%%%%%%%%%%%%%%%%%%%%%%%%%
%%%% Introduction
%%%%%%%%%%%%%%%%%%%%%%%%%%%%%%%%%%%%%%%%%%%%

\section{Introduction}

Near Earth Objects (NEOs) are minor bodies (asteroids, comets, meteoroids) on orbits with perihelia distances
q$\textless$1.3~AU. As of April 2016, 14,263 NEOs have been discovered\footnote{Numbers from the Minor Planet Center: \url{http://www.minorplanetcenter.net/}}. About 90$\%$ of NEOs originated in the asteroid belt and have a rocky nature \citep{Jewitt2002, DeMeo2008}. Despite the impressive number of discovered NEOs, physical information for these objects remains limited. Rotational light curves are one tool to constrain the physical evolution of these objects. The rotational states of asteroids provide information about  physical properties such as a lower limit to density, internal structure, cohesion, and shape or surface heterogeneity \citep{Pravec2000, Holsapple2001, Holsapple2004}.  

Large objects (diameter greater than 1~km) have been well-studied with photometric, spectroscopic, and/or radar techniques \citep{Benner2015, Thomas2014, Thomas2011, Warner2009, Pravec2002, Binzel2002}, but small objects are also of interest for a number of reasons. First, objects in the meter to decameter size regime can impact the Earth on human timescales, as opposed to the 10$^{6}$~years impact interval of km-scale objects \citep{Harris2015}. As evidenced in Chelyabinsk, Russia in 2013 \citep{Popova2013} relatively small objects can pose a modest impact hazard. In addition, these small NEOs are the immediate parent bodies of meteorites. To interpret meteorites in an astrophysical context requires that we better understand their source population. In addition, studying these small objects can provide deeper insight into size-dependent evolutionary processes such as the radiative Yarkovsky and Yarkovsky-O'Keefe-Radzievskii-Paddack (YORP) effects \citep{Bottke2006}. Finally, the much greater number of NEOs with sizes of $\sim$100~m compared to km-size objects provides more opportunities for detailed physical study. This includes increased possibilities for a variety of exploration mission scenarios (e.g., \citet{Abell2009}) as well as more frequent near-Earth encounters to study physical changes associated with gravitational perturbation events (e.g., \citet{Scheeres2005}; \citet{Binzel2010}).

Because NEOs have their origin in the Main Belt of asteroids, and are the result of multiple collisions, their shape as well as rotational properties are valuable tracers of their evolution. \citet{Binzel2002} suggested that NEOs should be similar in rotation and shape to similar sized Main Belt asteroids, and spin distribution of km-size Main Belt Asteroids and similar sized NEOs supports this assertion \citep{Pravec2008, Polishook2009}. 
Therefore, the study of small NEOs (sub-km objects) allow us to infer the properties of small Main Belt asteroids, which remain unobservable with current facilities. 

We present here a study focused on the rotational properties of sub-km NEOs. Our ultimate goal is to obtain the most comprehensive data-set of sub-km NEOs to date, allowing a homogeneous and detailed study of the shape, surface and rotational properties of these objects. This paper is divided into six sections. In the next section, we introduce briefly the Mission Accessible Near-Earth Objects Survey (MANOS). Then, we describe the observations and the data set analyzed. In Section 4, we present our main results regarding rotational period and lightcurve amplitudes of our targets. In Section 5, we discuss our results and compare them to the literature. In Section 6, we put constraints on the internal structure of NEOs. Finally, Section 7 is dedicated to the summary and conclusions of this work.
 
%%%%%%%%%%%%%%%%%%%%%%%%%%%%%%%%%%%%%%%%%%%%
%%%% MANOS survey
%%%%%%%%%%%%%%%%%%%%%%%%%%%%%%%%%%%%%%%%%%%%

\section{What is MANOS?}

The Mission Accessible Near-Earth Object Survey (MANOS) began in August 2013 as a multi-year survey program (2013B-2016B) awarded by the National Optical Astronomy Observatory (NOAO), and funded through NASA NEOO (Near-Earth Object Observations) office. MANOS is a physical characterization survey of NEOs providing physical data for several hundred mission accessible NEOs across visible and near-infrared wavelengths (Moskovitz et al., In prep.). This comprehensive study aims to provide lightcurves, astrometry, and reflectance spectra. MANOS primarily focuses on newly discovered objects. Targets for MANOS are selected based on two other criteria (besides observability): i) mission accesibility (i.e., $\Delta v$$^{SH}$$<$7~km~s$^{-1}$ (see below for more details)), and ii) absolute magnitude greater than 20 (i.e. object with a diameter smaller than $\sim$300~m assuming an albedo of 0.2). Typical NEOs are discovered at V$\sim$20 and fade by 3.5~mag after 1 month, thus their characterization requires a challenging set of rapid response observations \citep{Galache2015}. For such a rapid response, MANOS employs 1- to 8-m class facilities in the Northern and Southern hemispheres through queue, remote or in-situ observations. Currently, we have the capacity to characterize between 5 to 10 newly discovered objects per month. Large telescopes allow us to obtain rotational lightcurves for objects down to V$\sim$22~mag, and visible spectra down to V$\sim$~20.5~mag.     

MANOS was initially awarded time on the 8.1~m Gemini Telescopes (Northern and Southern hemispheres), the 4.1~m Southern Astrophysical Research (SOAR) telescope, the 4~m Mayall Telescope (Kitt Peak Observatory), and the 1.3~m Small and Moderate Aperture Research Telescope System (SMARTS) telescope. We have also employed facilities at Lowell Observatory and the University of Hawaii. Gemini and University of Hawaii facilities are dedicated to spectroscopic observations and will not be presented here. 

\begin{figure}
\includegraphics[width=15cm, angle=180]{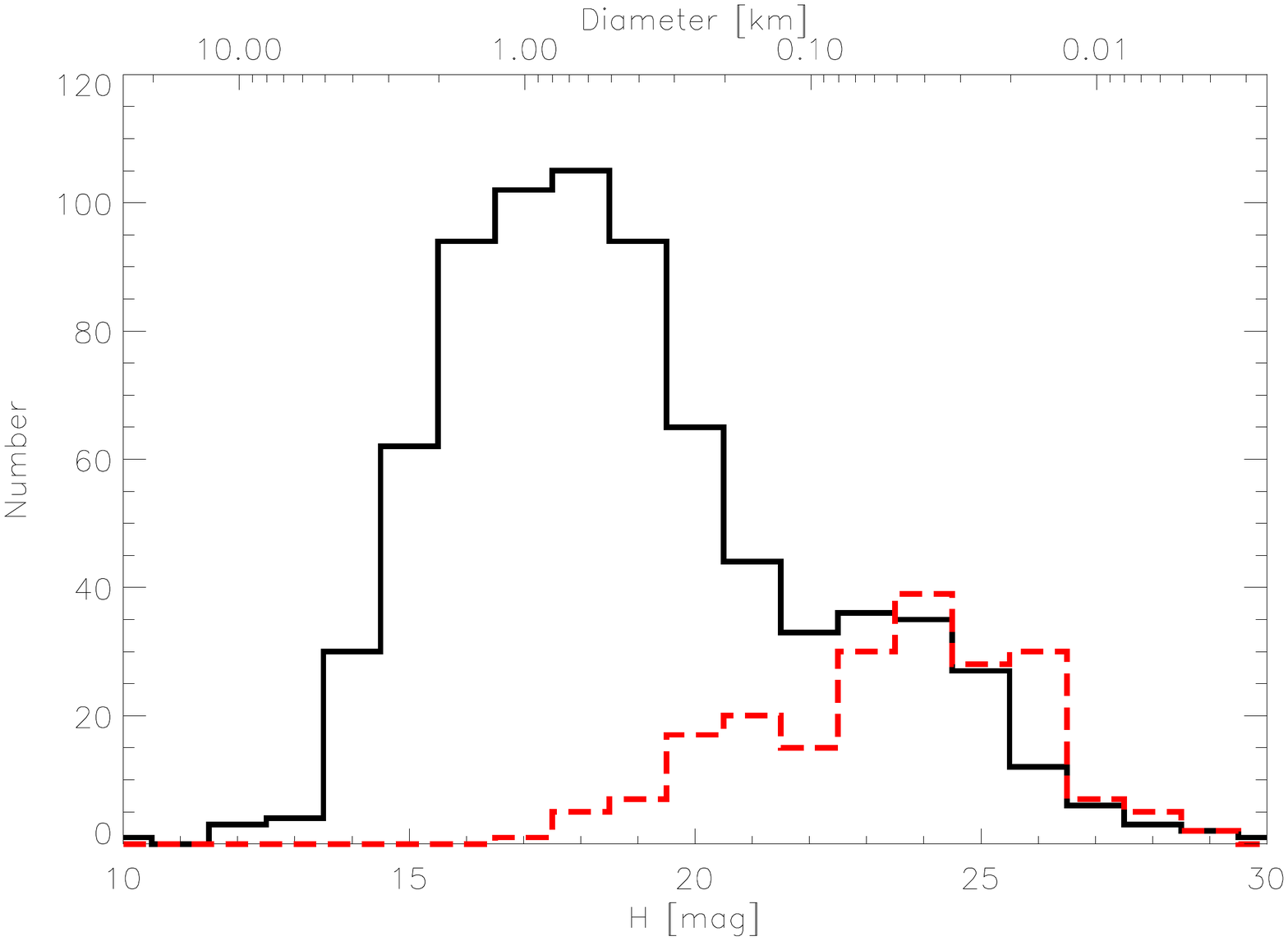}
\caption {\textit{Number of objects versus absolute magnitude}: Absolute magnitude (H) distribution of all NEOs with previously obtained lightcurve (continuous black line: data from \citet{Warner2009} on July 2015), and MANOS objects observed for lightcurves over $\sim$2 years (discontinuous red line). Here we report results for 86 MANOS targets. Our full sample will be published at a later date. Diameter was computed assuming an albedo of 0.2. }
\label{fig:Histo1}
\end{figure}

Figure~\ref{fig:Histo1} shows all published NEO lightcurves reported in the lightcurve database of \citet{Warner2009} as of July 2015. The peak of the distribution is at an absolute magnitude H$\sim$17-18, i.e objects with diameter of approximately 1~km (assuming an albedo of 0.2; albedo value used for this entire paper). At greater H, the number of objects studied for rotational properties is low. In the first $\sim$2~years of our survey, we have doubled the number of photometrically characterized objects with an absolute magnitude H=23.5-25.5, and increased by 300\% the number of objects in the range of H=25.5-26.5.

As mentioned, one of our main selection criteria is the mission accesibility. The Asteroid Redirect Mission (ARM) is a potential future space mission proposed by NASA \citep{Mazanek2015}. As of mid-2016 the outline of this mission is to rendez-vous with a ``large" near-Earth asteroid, use robotic arms to retrieve a boulder up to 4~m in size from the surface and then return it to cis-lunar orbit where it can be studied in-situ by astronauts. 

A key parameter for a mission to a NEO (and by extension all missions) is the delta-v ($\Delta v$) required to reach the orbit of the object. This parameter is the change in velocity needed to go from Low Earth Orbit (LEO) to a NEO rendez-vous using a Hohman transfer orbit. In first approximation, the LEO-NEO $\Delta v$ values are computed using the \citet{Shoemaker1978} formalism ($\Delta v$$^{SH}$). On April 2016, 14,263 NEOs are known, and only 13 objects have a $\Delta v$$^{SH}$$<$4~km~s$^{-1}$, 145 have a $\Delta v$$^{SH}$$<$4.5~km~s$^{-1}$, and 625 with a $\Delta v$$^{SH}$$\leq$5~km~s$^{-1}$ \footnote{\url{http://echo.jpl.nasa.gov/$\sim$lance/delta$\_$v/delta$\_$v.rendezvous.html}}. 

\begin{figure}
\center
\includegraphics[width=18cm, angle=0]{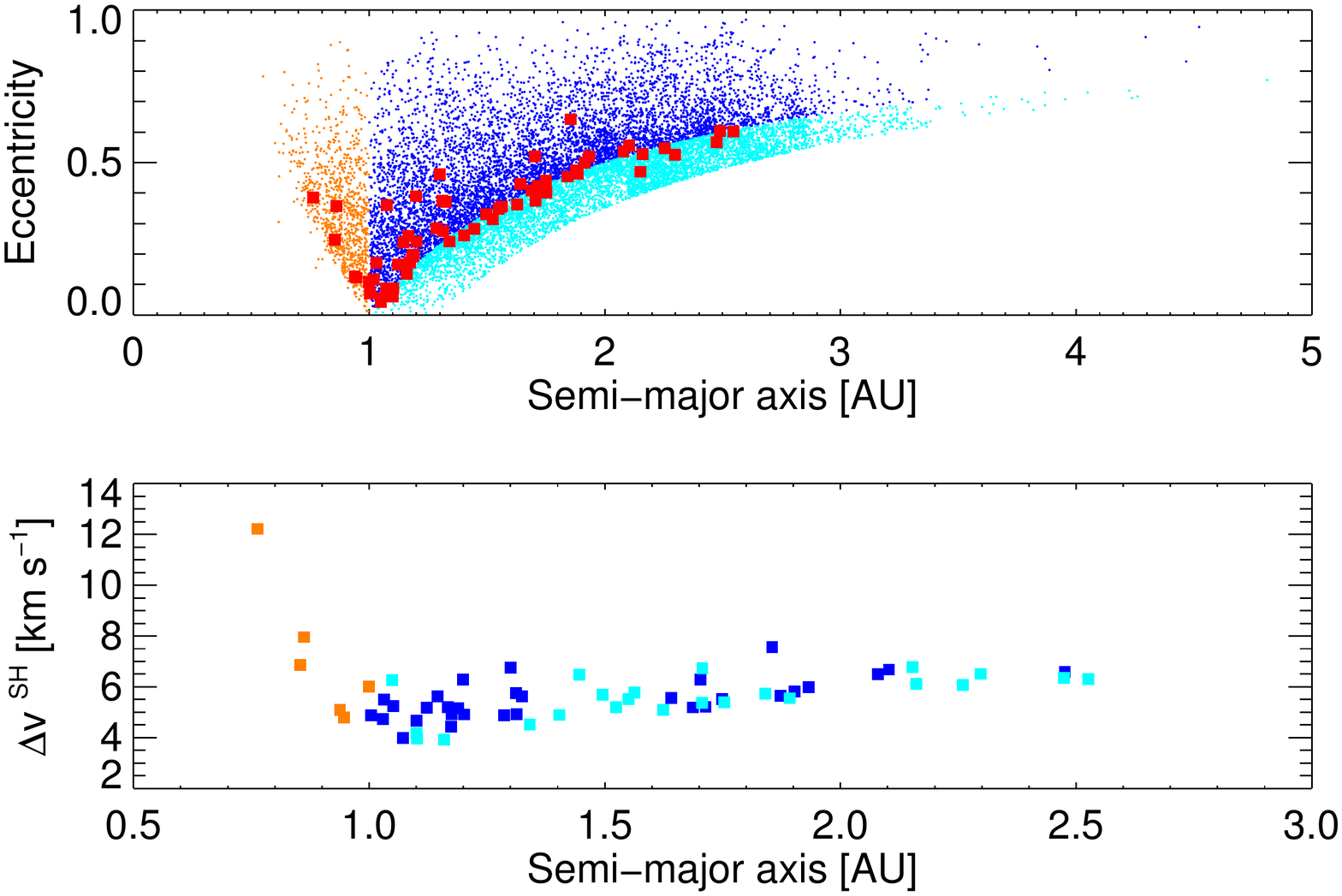}
\caption {\textit{Eccentricity and Delta-v$^{SH}$ versus semi-major axis}: All known NEOs (data from the MPC webpage) are plotted. Different colors correspond to different dynamical classes: Aten in orange, Apollo in blue, and Amor in cyan. MANOS objects with a rotational period and lightcurve amplitude are plotted (red squares upper plot). MANOS targets are plotted in the lower plot according to their dynamical class (Aten in orange, Apollo in blue, and Amor in cyan). The $\Delta$$_{v}$$^{SH}$ values are computed using the \citet{Shoemaker1978} formalism. MANOS observed 5 objects with a $\Delta v$$^{SH}$ lower than 4~km~s$^{-1}$, 4 objects with a $\Delta v$$^{SH}$ between 4~km~s$^{-1}$ and 4.5~km~s$^{-1}$, and 23 objects with a $\Delta v$$^{SH}$ between 4.5~km~s$^{-1}$ and 5.5~km~s$^{-1}$.
 }
\label{fig:dv}
\end{figure}

In Table~\ref{Summary_photo}, the $\Delta v$$^{SH}$ of MANOS objects are reported. MANOS observed 5 objects with a $\Delta v$$^{SH}$ lower than 4~km~s$^{-1}$, 4 objects with a $\Delta v$$^{SH}$ between 4~km~s$^{-1}$ and 4.5~km~s$^{-1}$, and 23 objects with a $\Delta v$$^{SH}$ between 4.5~km~s$^{-1}$ and 5.5~km~s$^{-1}$ (i.e. 63 MANOS objects have a $\Delta v$$^{SH}$$>$5.5~km~s$^{-1}$). Figure~\ref{fig:dv} shows MANOS objects reported in this work. 
 
The \citet{Shoemaker1978} formalism is only a first approximation to estimate if an object is truly spacecraft accessible. In fact, full orbital integrations are needed to calculate accurate $\Delta v$. The Near-Earth Object Human Space Flight Accessible Targets Study (NHATS) performs more accurate $\Delta v$$^{NHATS}$ calculations that take into account specific launch windows, and the duration of the mission\footnote{for more details, see \url{http://neo.jpl.nasa.gov/nhats/}}. Several MANOS objects are listed on the NHATS webpage. For example, 2014~UV$_{210}$ with a $\Delta v$$^{SH}$ of 3.93~km~s$^{-1}$ using \citet{Shoemaker1978}, has a $\Delta v$$^{NHATS}$ of 5.902~km~s$^{-1}$ according to NHATS. 

The cut-off for NHATS is for a $\Delta v$$^{NHATS}$ of 12~km~s$^{-1}$. A total of 33 MANOS objects have a $\Delta v$$^{NHATS}$$\leq$ 12~km~s$^{-1}$ and are mission accessible according to NHATS calculations: 5 Aten, 9 Amor, and 19 Apollo (Figure~\ref{fig:dv}, and Table~\ref{Tab:candidates}). We report complete lightcurve for 30 of these 33 objects (2 have flat lightcurves, 1 with partial lightcurve), and 26 of them are characterized with lightcurve and spectra in the visible (5 of them have also infrared spectra). \citet{Abell2009} consider a rotational period of 1~h as a practical limit independent of the nature of the future mission (robotic or human mission). Only 7 objects of the objects presented here meet the NHATS dynamical criteria and the 1~h rotation limit (Table~\ref{Tab:candidates}).

%%%%%%%%%%%%%%%%%%%%%%%%%%%%%%%%%%%%%%%%%%%%
%%%% Observations
%%%%%%%%%%%%%%%%%%%%%%%%%%%%%%%%%%%%%%%%%%%%

\section{Observations and data reduction}
\subsection{Telescope Resources}

Here, we present photometric results for 86 MANOS targets, representing a statistically significant subset of the overall MANOS sample. In approximately 2~years, a total of 207 objects have been observed for lightcurves. The remaining objects as well as future observations will be published at a later time. The data presented here were obtained with the 4.3~m Lowell Discovery Channel Telescope (DCT), the 4.1~m Southern Astrophysical Research (SOAR) telescope, the 4~m Nicholas U. Mayall Telescope, the 2.1~m at Kitt Peak Observatory, the 1.8~m Perkins telescope, the 1.5~m Sierra Nevada Observatory (OSN), and the 1.3~m SMARTS telescope between August 2013 and October 2015. 

The DCT is forty miles southeast of Flagstaff at the Happy Jack site (Arizona, United States of America). Images were obtained using the Large Monolithic Imager (LMI), which is a 6144$\times$6160 CCD \citep{Levine2012}. The total field of view is 12.5$\arcmin$$\times$12.5$\arcmin$ with a plate scale of 0.12$\arcsec$/pixel (unbinned). Images were obtained using the 3$\times$3 or 2$\times$2 binning modes. Observations were carried out in-situ. 

The SOAR telescope is located on Cerro Pach\'{o}n, Chile. Images were obtained using the Goodman High Throughput Spectrograph (Goodman-HTS) instrument in its imaging mode. The instrument consists of a 4096$\times$4096 Fairchild CCD, with a 7.2$\arcmin$ diameter field of view (circular field of view) and a plate scale of 0.15$\arcsec$/pixel. Images were obtained using the 2$\times$2 binning mode. Observations were conducted remotely.

The Mayall telescope is a 4~m telescope located at the Kitt Peak National Observatory (KPNO), Tucson, Arizona, USA. The NOAO CCD Mosaic-1.1 is a wide field imager composed of an array of eight CCD chips. The field of view is 36$\arcmin$$\times$36$\arcmin$, and the plate scale is 0.26~$\arcsec$/pixel. Observations were performed remotely.      

The 2.1~m at Kitt Peak Observatory was operated with the STA3 2k$\times$4k CCD, which has a plate scale of 0.305~$\arcsec$/pixel and a field of view of 10.2$\arcmin$$\times$6.6$\arcmin$. The instrument was binned 2$\times$2 and the observations were conducted in-situ.

The Perkins 72" telescope is located at the Anderson Mesa station at Lowell Observatory (Flagstaff, Arizona, USA). We used the PRISM (Perkins ReImaging SysteM) instrument, a 2$\times$2~k Fairchild CCD. The PRISM plate scale is 0.39$\arcsec$/pixel for a field of view of 13$\arcmin$$\times$13$\arcmin$. Observations were performed in-situ. 
 
The 1.5~m telescope located at the Observatory of Sierra Nevada (OSN) at Loma de Dilar in the National Park of Sierra Nevada (Granada, Spain) was operated in-situ. Observations were carried out with a 2k$\times$2k CCD, with a total field of view of 7.8$\arcmin$$\times$7.8$\arcmin$. We used 2$\times$2 binning mode, resulting in an effective plate scale of 0.46$\arcsec$/pixel.

The 1.3~m SMARTS telescope is located at the Cerro Tololo Inter-American Observatory (Coquimbo region, Chile). This telescope is equipped with a camera called ANDICAM (A Novel Dual Imaging CAMera). ANDICAM is a Fairchild 2048$\times$2048 CCD. The pixel scale is 0.371$\arcsec$/pixel, and the field of view is 6$\arcmin$$\times$6$\arcmin$. Observations were carried out in queue mode.

%%%%%%%%%%%%%%%%%%%%%%%%%%%%%%%%%%%%%%%%%%%%
%%%% Observing strategy
%%%%%%%%%%%%%%%%%%%%%%%%%%%%%%%%%%%%%%%%%%%%

\subsection{Observing strategy, data reduction and analysis}
\label{sec:obs}
 
Exposure times were chosen based on two competing factors: i) the exposure had to be long enough to achieve sufficient signal-to-noise ratio (S/N) to resolve typical light urve variability (i.e. S/N$>$20); and ii) the exposure had to be short enough to avoid significant elongation of sources due to the non-sidereal motion of the target. We always elected to track the telescopes at sidereal rates, mainly to avoid significant elongation of the sources, and because we can use the same reference stars for the photometry. Exposure times between 1 to 200 seconds were used according to the sky motion and brightness of the object, and the telescope aperture. It is important to point out that the use of long exposure times is a problem in case of fast or ultra-rapid rotators. In fact, if the exposure time (+ the read-out time) are consistent or longer than the rotational period of the object, we will not be able to detect the fast rotation of the object and the lightcurve will be flat. Here, we report several flat lightcurves. Most of the objects are large and we are not expecting them to be fast rotators. However, several objects are small and are potential fast rotators whose rotational period get undetected because of the too long exposure time used (3 objects reported here: 2014~YD$_{42}$, 2015~EQ, and 2015~HS$_{1}$). This topic will be studied in more details in a future work. 

Broad band photometric filters were chosen to maximize S/N, to minimize fringing at long wavelengths, and to choose the band pass to sky brightness conditions dictated by lunar phase. Observations at the OSN were performed without filter. We used the V, R, open and r' filters at SOAR. With the DCT and the 1.8~m Perkins we used broad VR filters. The broad wh-filter (transmission from 0.4-0.9 $\mu$m) was used at Kitt Peak. Since these observations focused on deriving relative photometric variations, the use of multiple filters and unfiltered images without absolute calibration did not affect our science goals.

Approximately 45$\%$ of sub-km NEOs have a rotational period $<$~3~h, whereas $\sim$88$\%$ of sub-100m NEOs have period less than 3~h \citep{Warner2009}. A MANOS goal is to characterize small objects thus we dedicated observing blocks of $\sim$3~h per object. With this strategy we were generally able to observe at least one full rotation, although this strategy does bias against the detection of slow rotators.  

As our strategy is designed for rapid response and building population statistics, we cannot and do not spend several nights per target. Therefore, shape modeling, which requires several epochs of data, is not feasible. However, we report three objects with two lightcurves obtained at different epochs, with the intent that these data will prove useful for future shape modeling efforts.

%%%%%%%%%%%%%%%%%%%%%%%%%%%%%%%%%%%%%%%%%%%%
%%%% Data reduction and analysis
%%%%%%%%%%%%%%%%%%%%%%%%%%%%%%%%%%%%%%%%%%%%

%\section{Data reduction and analysis}

During each observing night, a series of bias and flat fields (dome and/or twilight flats) were obtained to correct the images. We created a median bias and median flatfield for each night. Target images were bias subtracted and flatfielded. Relative photometry using up to 30 reference stars was carried out using Daophot routines \citep{Stetson1987}. Time-series photometry of each target was inspected for periodicity by means of the Lomb technique \citep{Lomb1976} as implemented in \citet{Press1992}. This method is a modified version of the Fourier spectral analysis. The main difference with the Fourier spectral analysis is the fact that this method takes into account irregularly spaced data. This method gives a weight to each data point instead of considering an interval time. We also verified our results by using the CLEAN \citep{Foster1995}, and the Phase Dispersion Minimization (PDM, \citet{Stellingwerf1978}) methods. 

\begin{figure}
\includegraphics[width=8cm, angle=0]{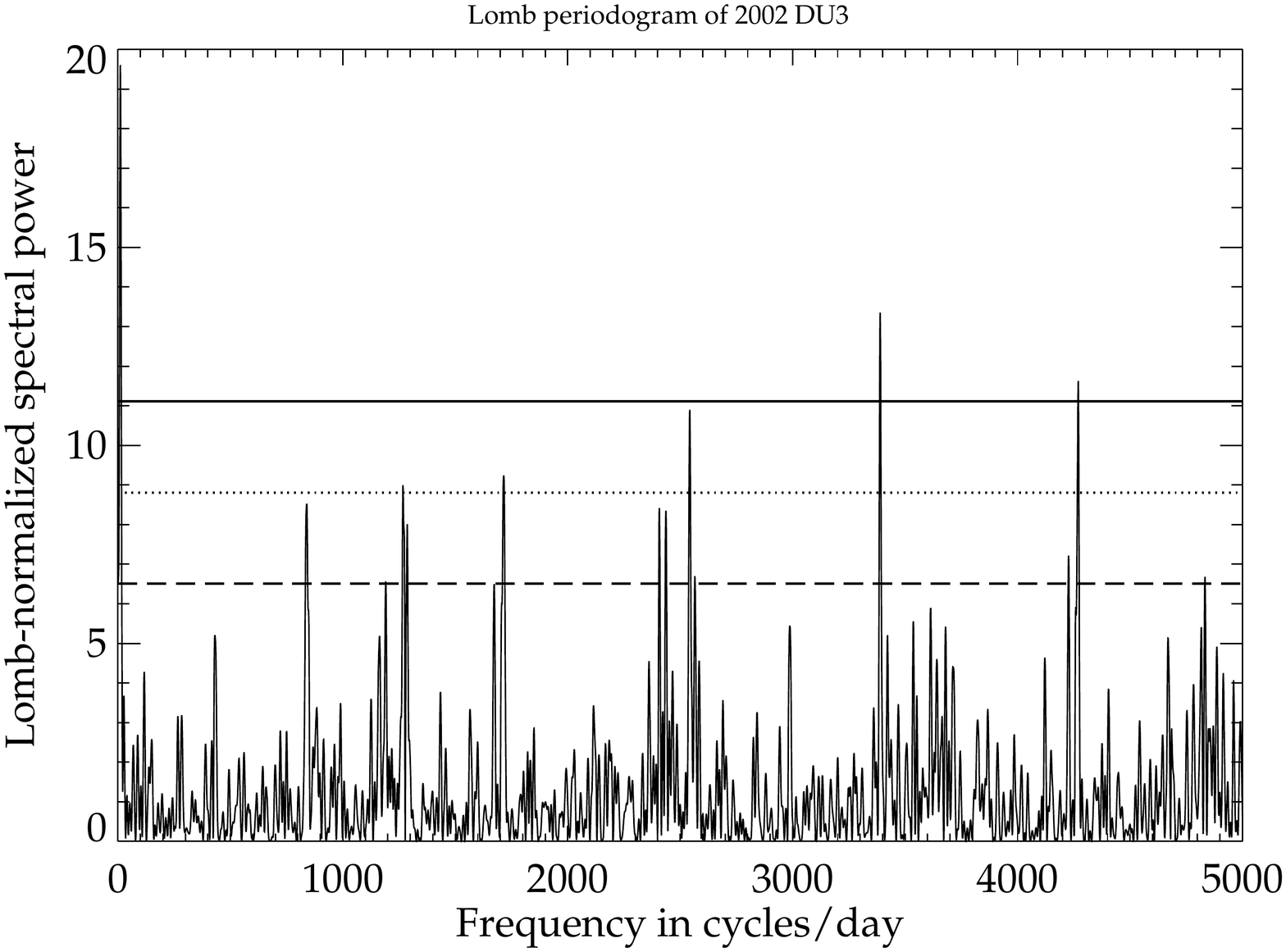}
\includegraphics[width=8cm, angle=0]{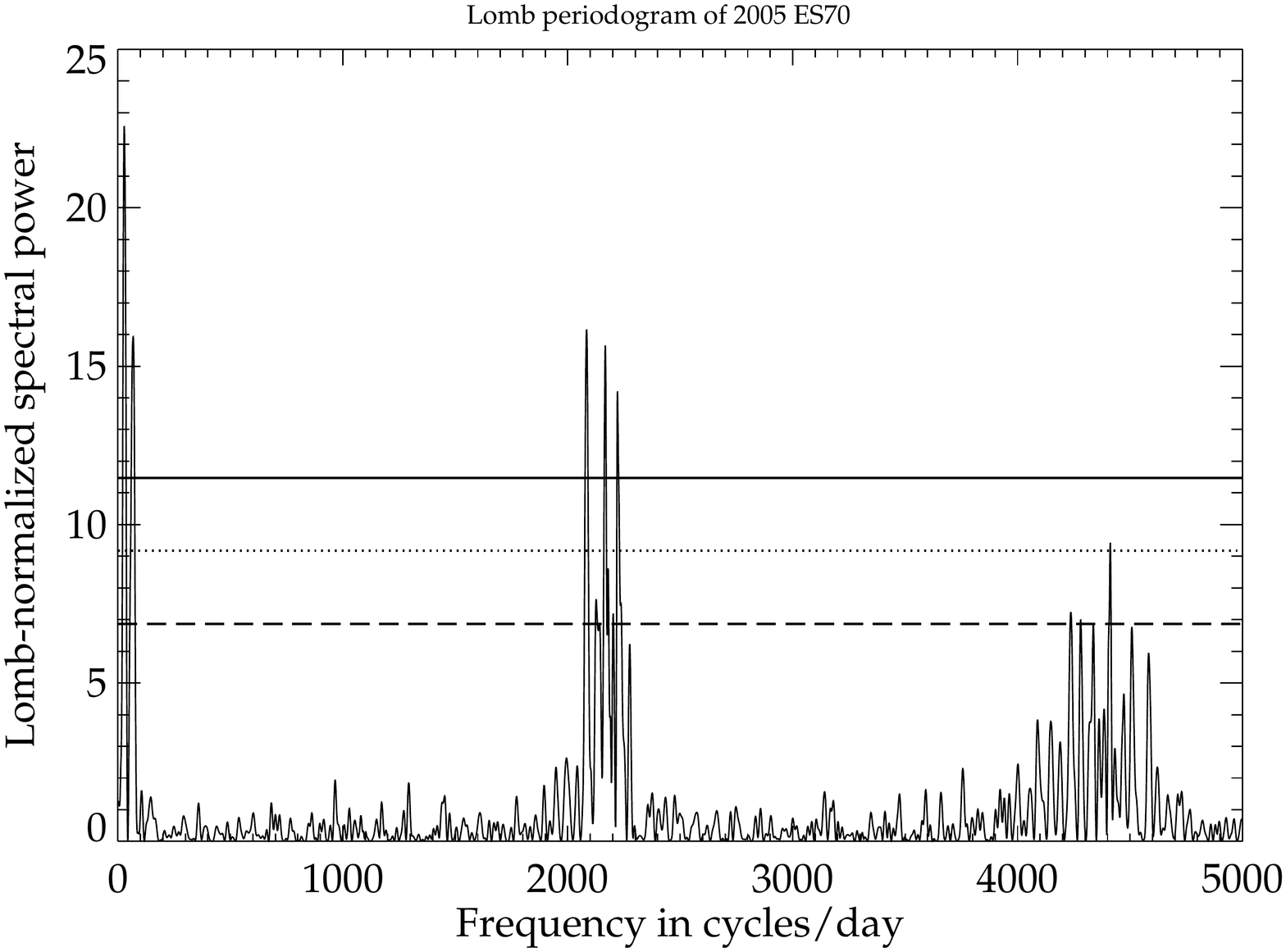}
\includegraphics[width=8cm, angle=0]{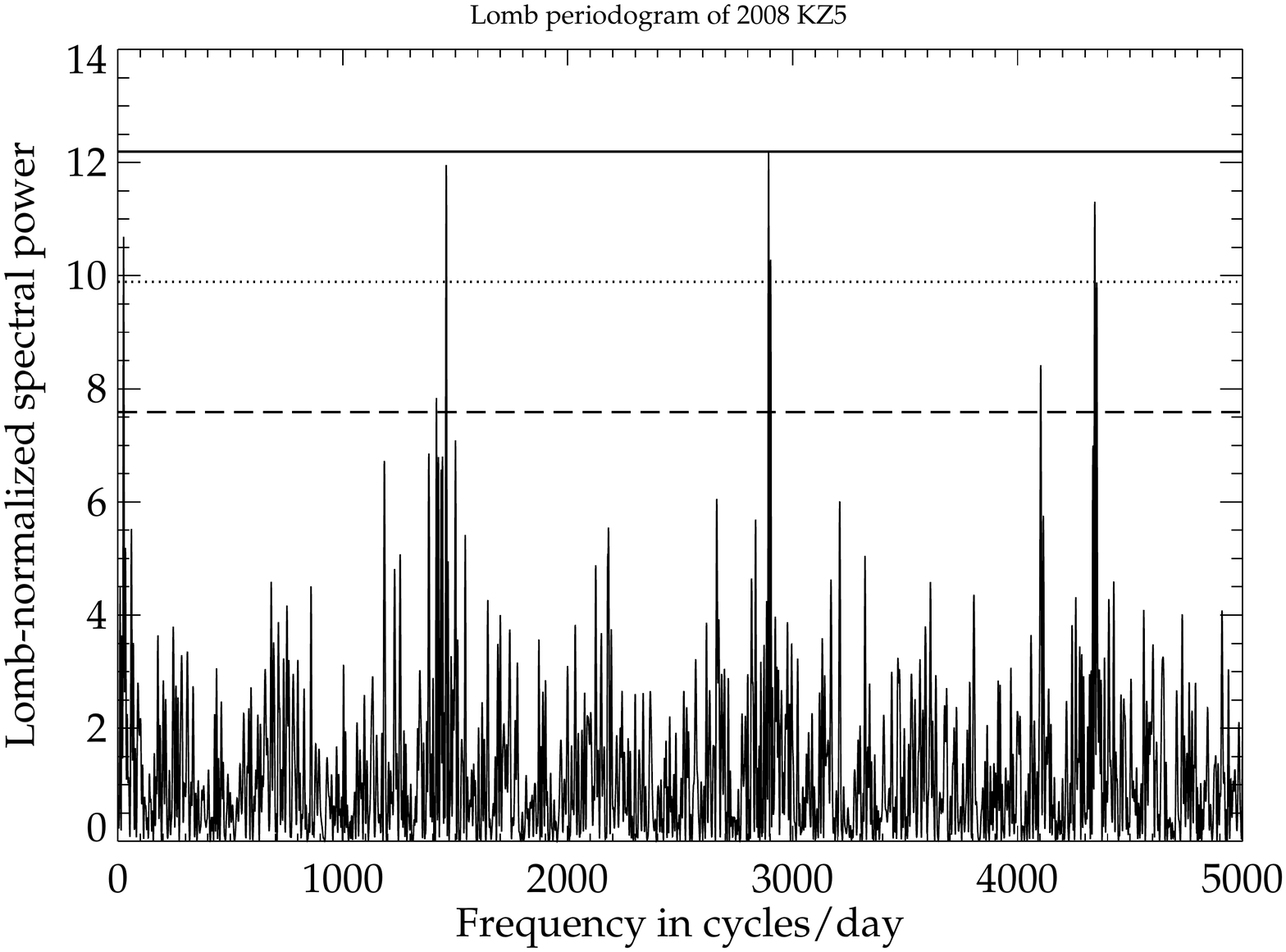}
\includegraphics[width=8cm, angle=0]{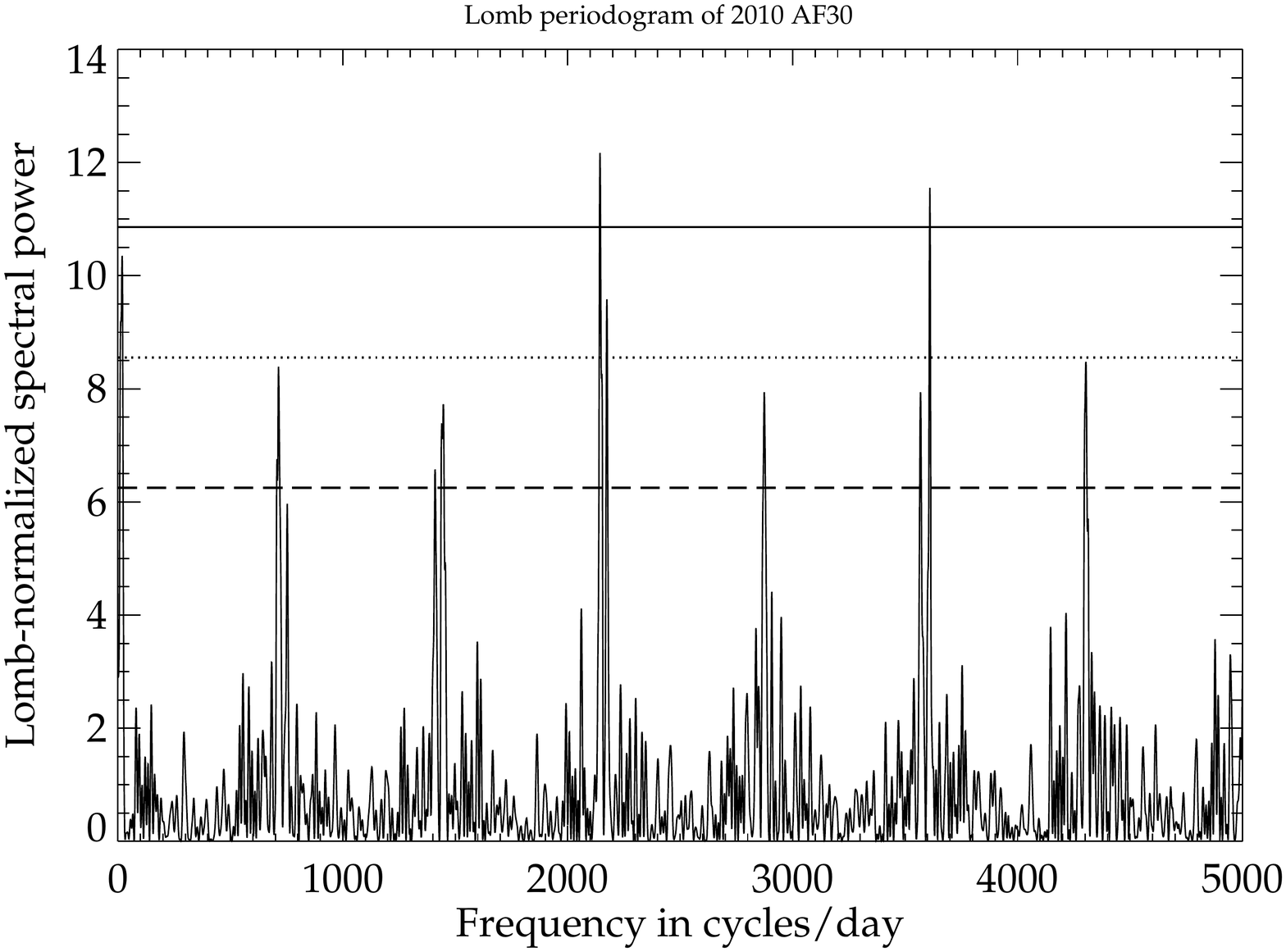}
\includegraphics[width=8cm, angle=0]{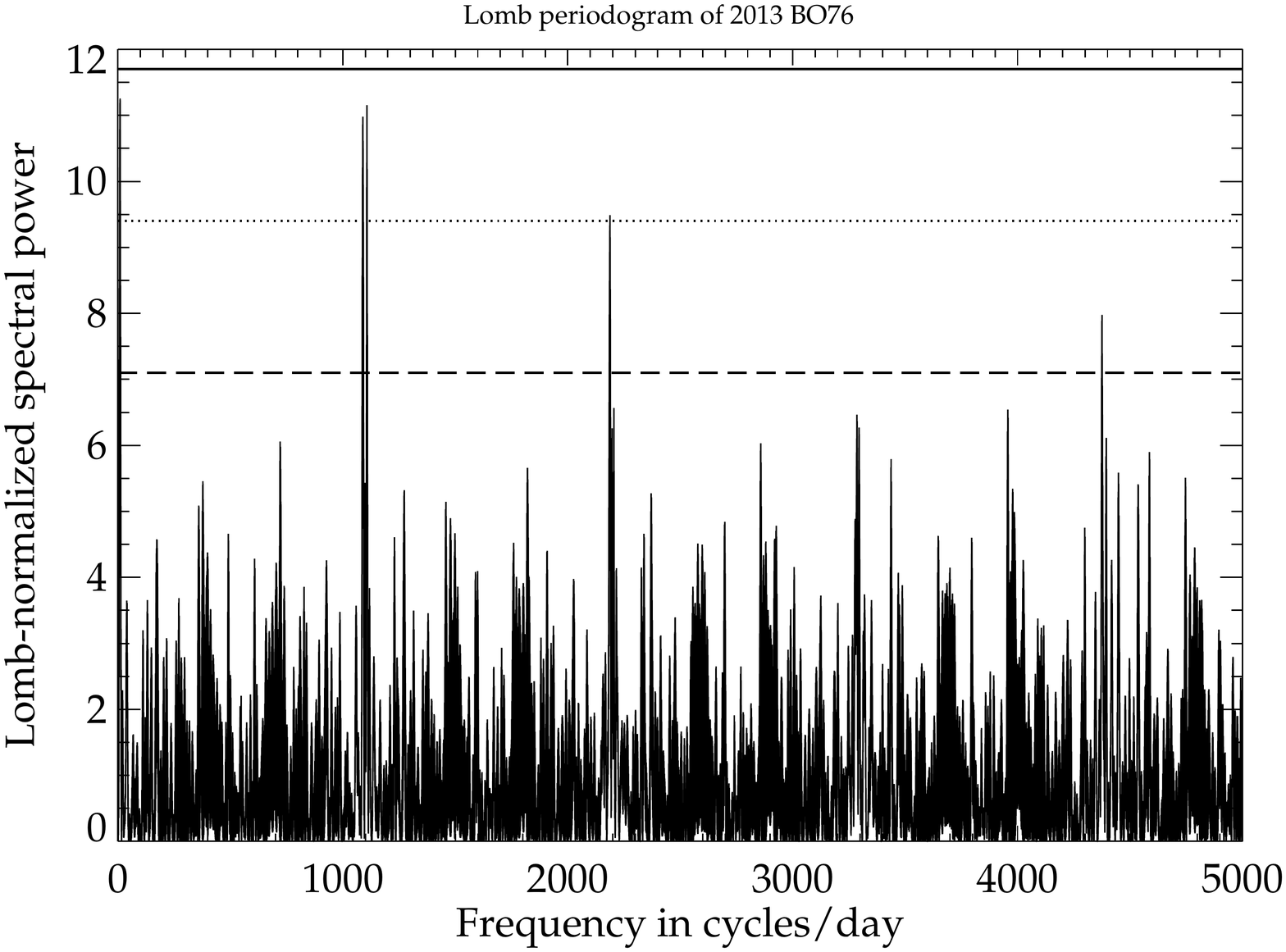}
\includegraphics[width=8cm, angle=0]{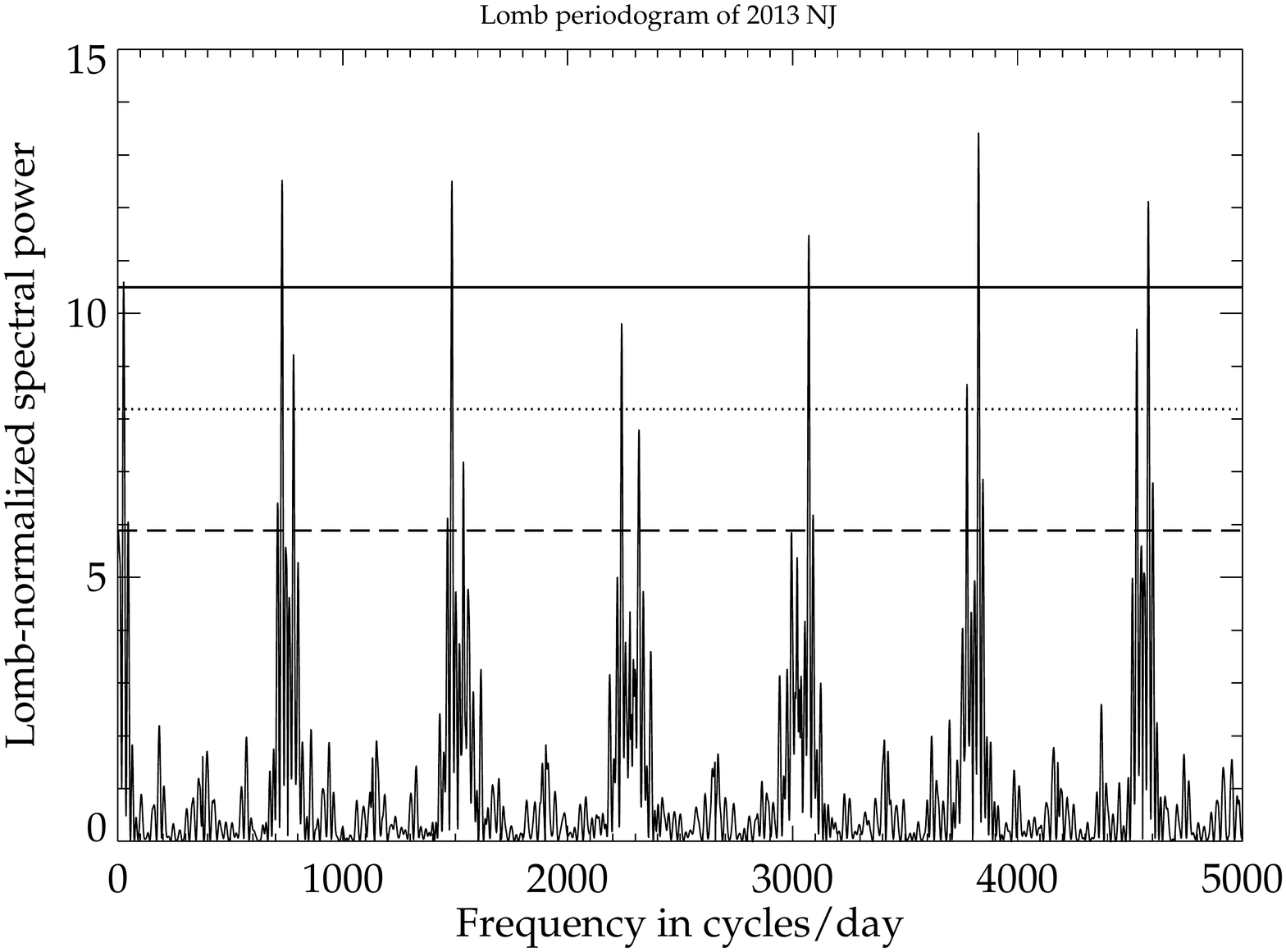}
\caption {\textit{Lomb-normalized spectral power versus Frequency}: Lomb periodograms of MANOS objects are plotted. Continuous line represents a 99.9$\%$ confidence level, dotted line a confidence level of 99$\%$, and the dashed line corresponds to a confidence level of 90$\%$.  }
\label{fig:Lomb1}
\end{figure}

\begin{figure}
\includegraphics[width=8cm, angle=0]{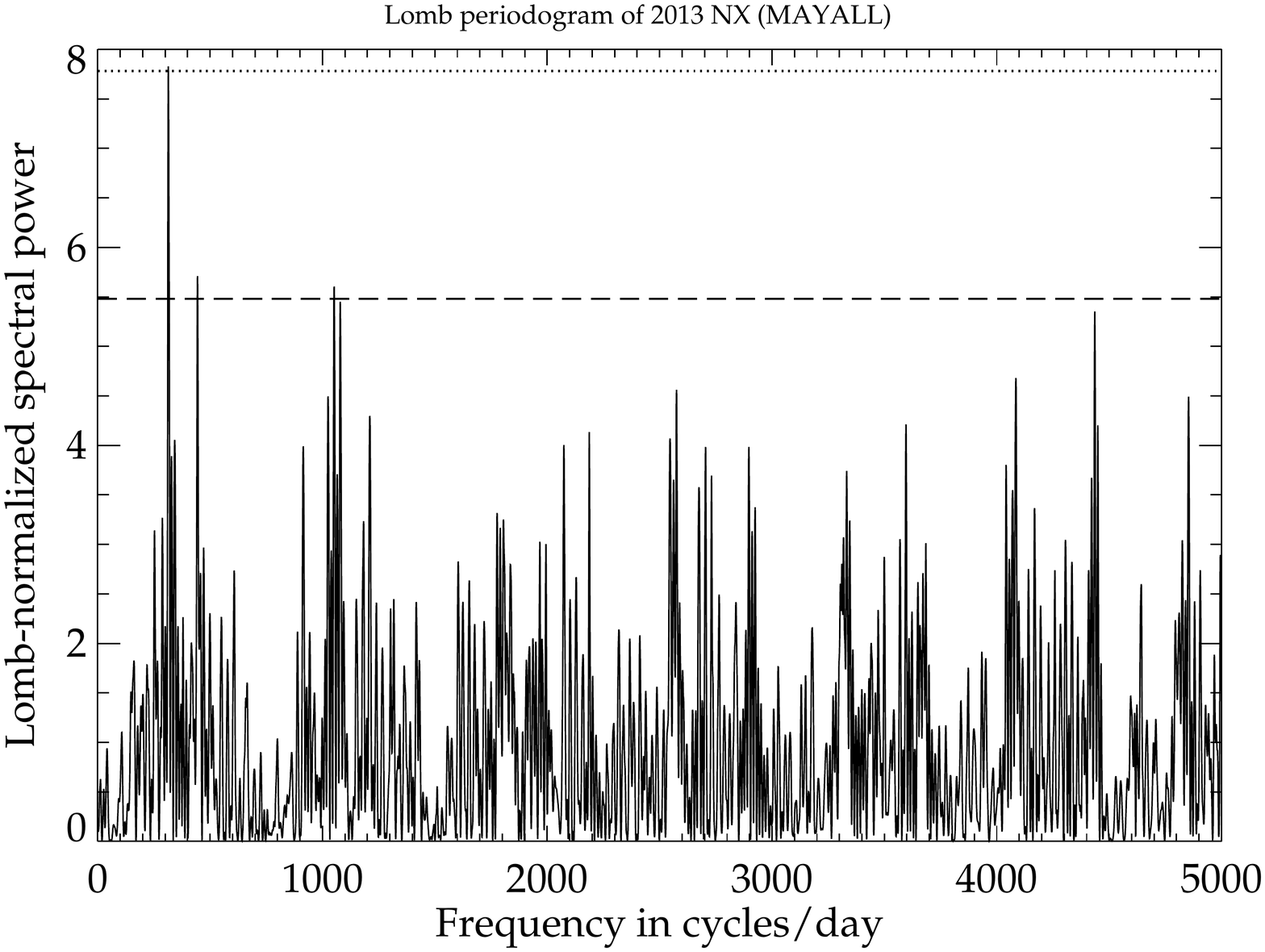}
\includegraphics[width=8cm, angle=0]{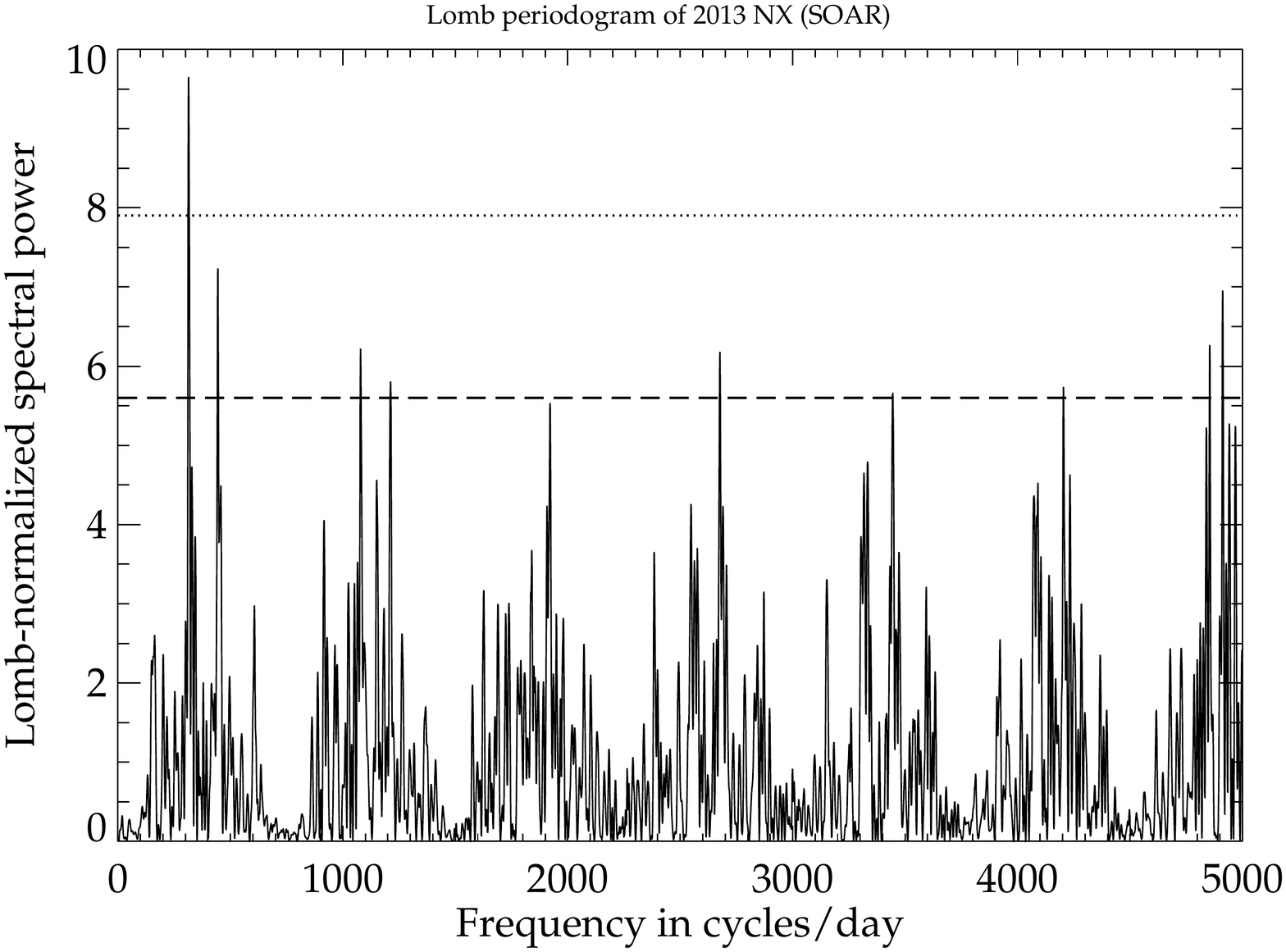}
\includegraphics[width=8cm, angle=0]{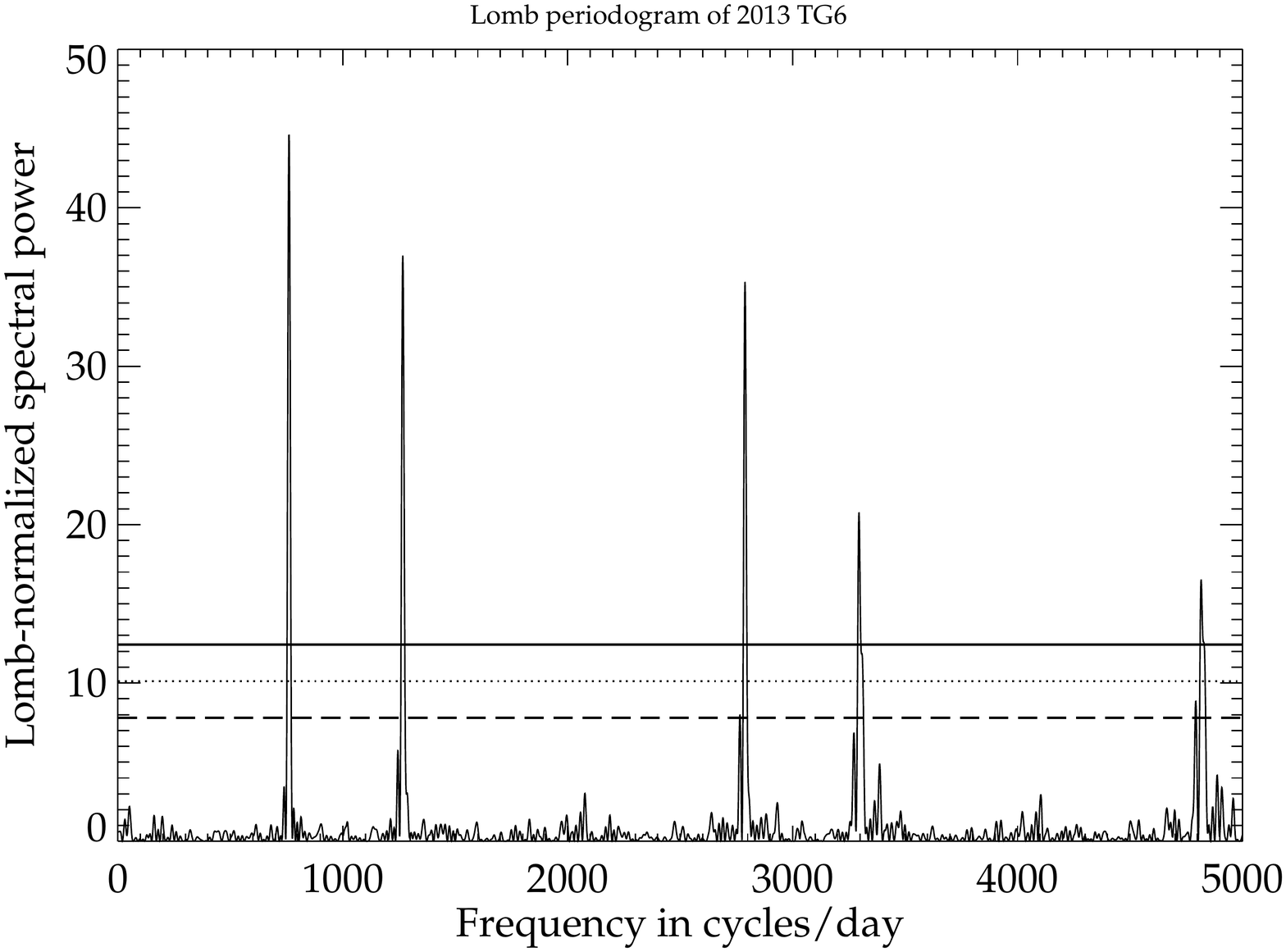}
\includegraphics[width=8cm, angle=0]{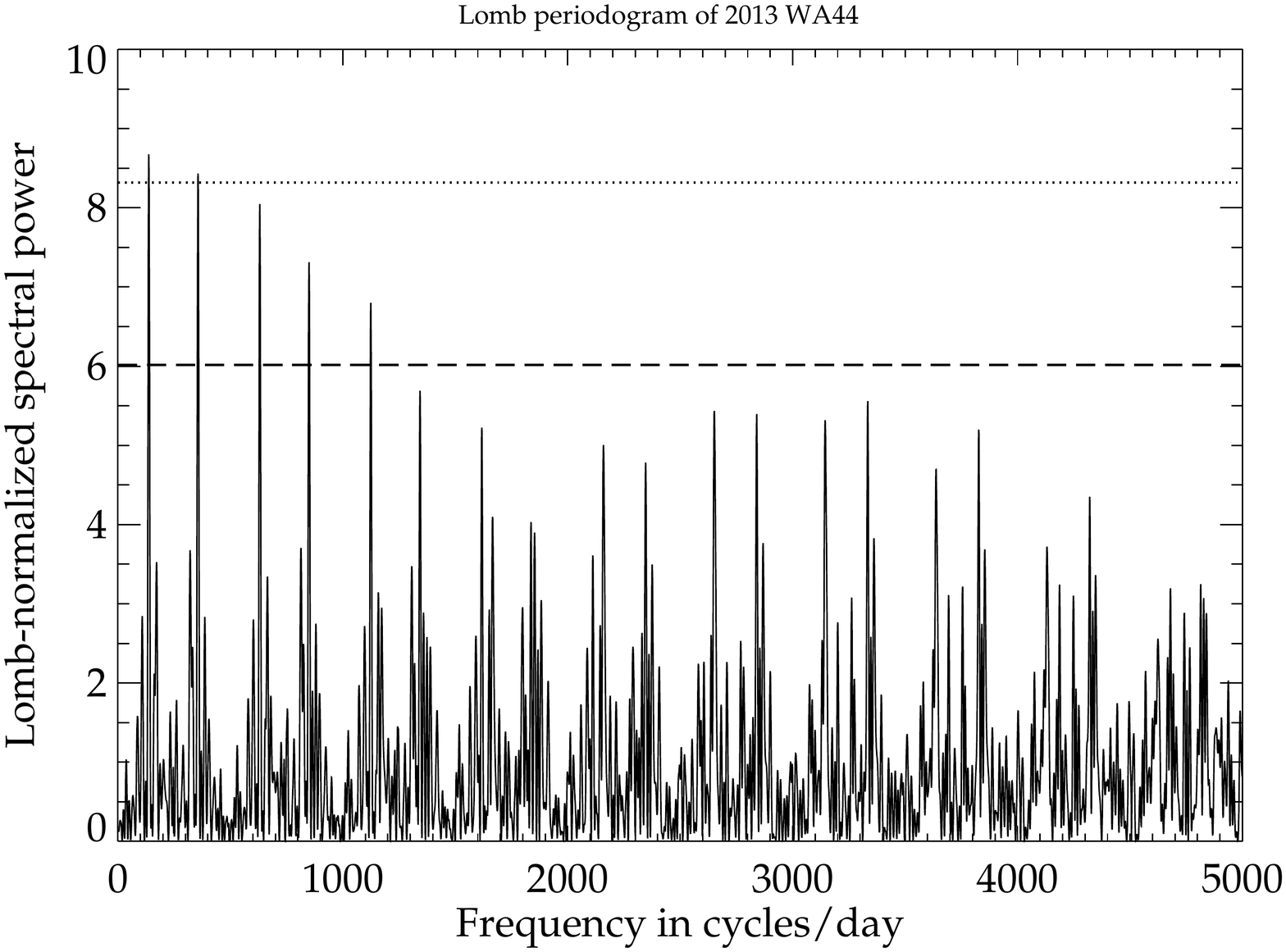}
\includegraphics[width=8cm, angle=0]{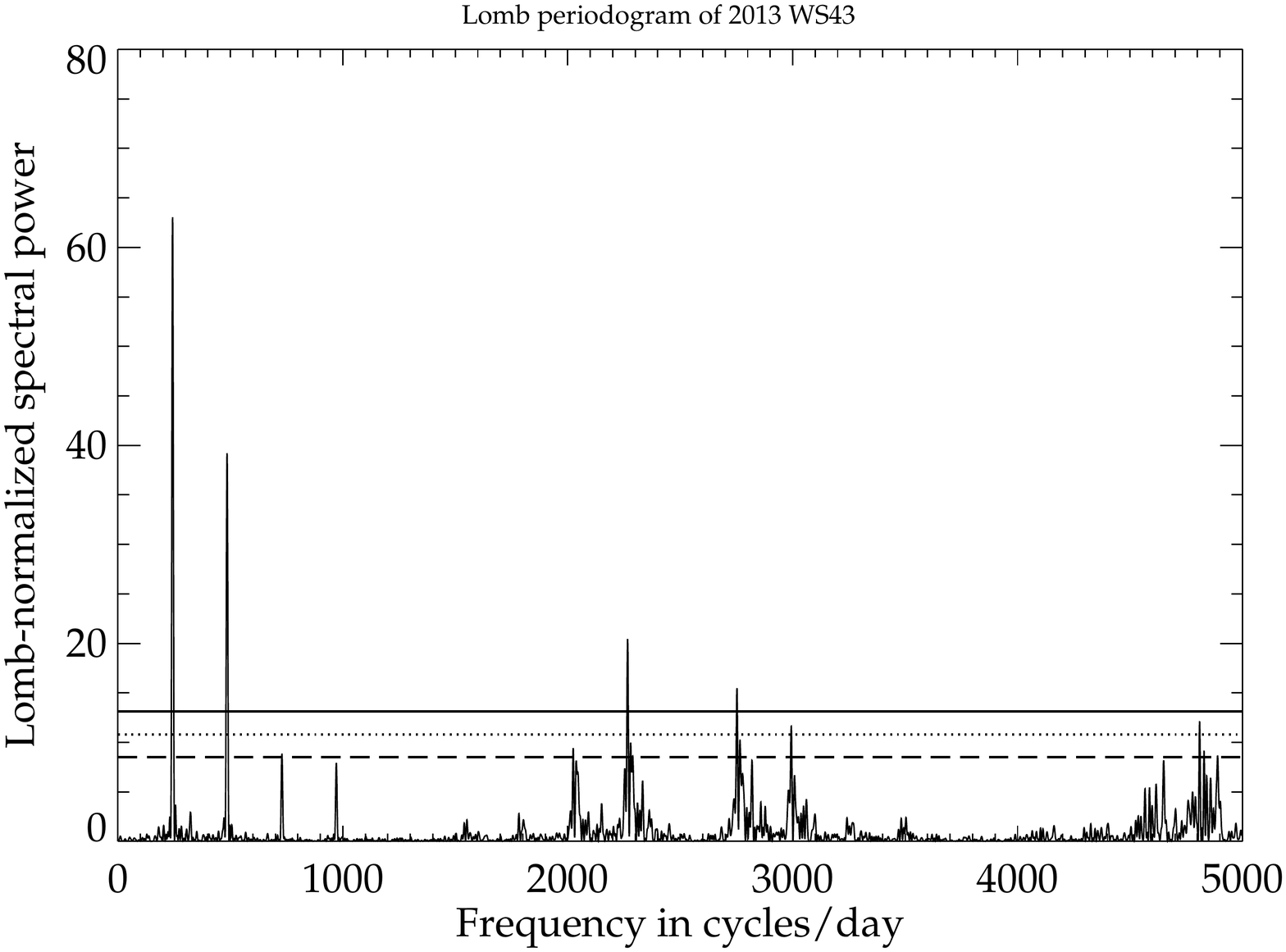}
\includegraphics[width=8cm, angle=0]{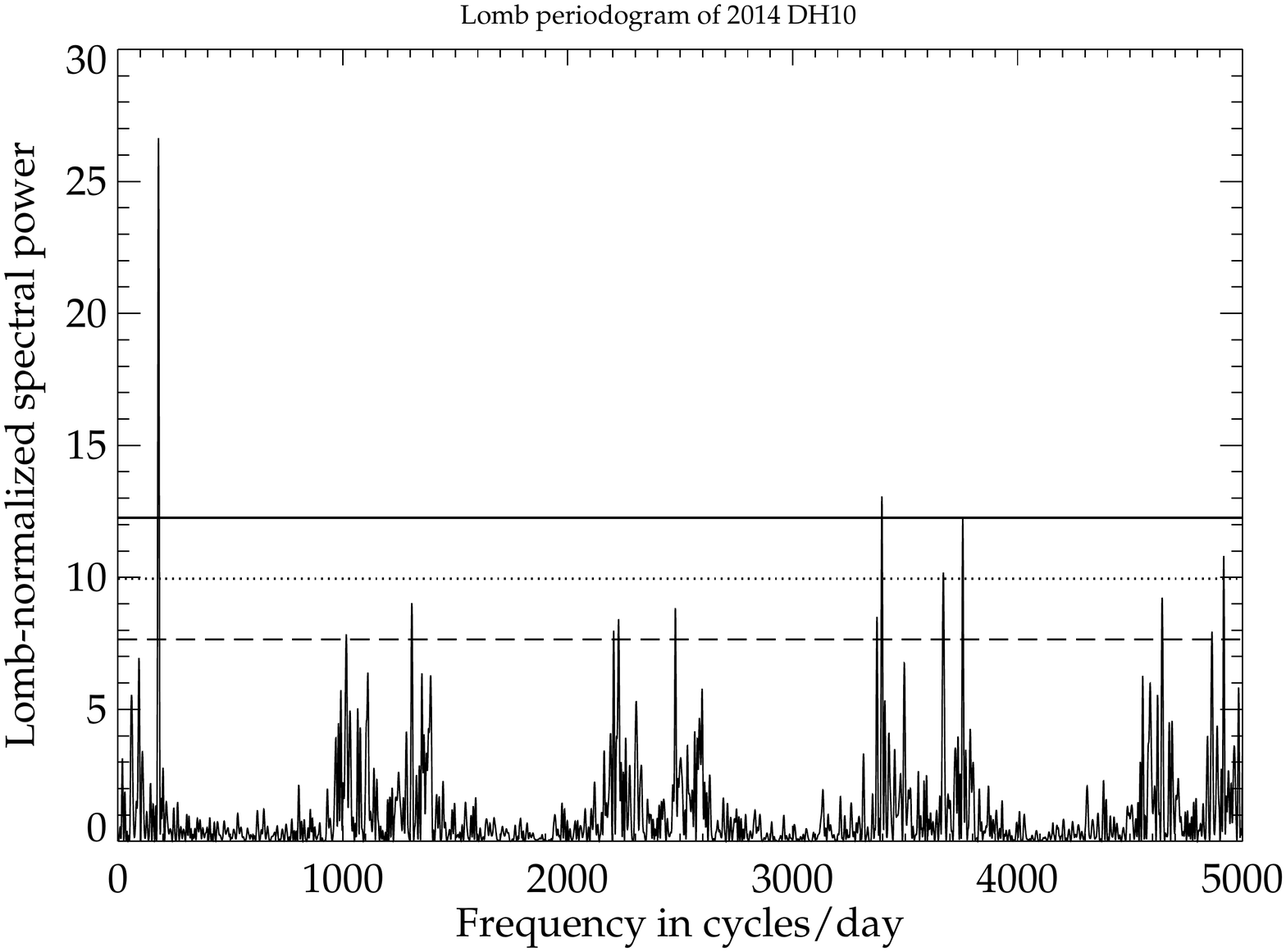}

\caption {\textit{Lomb-normalized spectral power versus Frequency}: Lomb periodograms of MANOS objects are plotted. Continuous line represents a 99.9$\%$ confidence level, dotted line a confidence level of 99$\%$, and the dashed line corresponds to a confidence level of 90$\%$.  }
\label{fig:Lomb2}
\end{figure}

\begin{figure}
\includegraphics[width=8cm, angle=0]{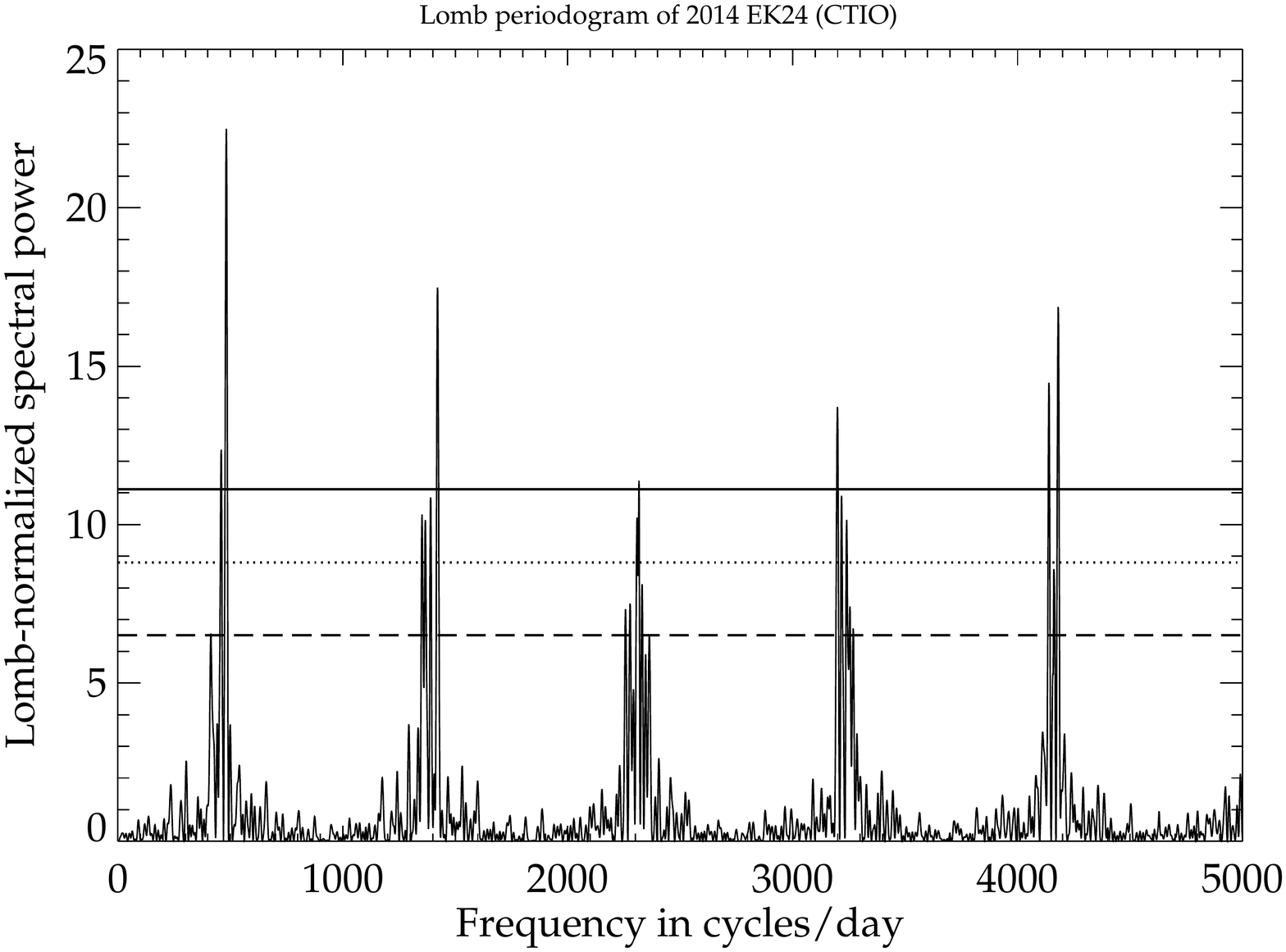}
\includegraphics[width=8cm, angle=0]{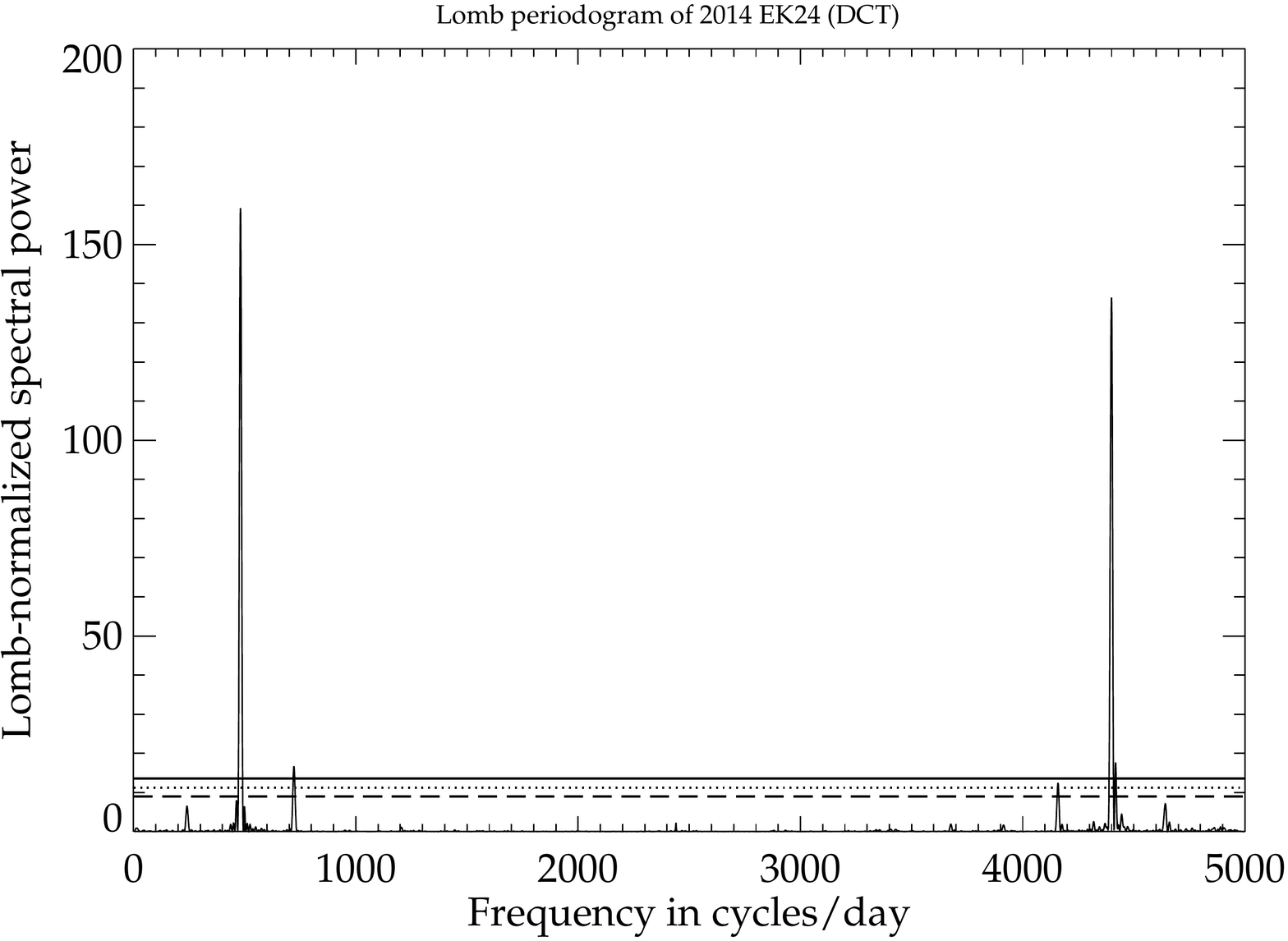}
\includegraphics[width=8cm, angle=0]{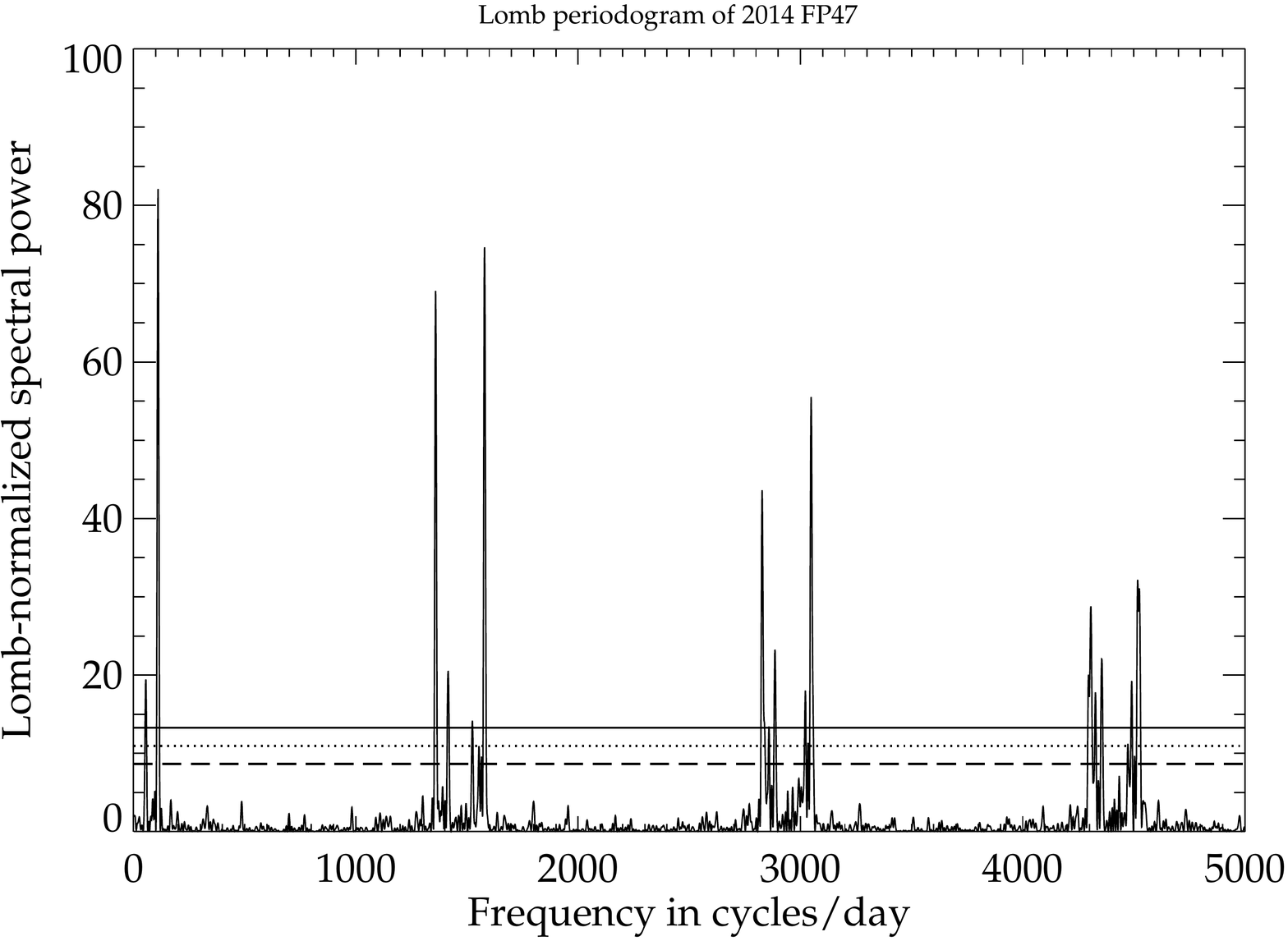}
\includegraphics[width=8cm, angle=0]{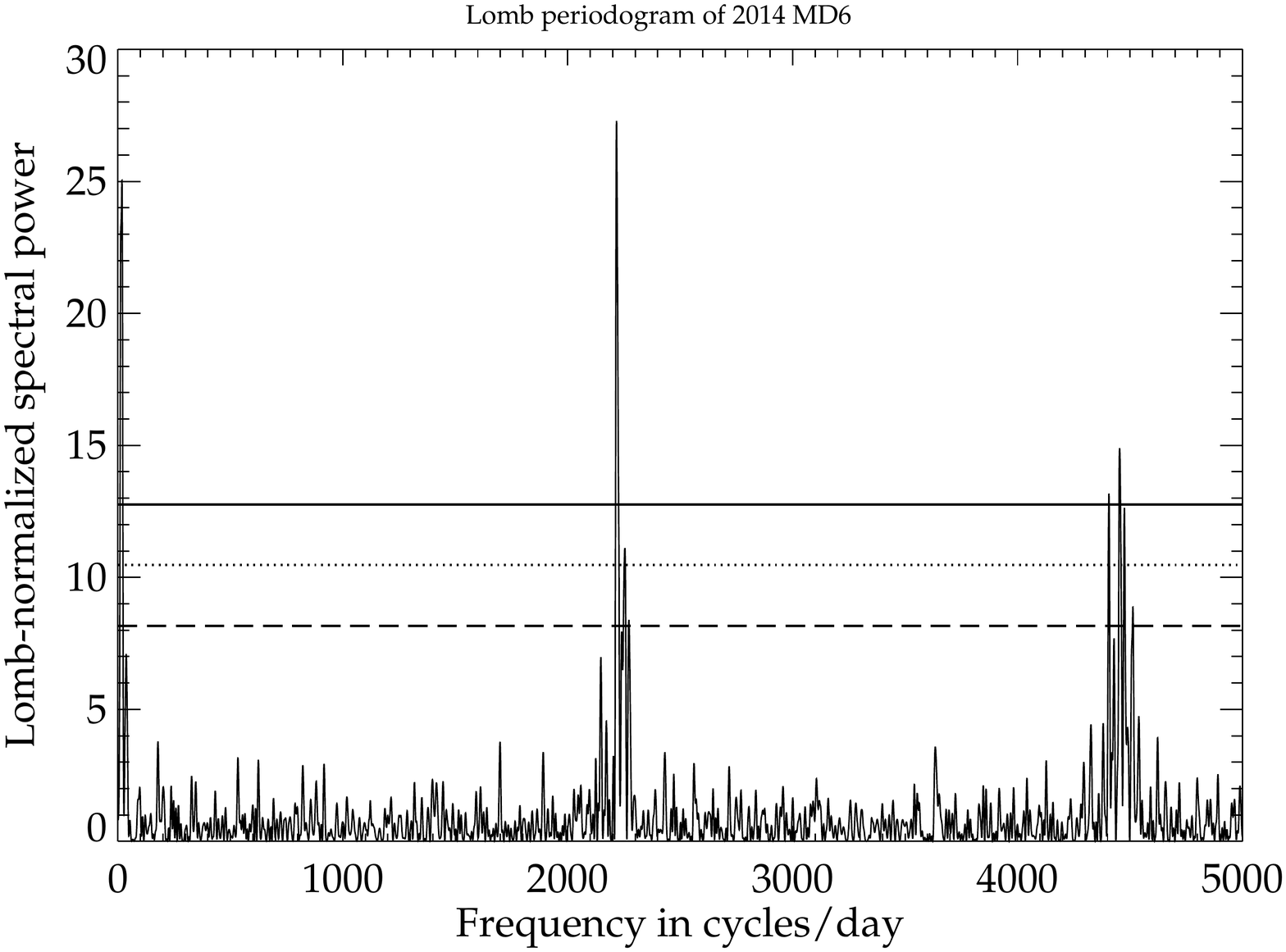}
\includegraphics[width=8cm, angle=0]{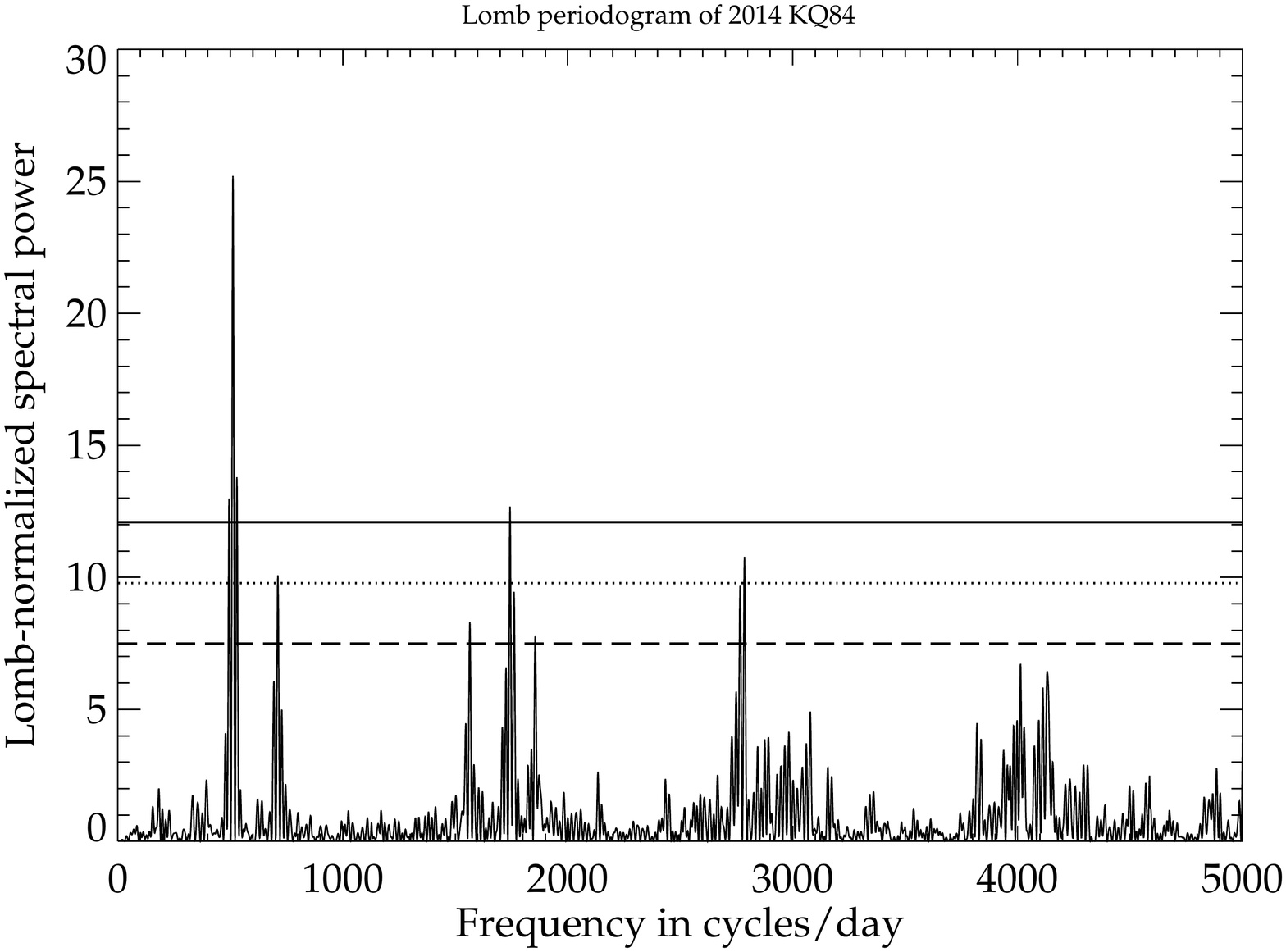}
\includegraphics[width=8cm, angle=0]{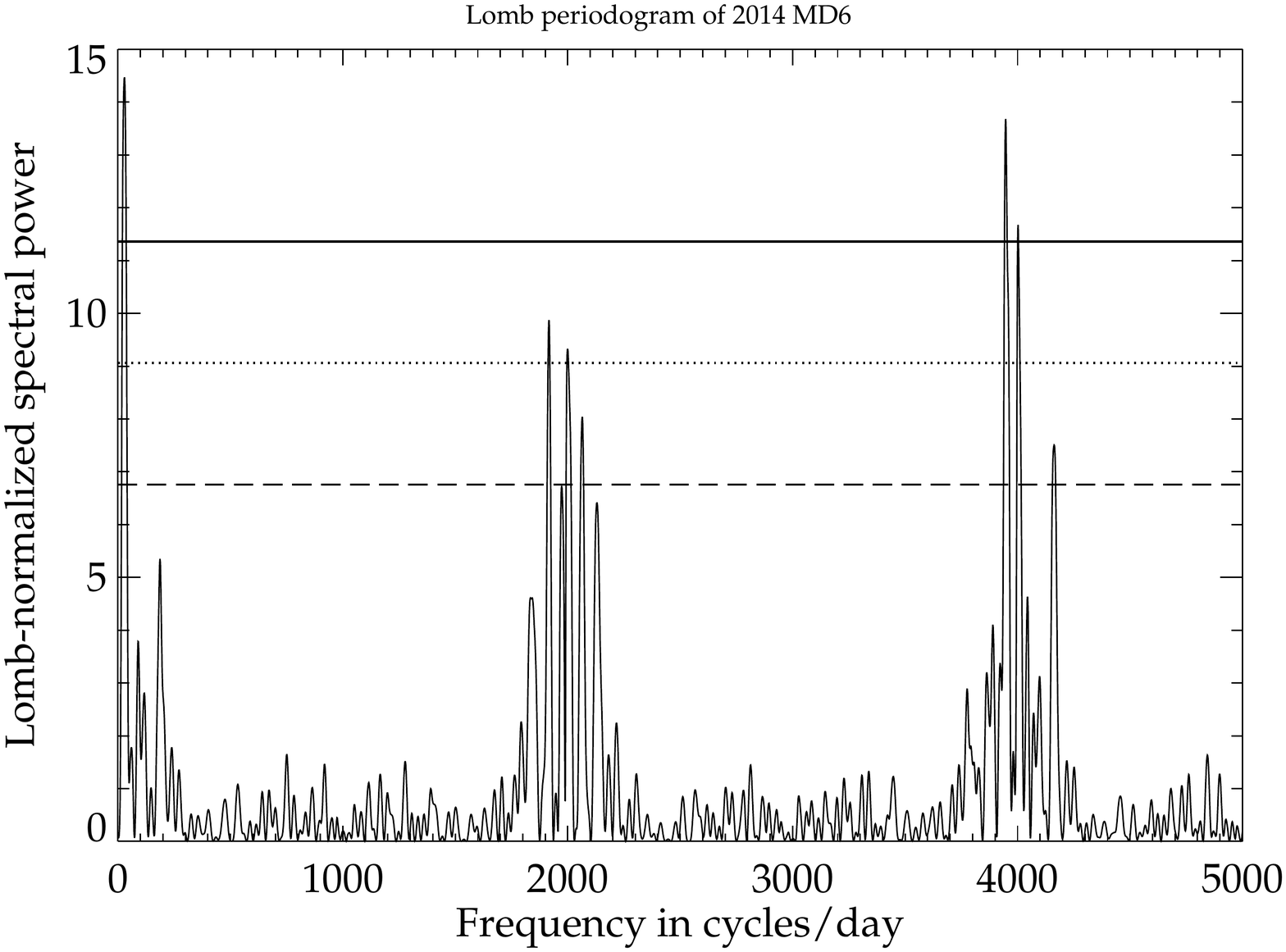}
\caption {\textit{Lomb-normalized spectral power versus Frequency}: Lomb periodograms of MANOS objects are plotted. Continuous line represents a 99.9$\%$ confidence level, dotted line a confidence level of 99$\%$, and the dashed line corresponds to a confidence level of 90$\%$.  }
\label{fig:Lomb3}
\end{figure}

\begin{figure}
\includegraphics[width=8cm, angle=0]{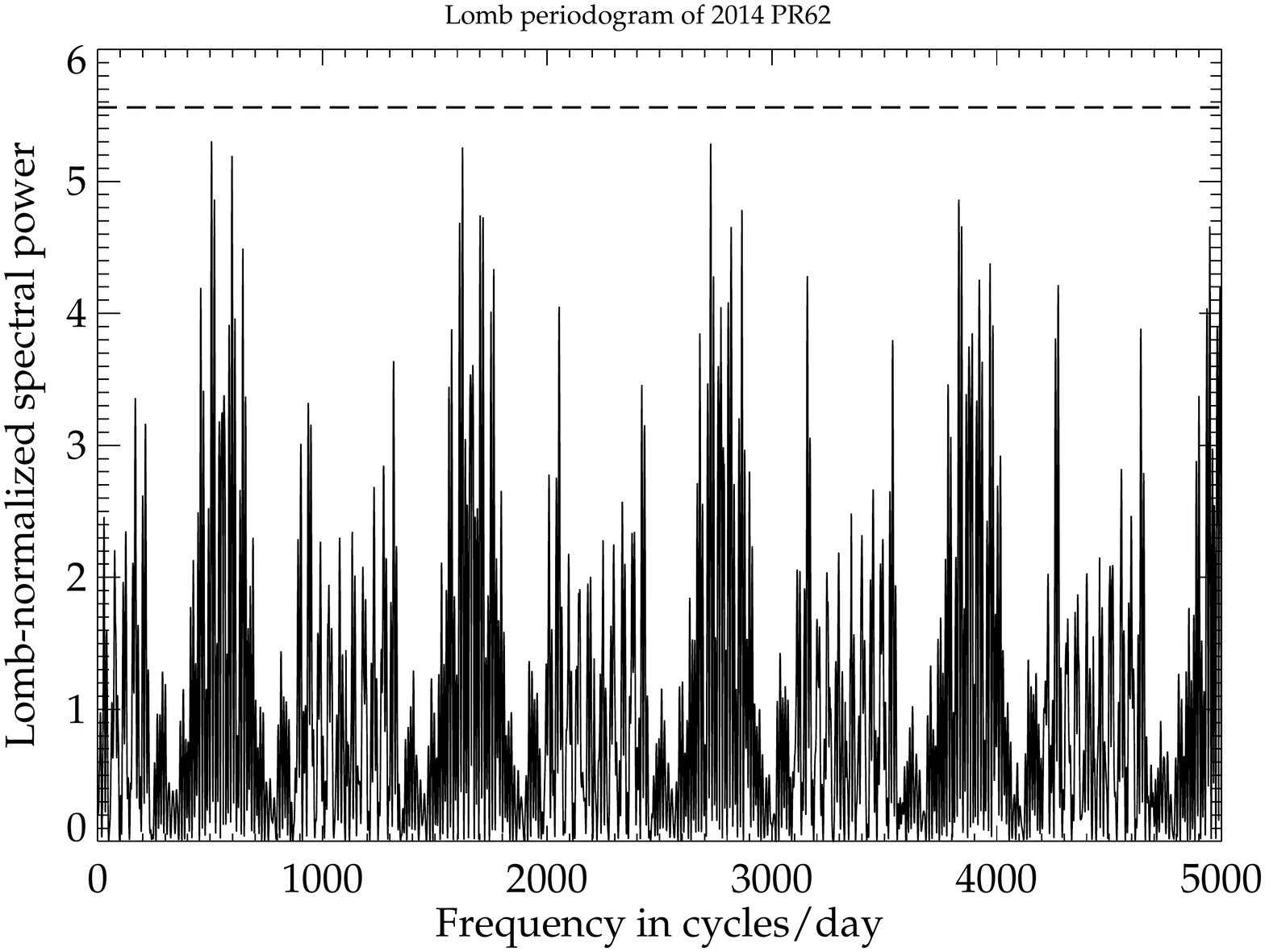}
\includegraphics[width=8cm, angle=0]{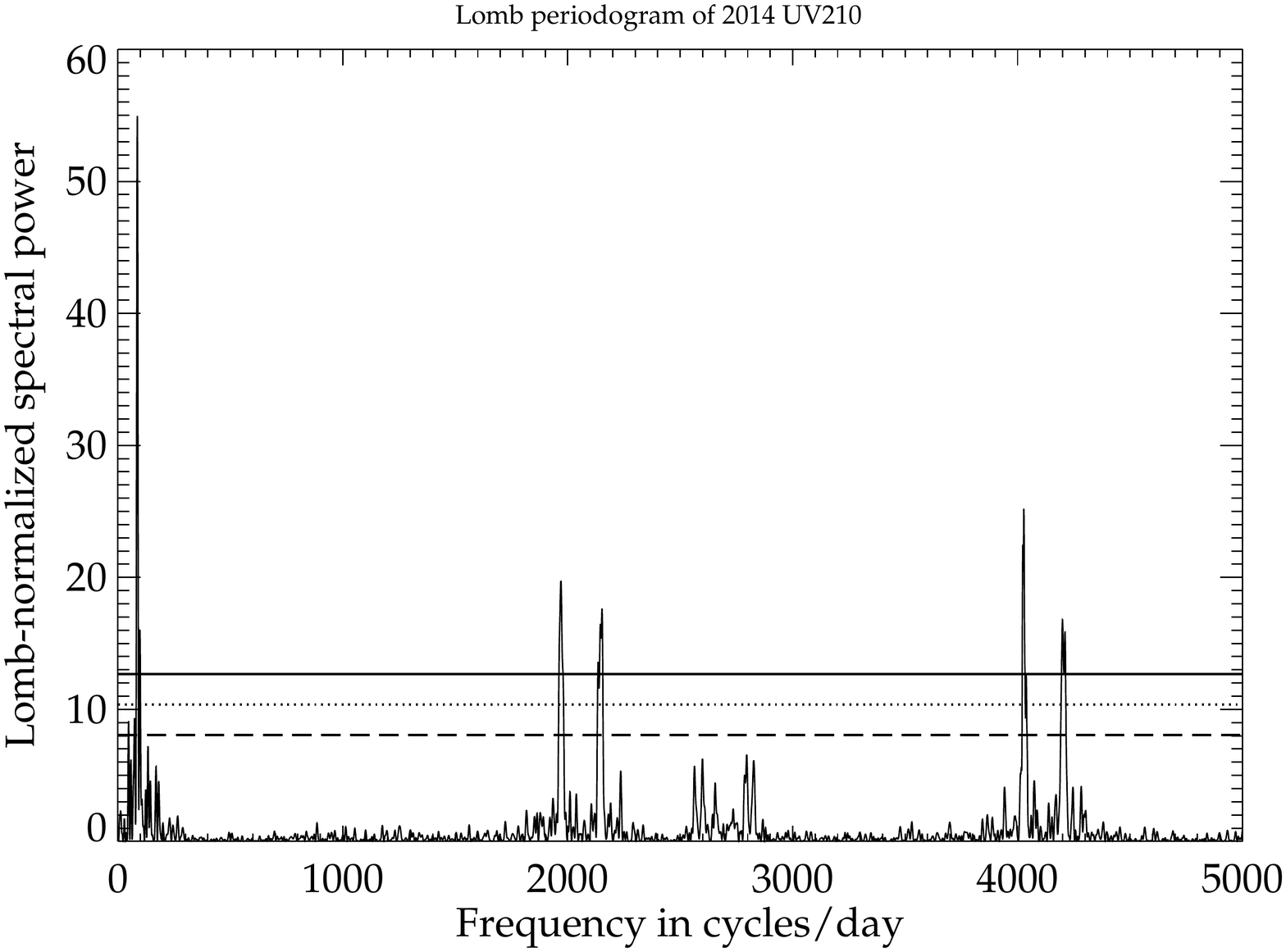}
\includegraphics[width=8cm, angle=0]{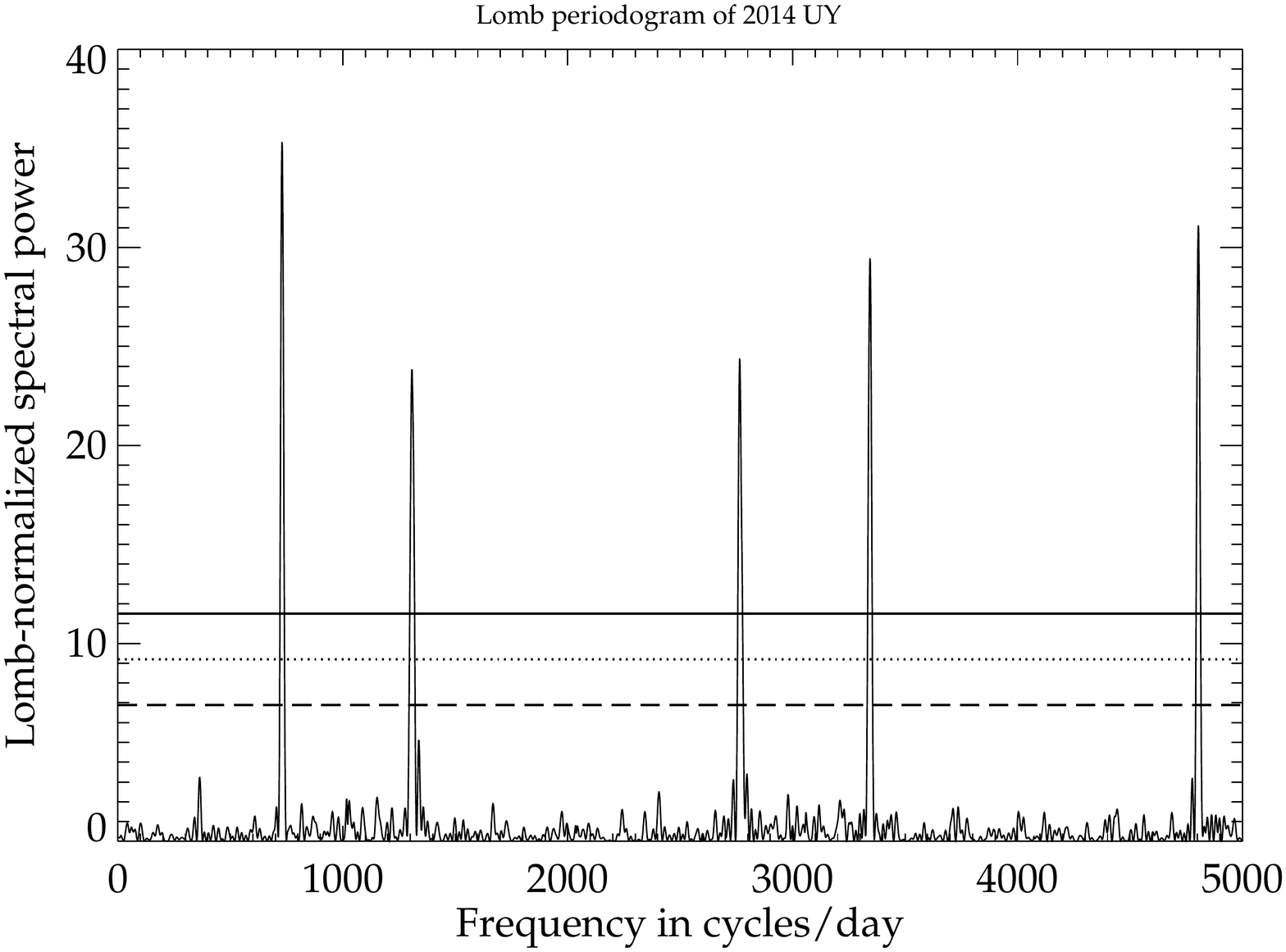}
\includegraphics[width=8cm, angle=0]{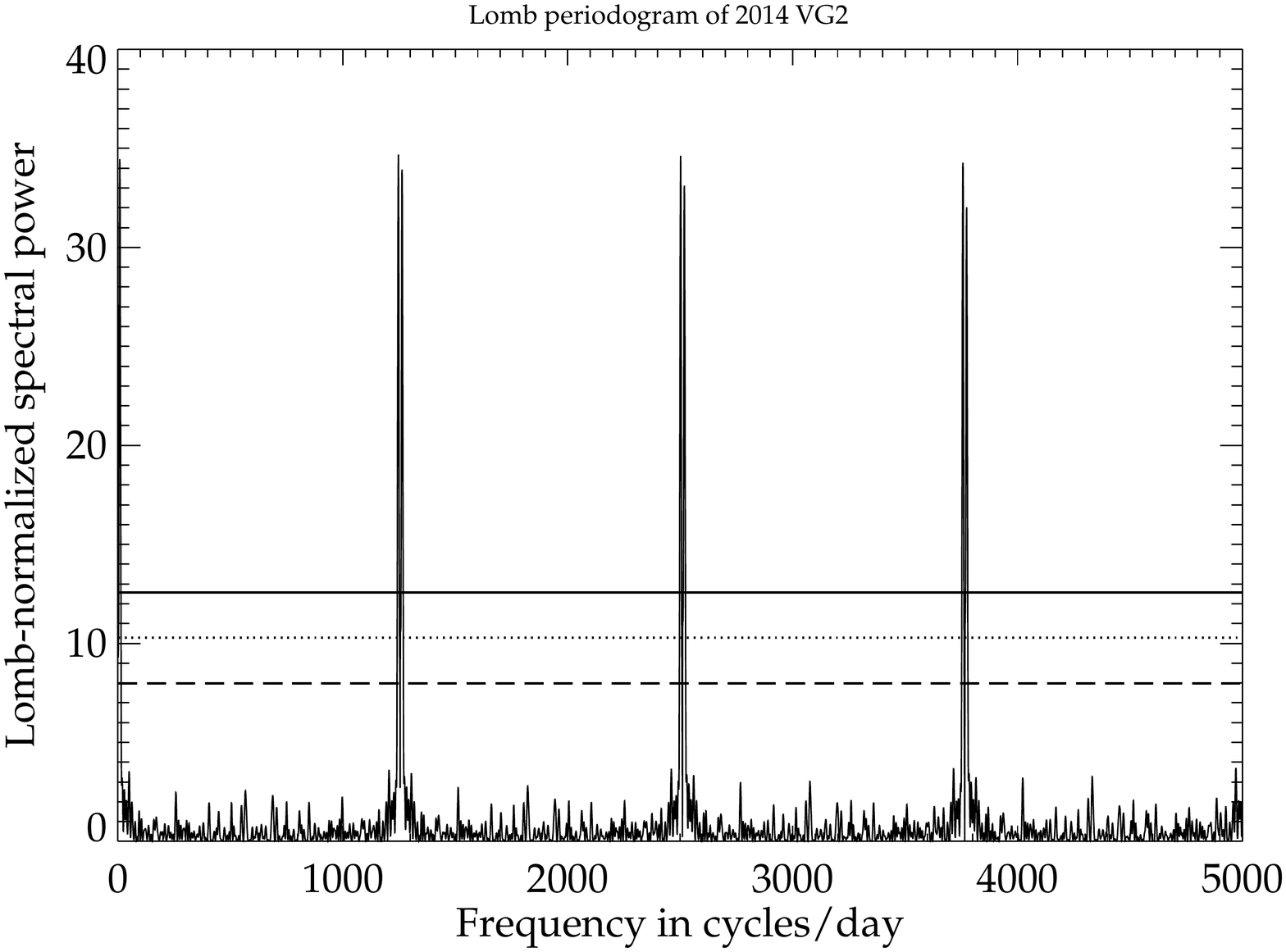}
\includegraphics[width=8cm, angle=0]{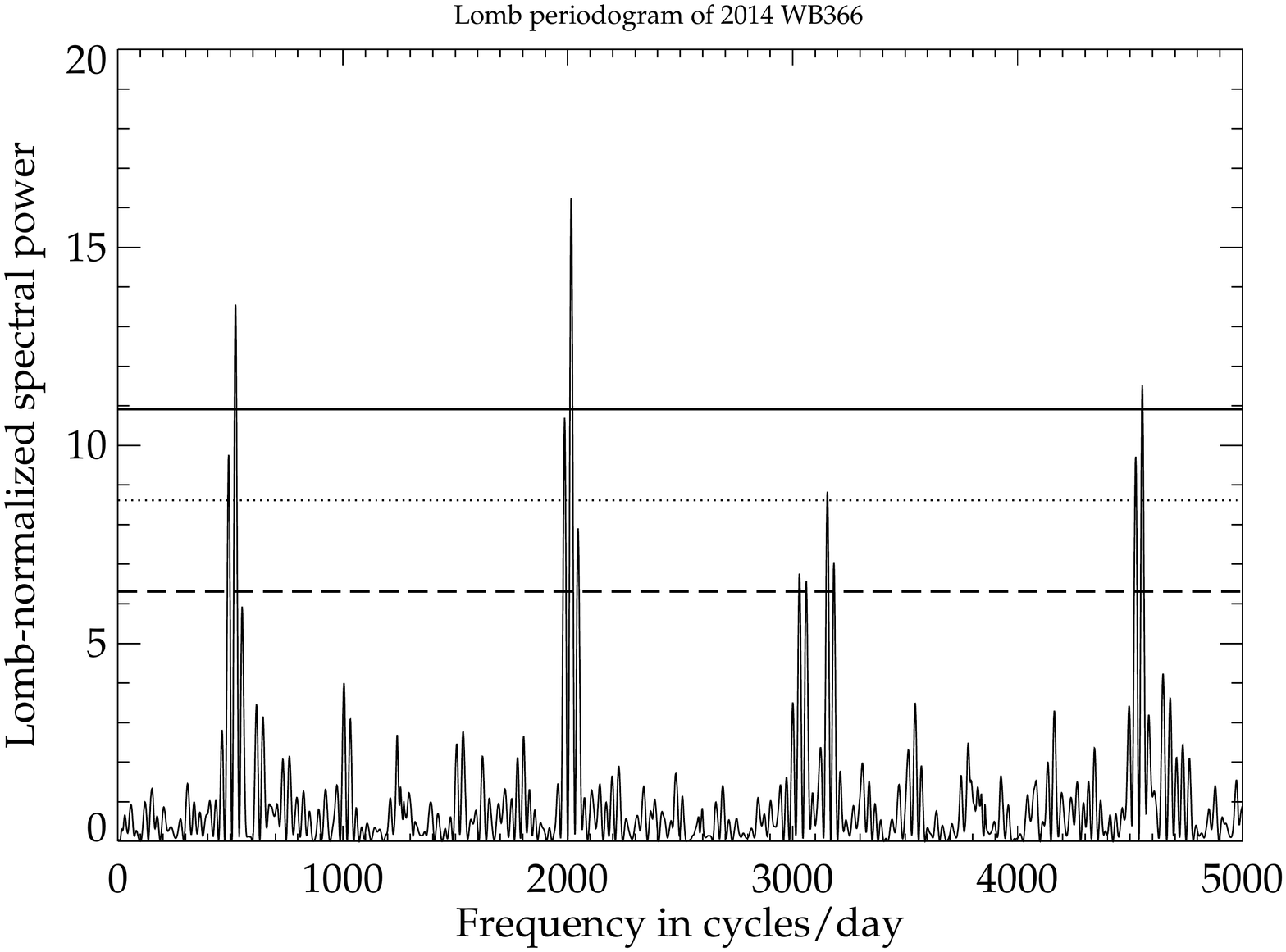}
\includegraphics[width=8cm, angle=0]{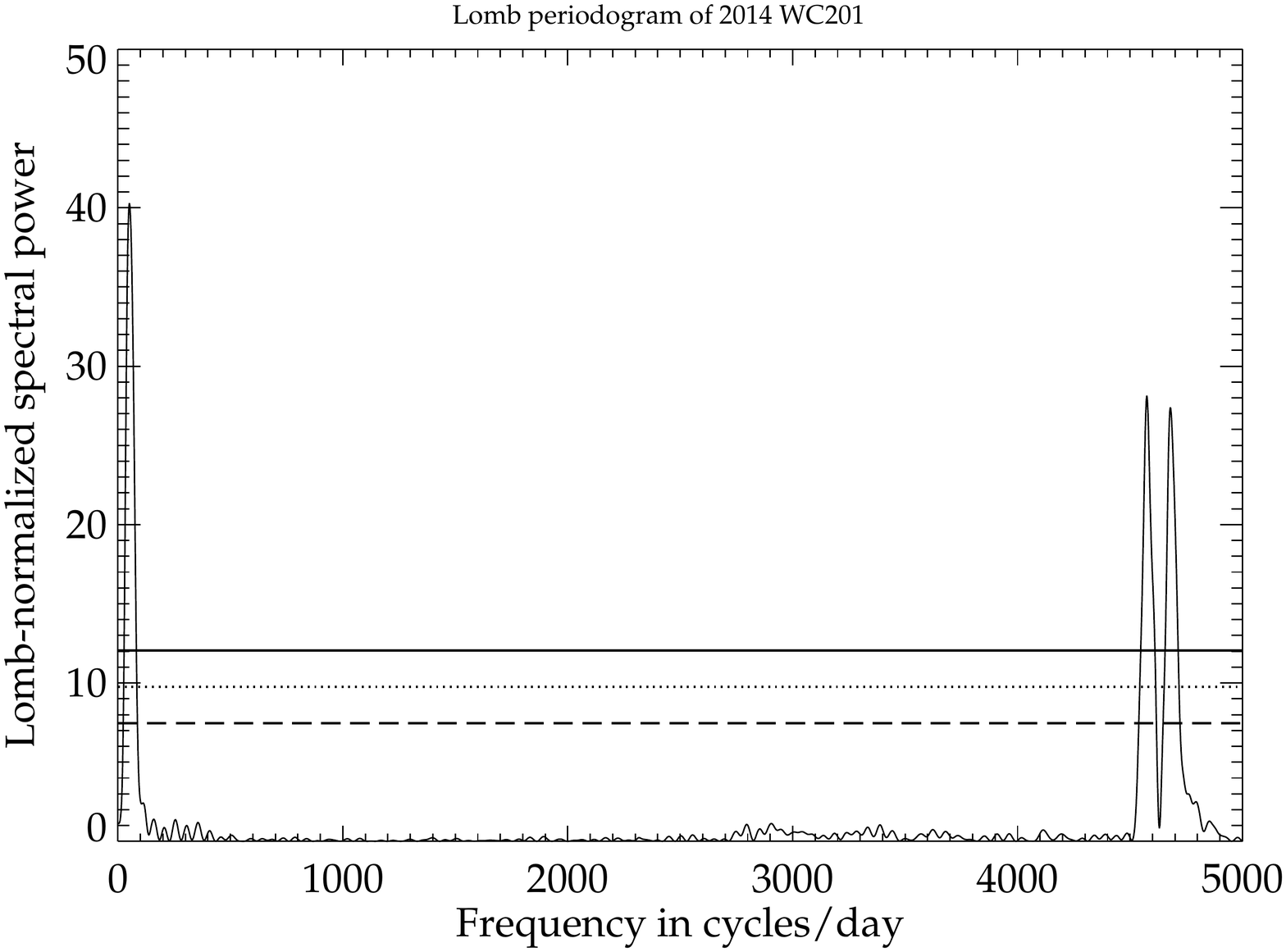}
\caption {\textit{Lomb-normalized spectral power versus Frequency}: Lomb periodograms of MANOS objects are plotted. Continuous line represents a 99.9$\%$ confidence level, dotted line a confidence level of 99$\%$, and the dashed line corresponds to a confidence level of 90$\%$.  }
\label{fig:Lomb4}
\end{figure}

\begin{figure}
\includegraphics[width=8cm, angle=0]{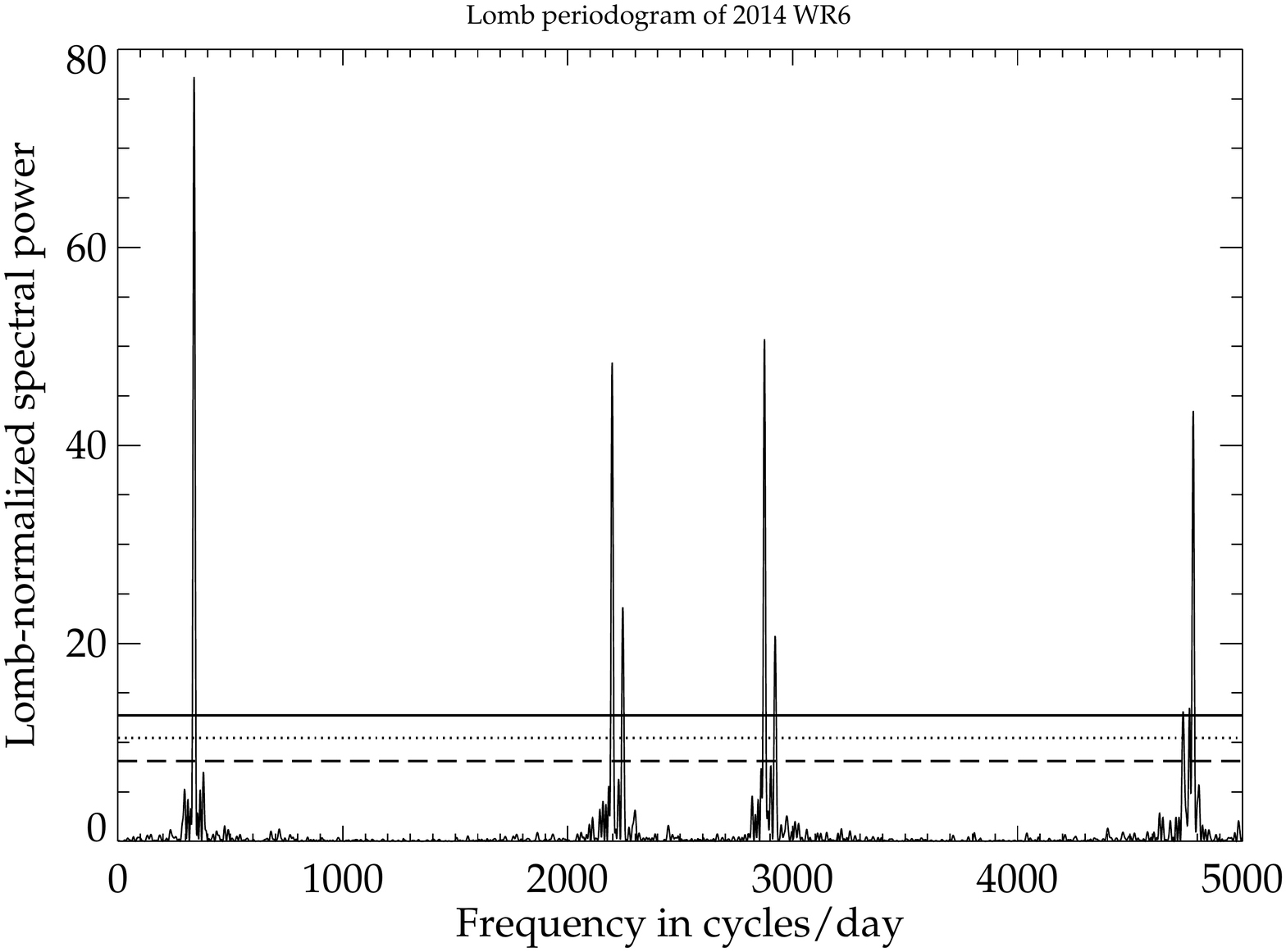}
\includegraphics[width=8cm, angle=0]{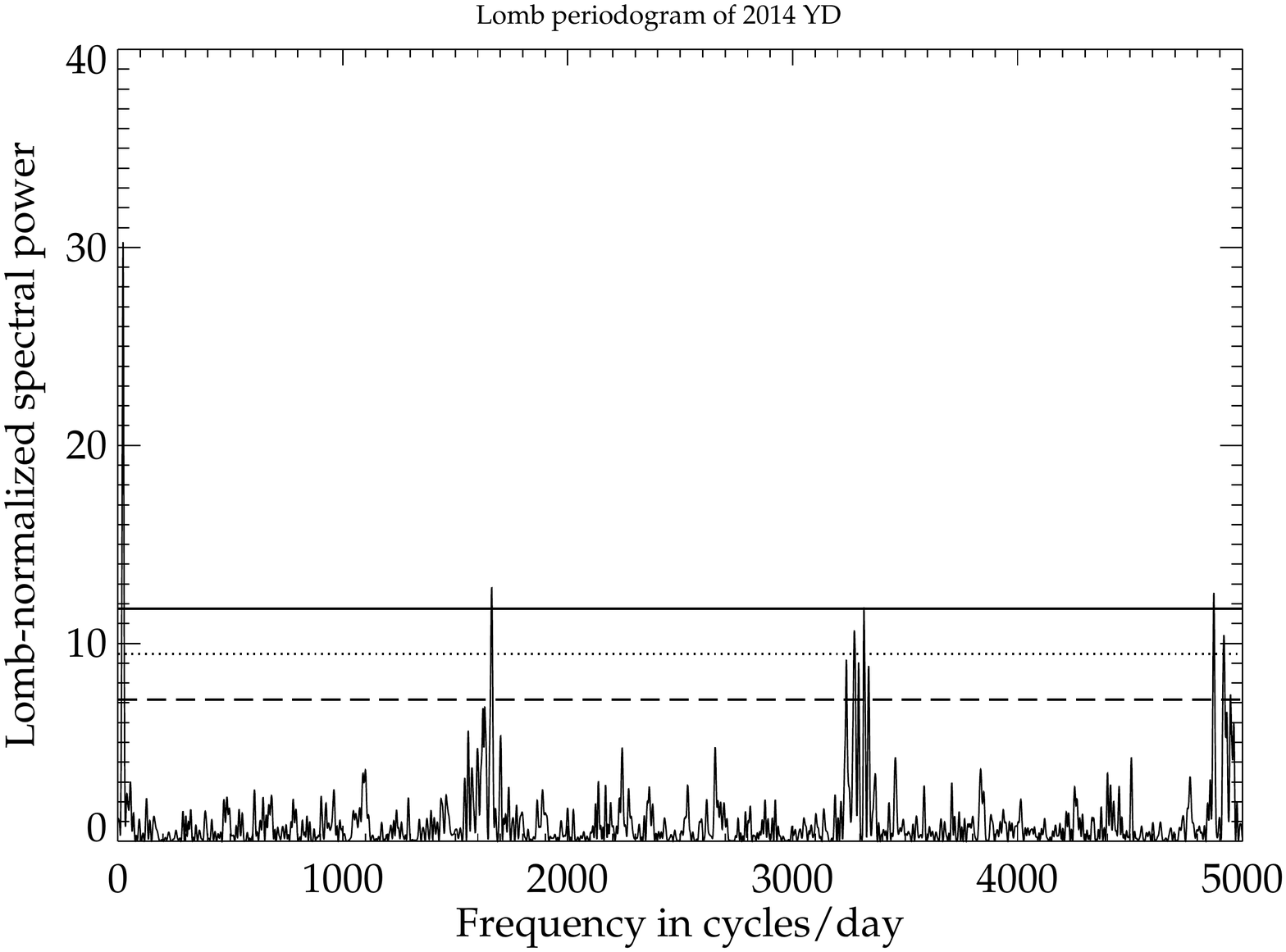}
\includegraphics[width=8cm, angle=0]{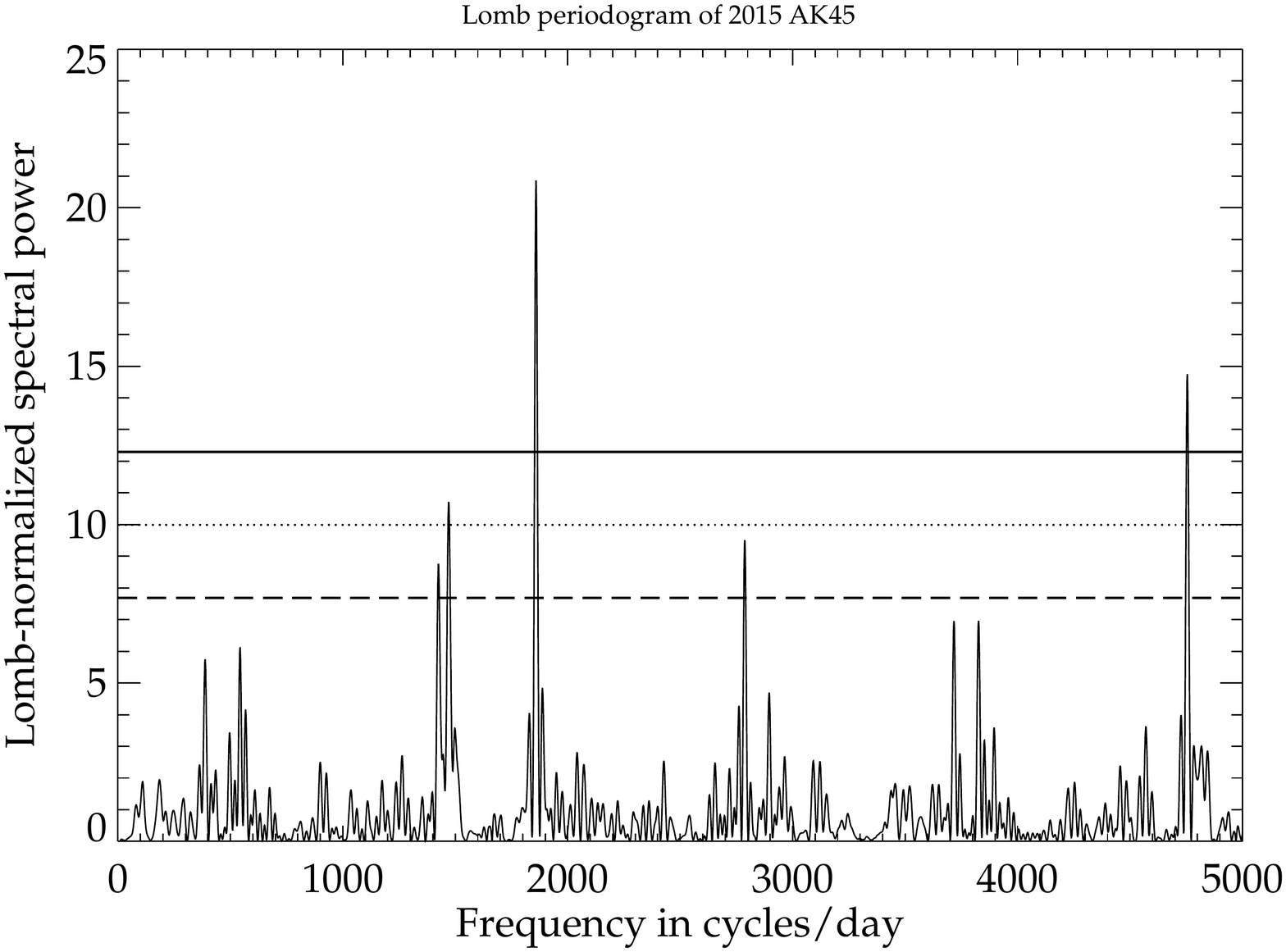}
\includegraphics[width=8cm, angle=0]{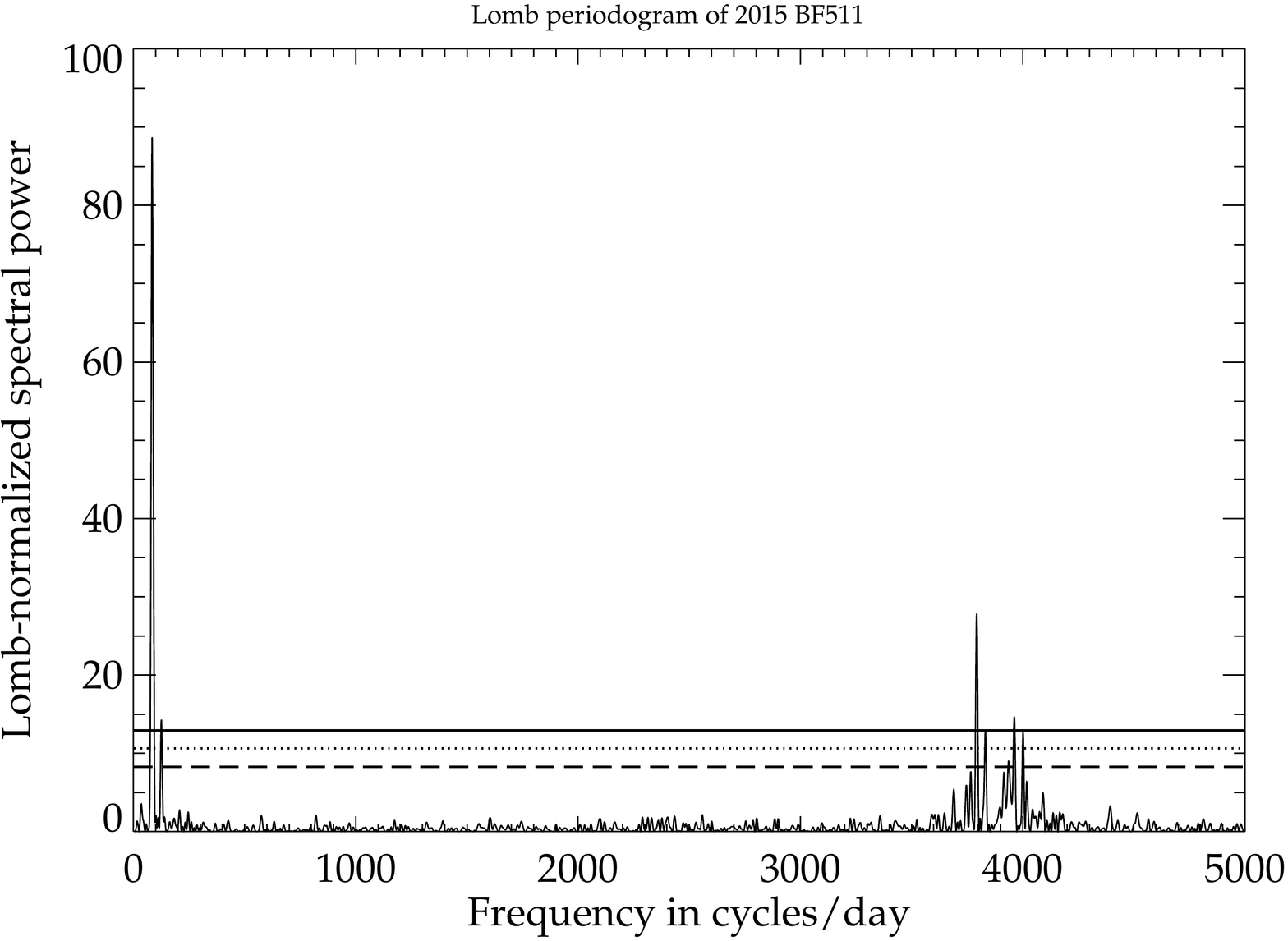}
\includegraphics[width=8cm, angle=0]{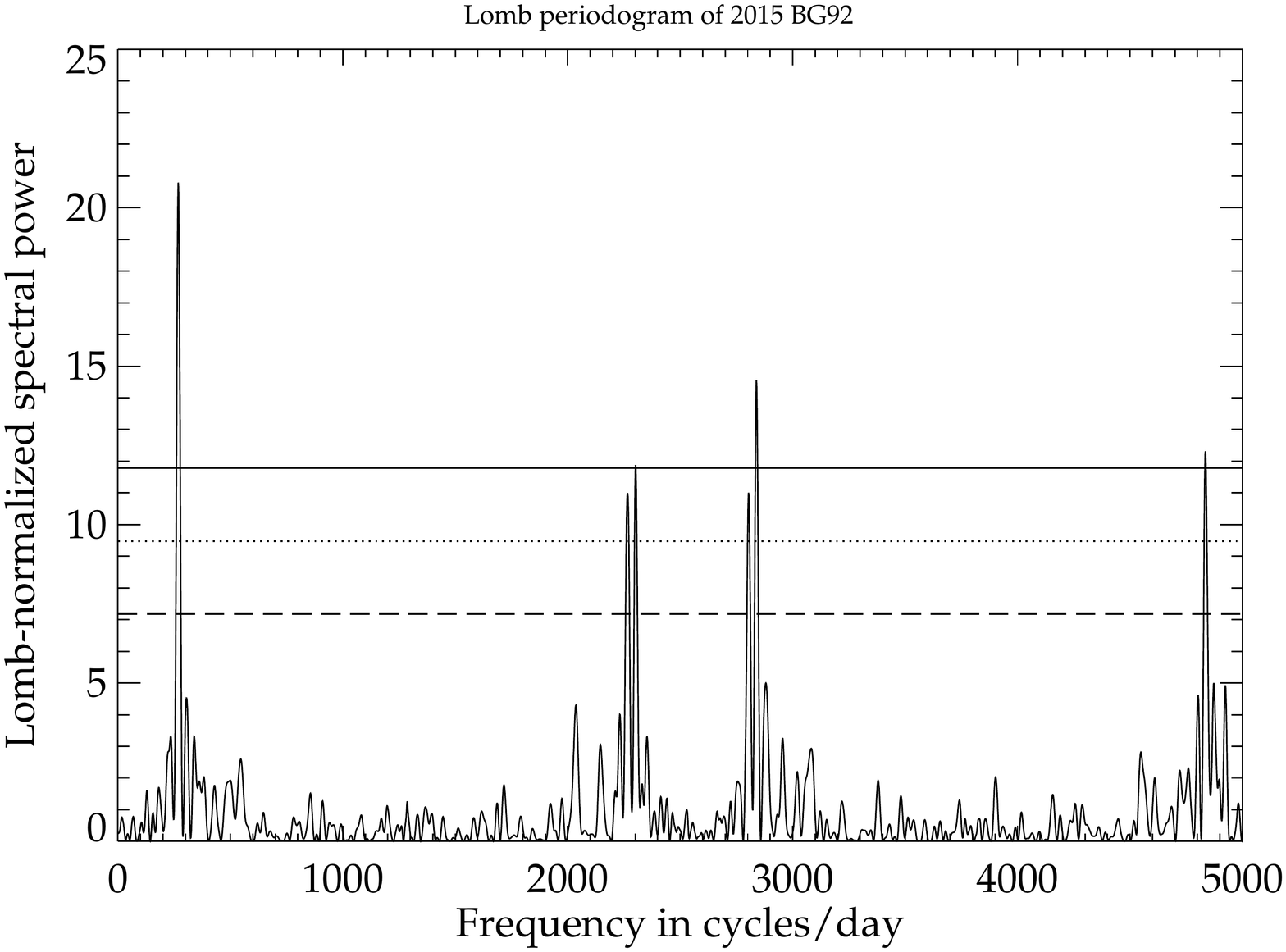}
\includegraphics[width=8cm, angle=0]{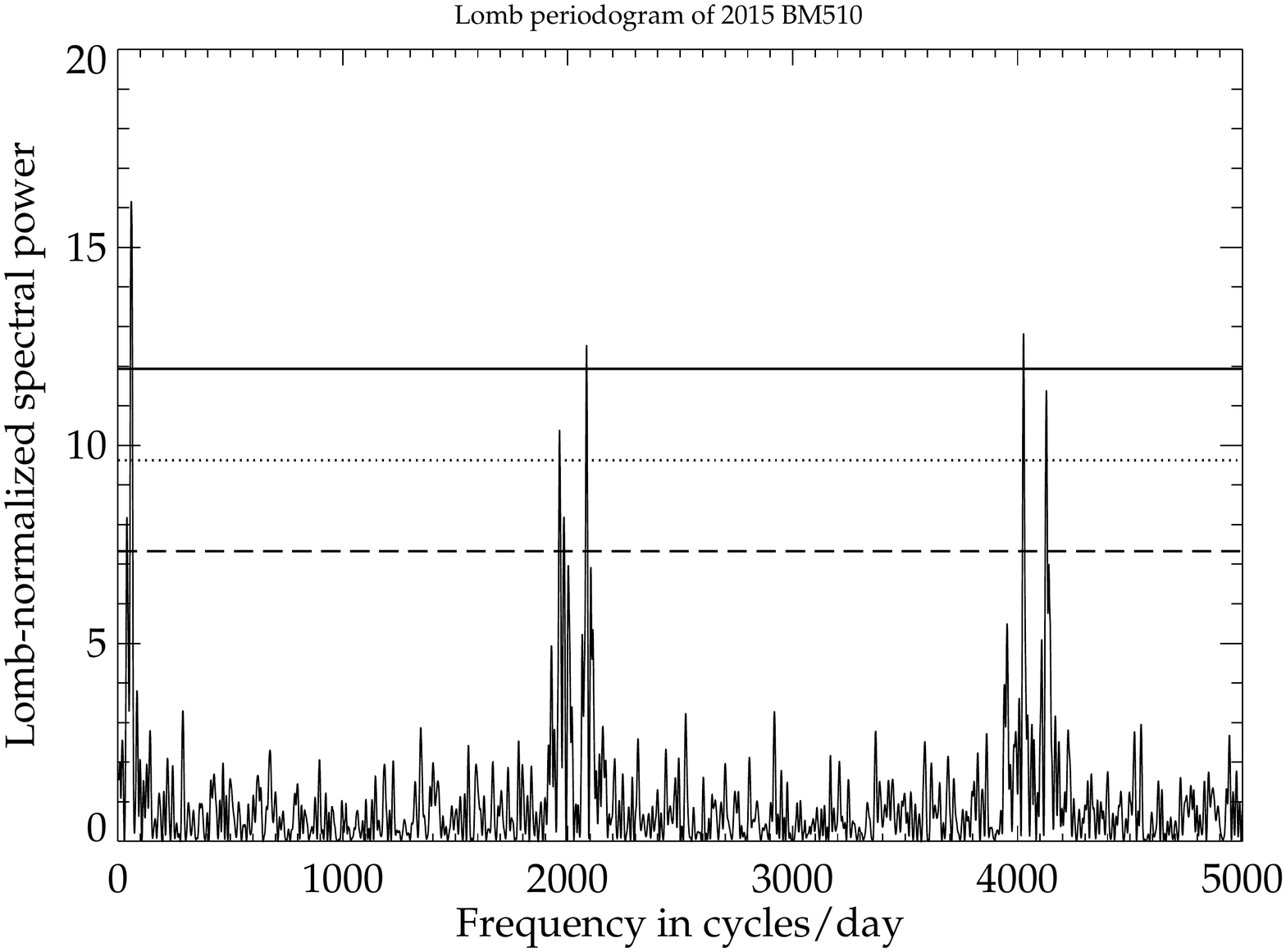}

\caption {\textit{Lomb-normalized spectral power versus Frequency}: Lomb periodograms of MANOS objects are plotted. Continuous line represents a 99.9$\%$ confidence level, dotted line a confidence level of 99$\%$, and the dashed line corresponds to a confidence level of 90$\%$.  }
\label{fig:Lomb5}
\end{figure}

\begin{figure}
\includegraphics[width=8cm, angle=0]{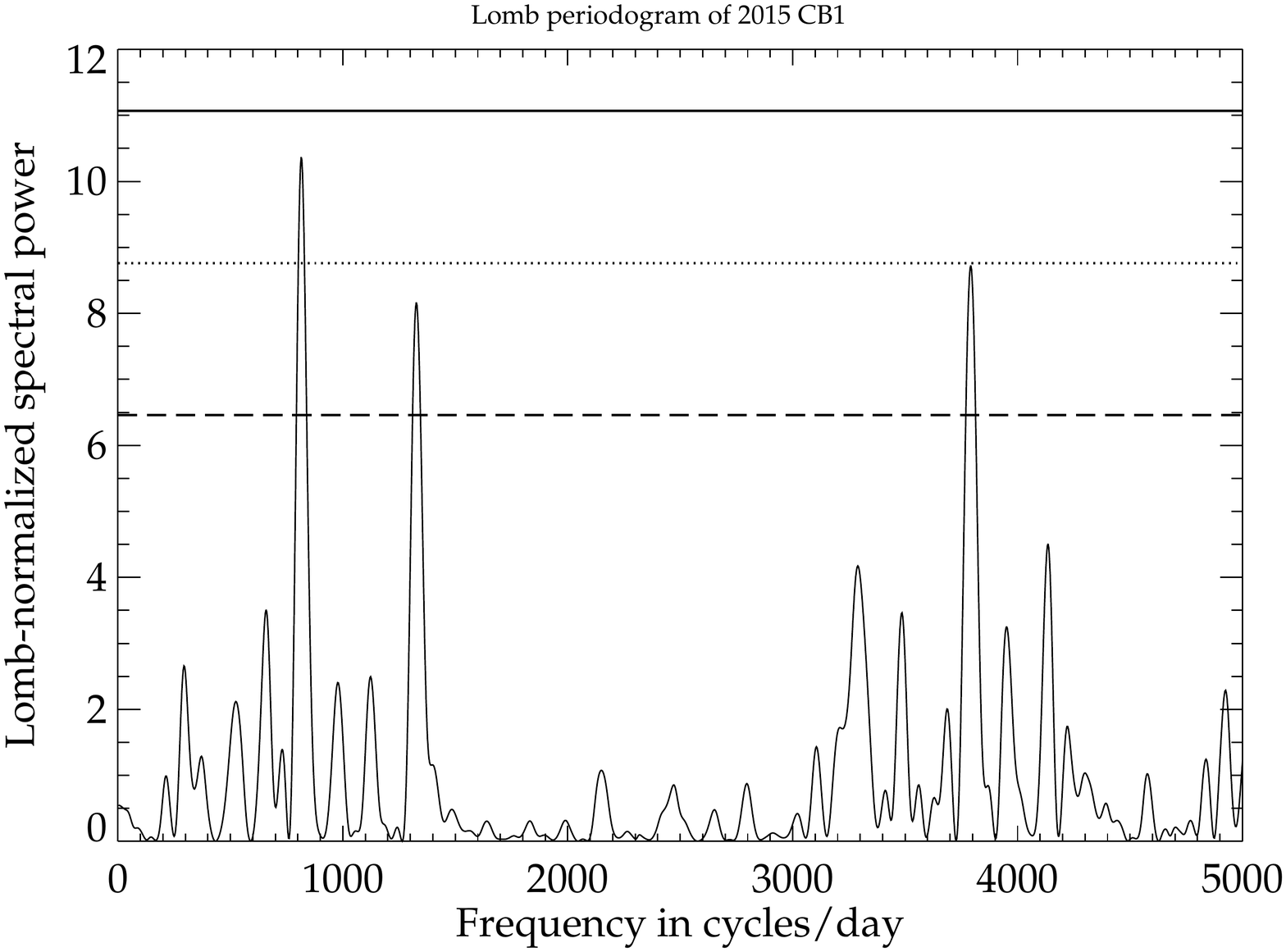}
\includegraphics[width=8cm, angle=0]{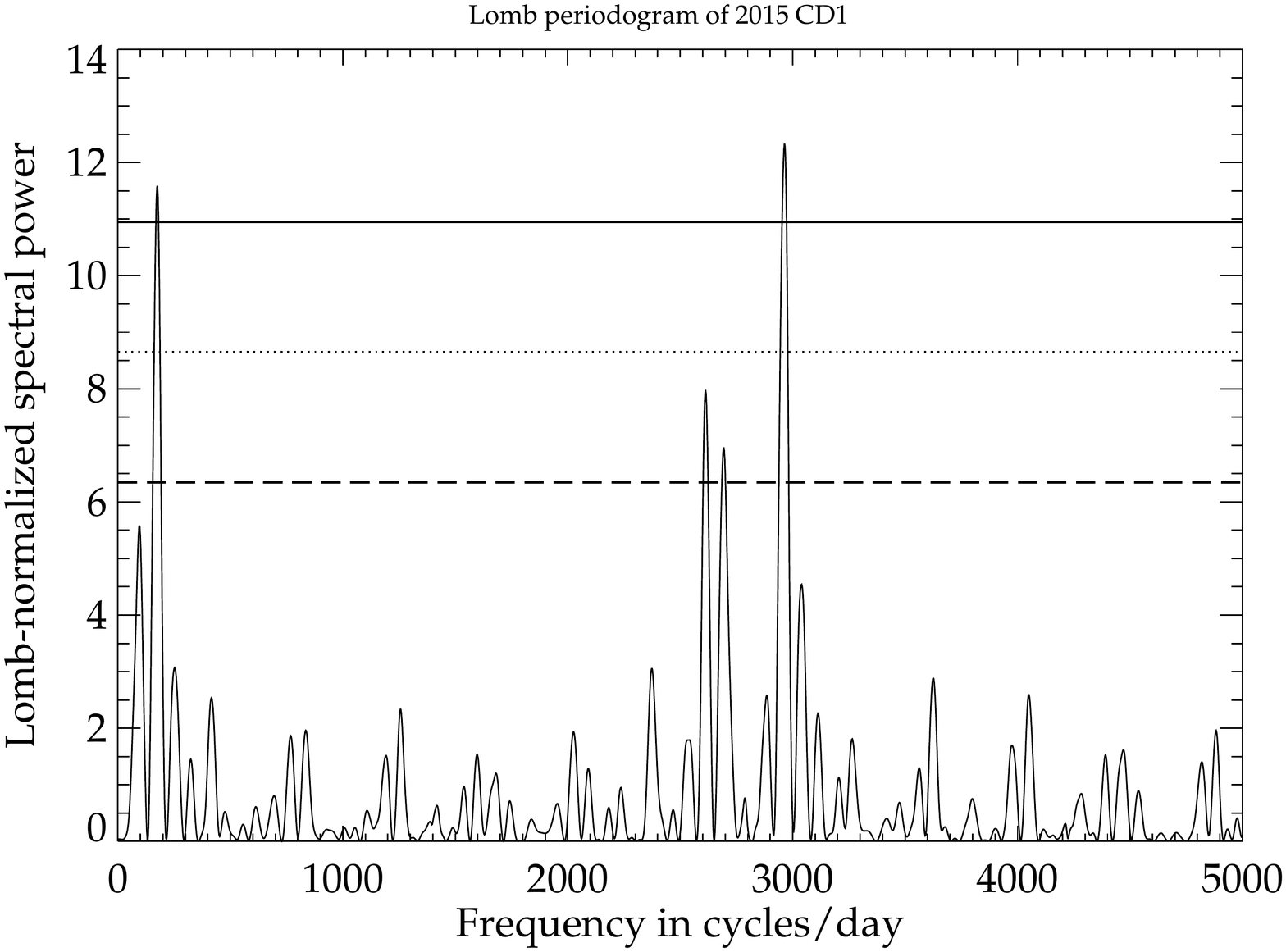}
\includegraphics[width=8cm, angle=0]{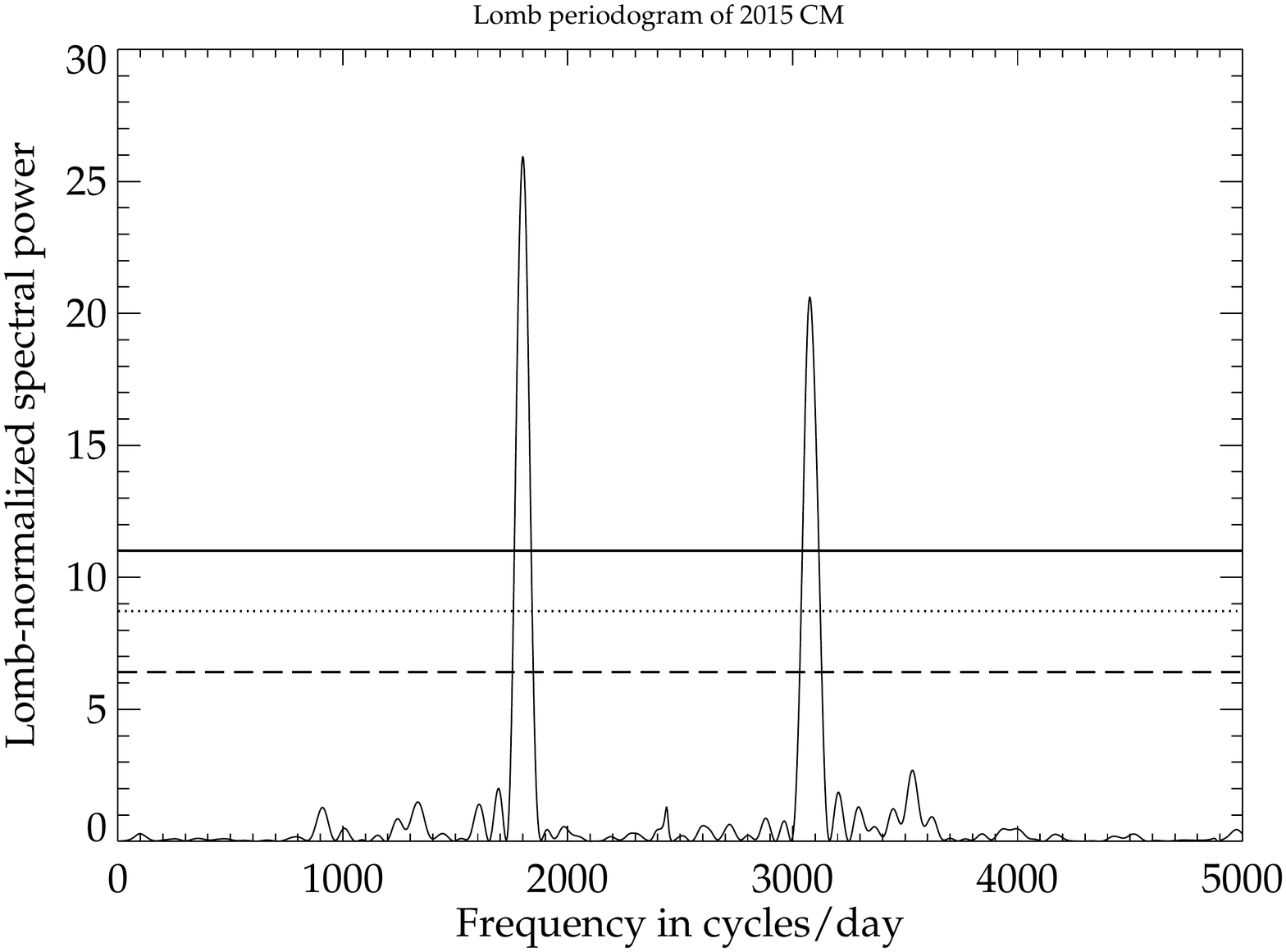}
\includegraphics[width=8cm, angle=0]{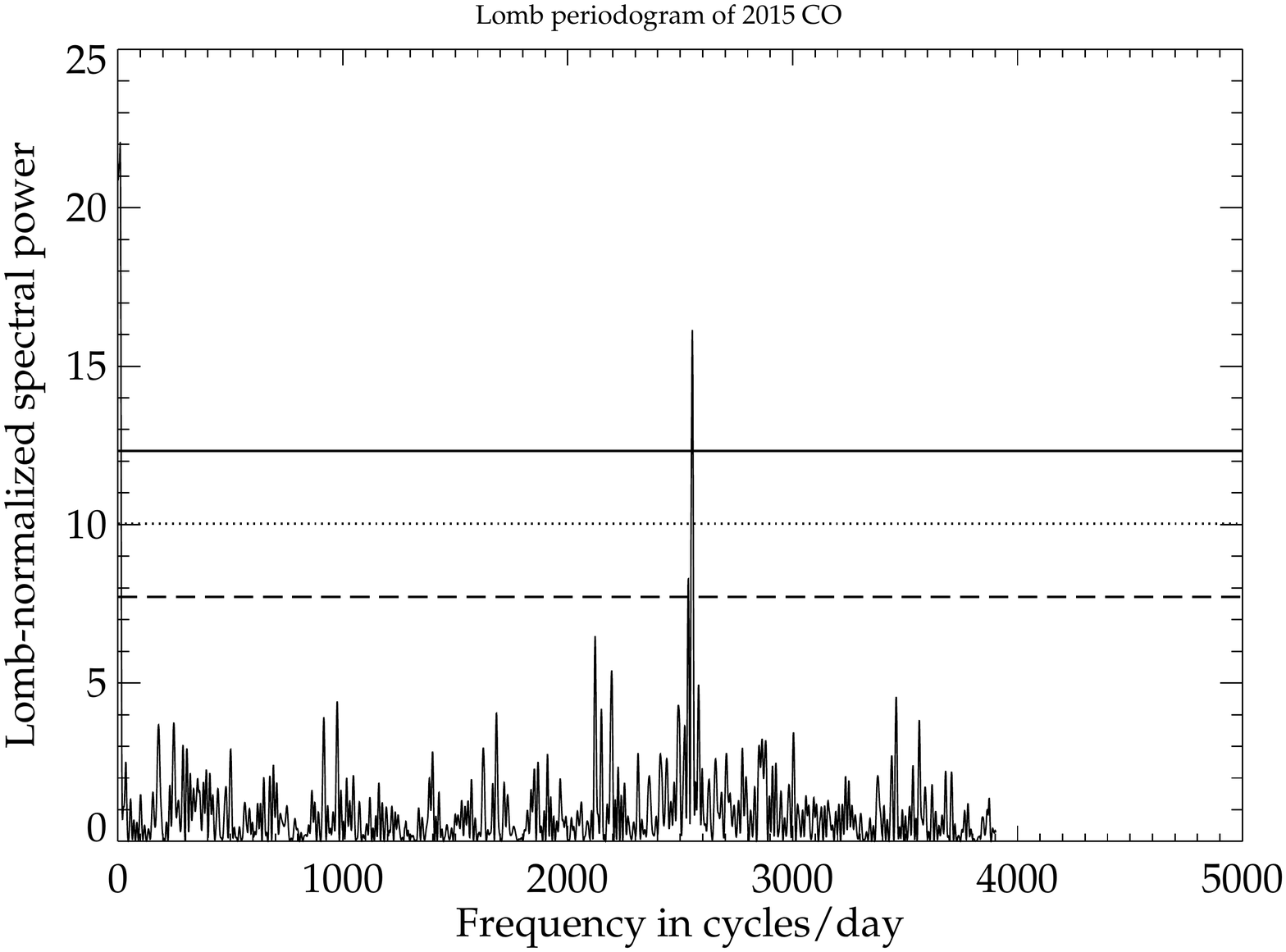}
\includegraphics[width=8cm, angle=0]{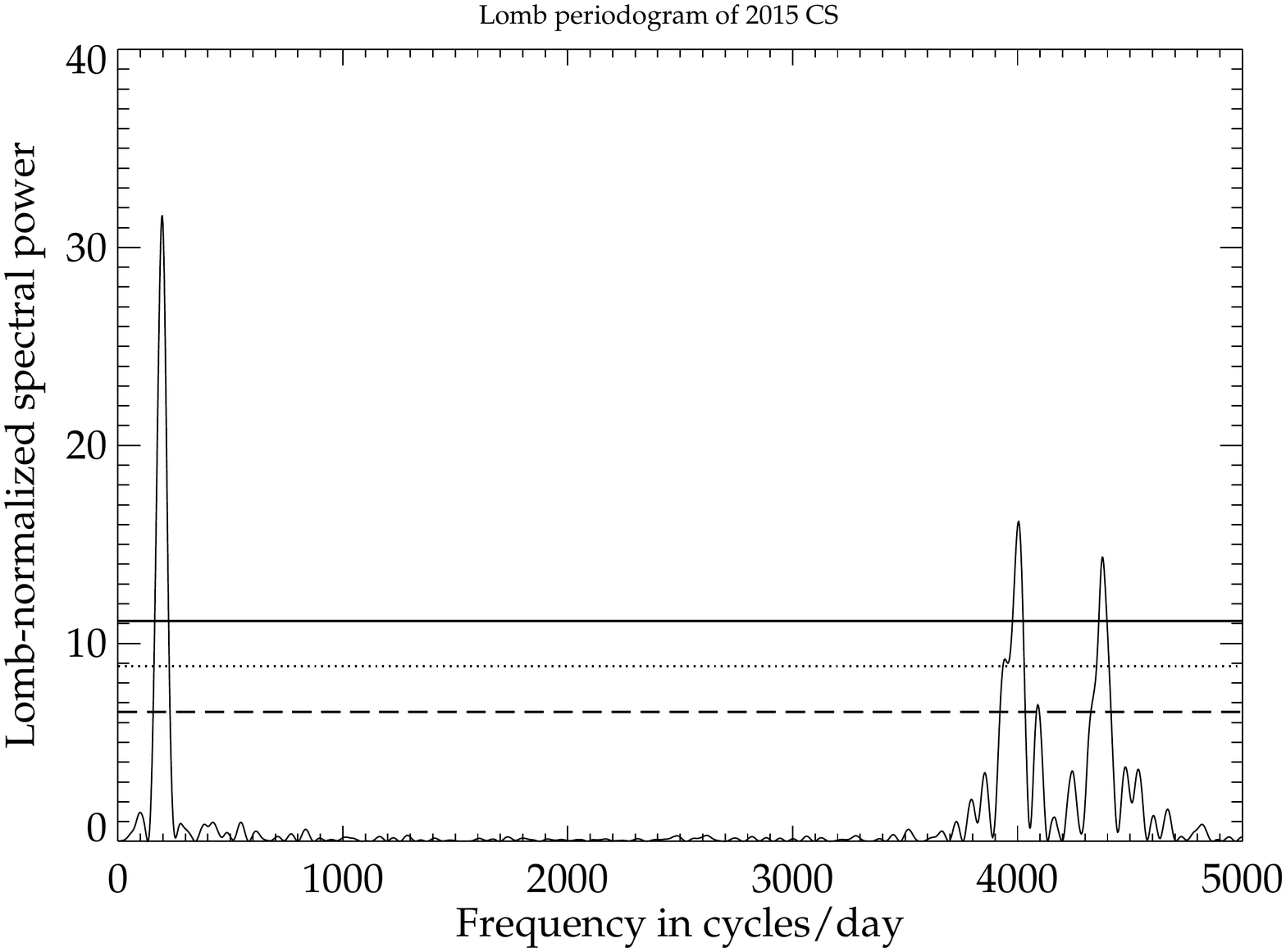}
\includegraphics[width=8cm, angle=0]{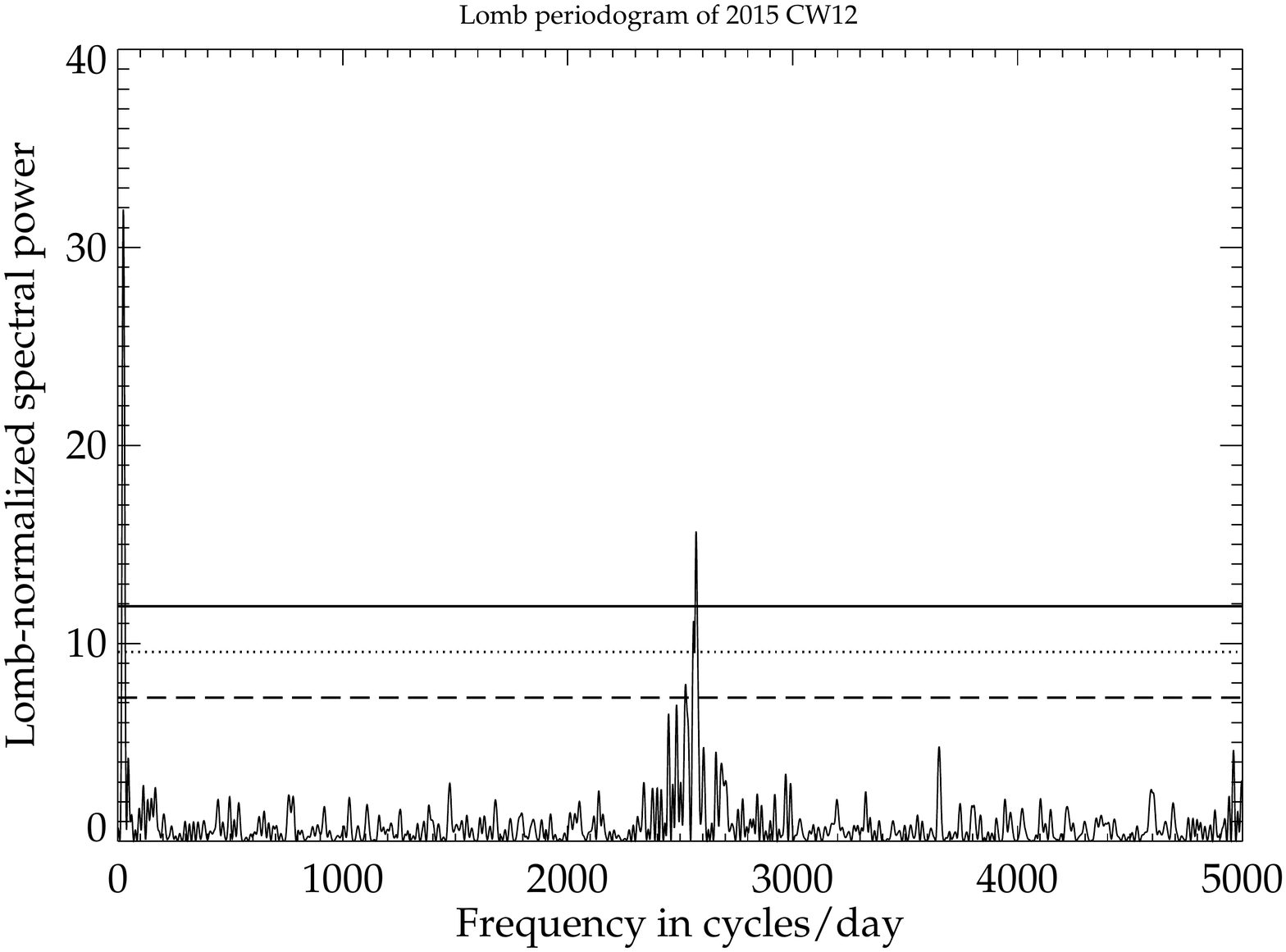}

\caption {\textit{Lomb-normalized spectral power versus Frequency}: Lomb periodograms of MANOS objects are plotted. Continuous line represents a 99.9$\%$ confidence level, dotted line a confidence level of 99$\%$, and the dashed line corresponds to a confidence level of 90$\%$.  }
\label{fig:Lomb6}
\end{figure}

\begin{figure}
\includegraphics[width=8cm, angle=0]{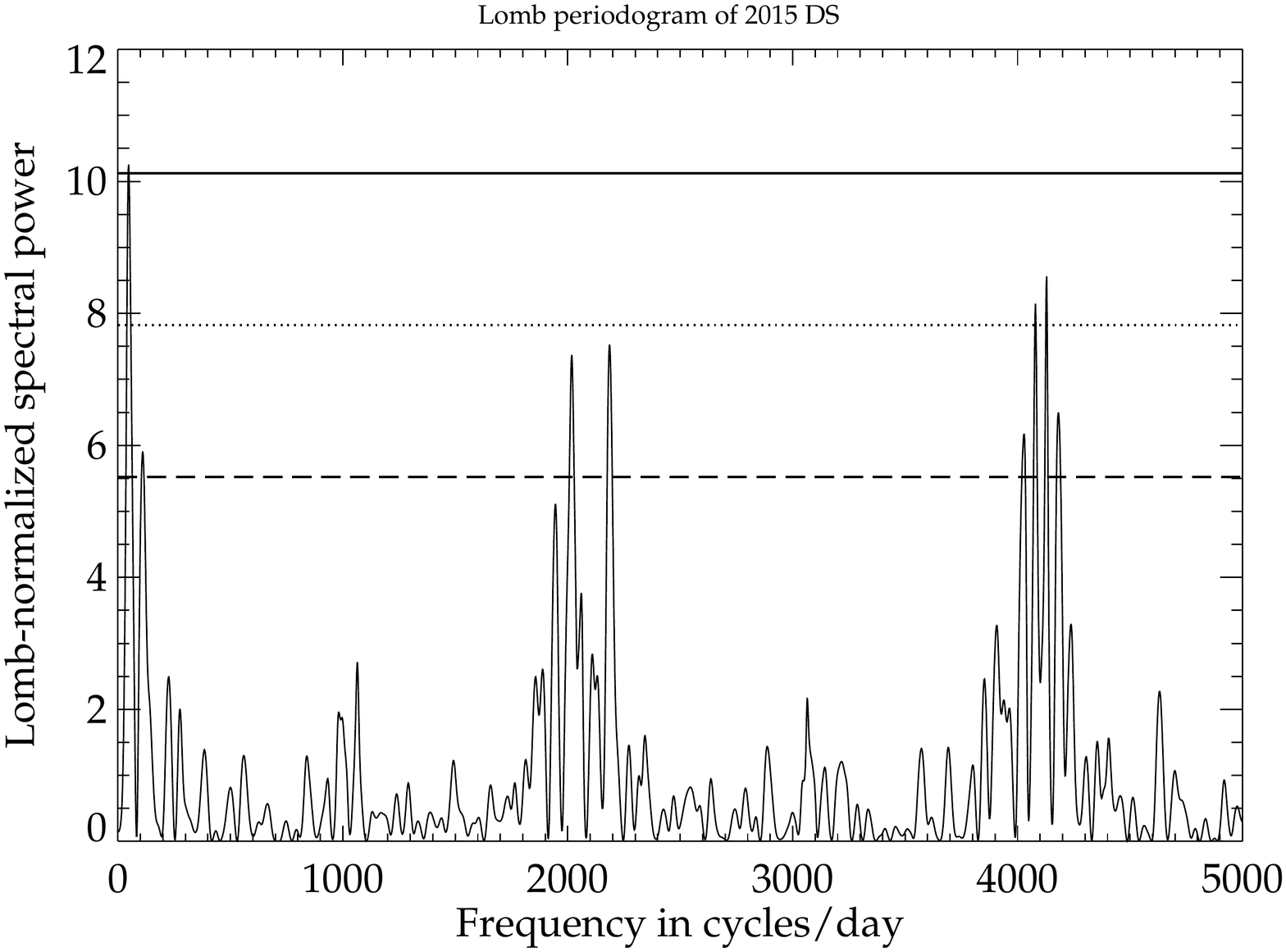}
\includegraphics[width=8cm, angle=0]{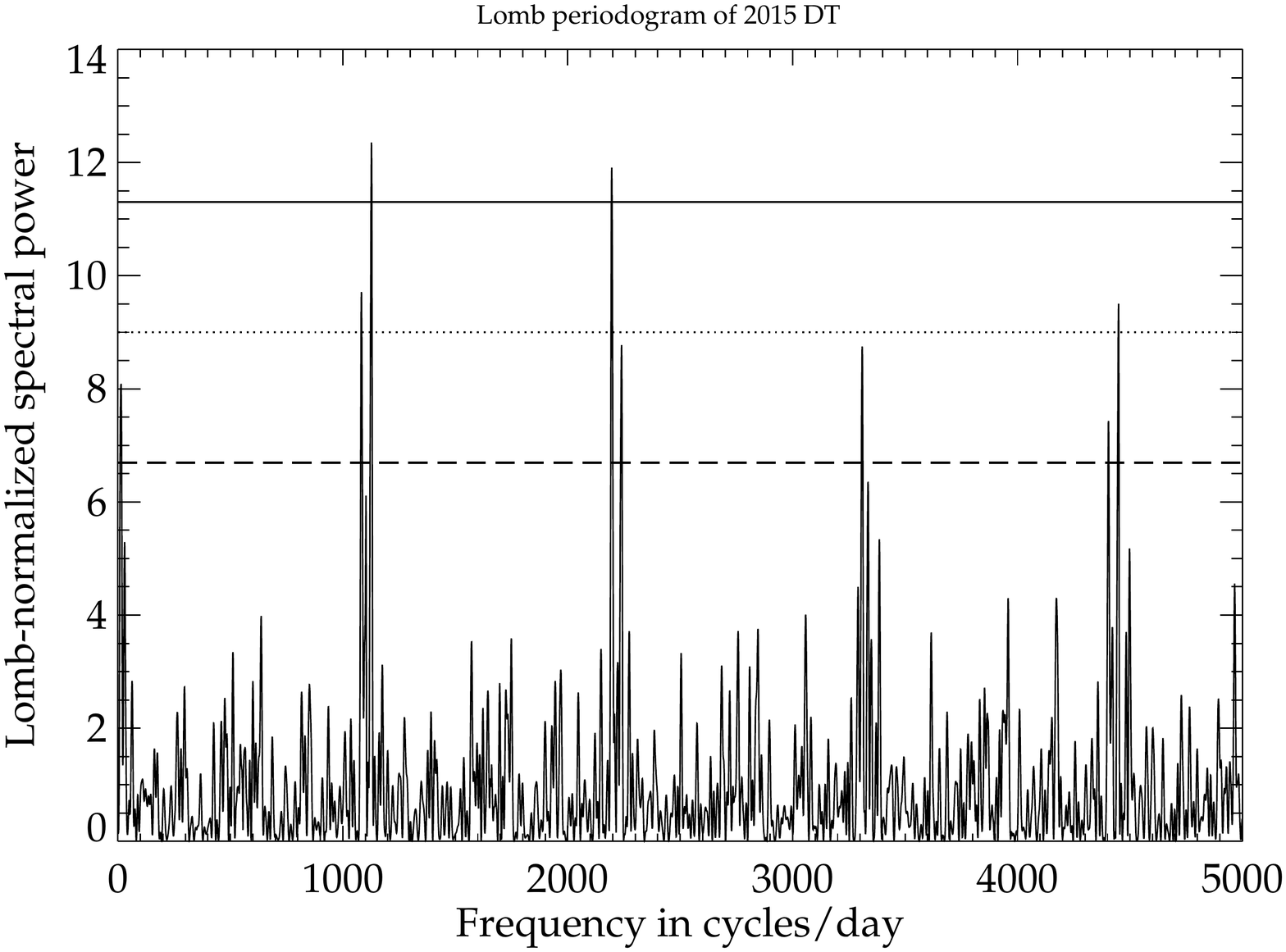}
\includegraphics[width=8cm, angle=0]{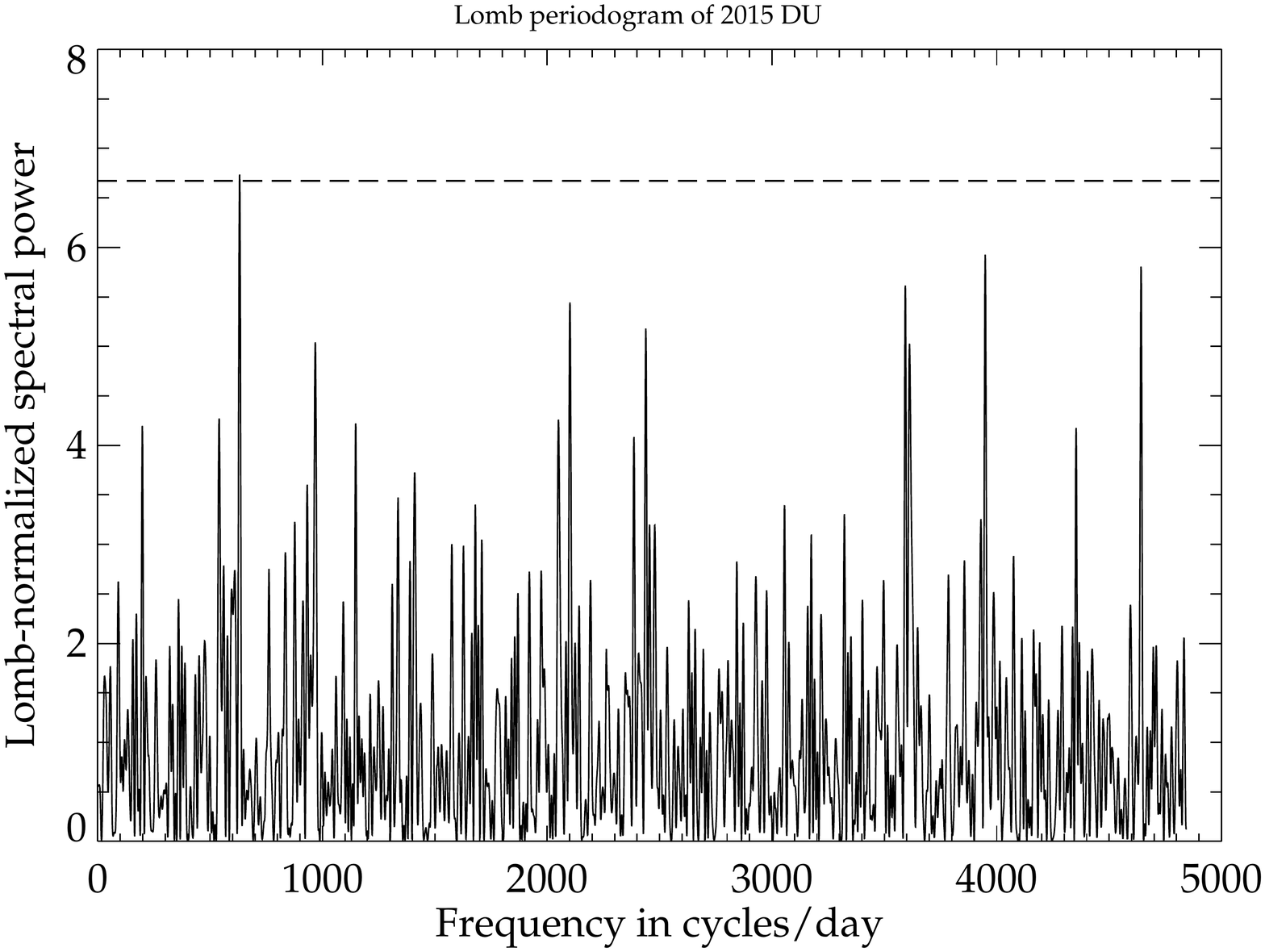}
\includegraphics[width=8cm, angle=0]{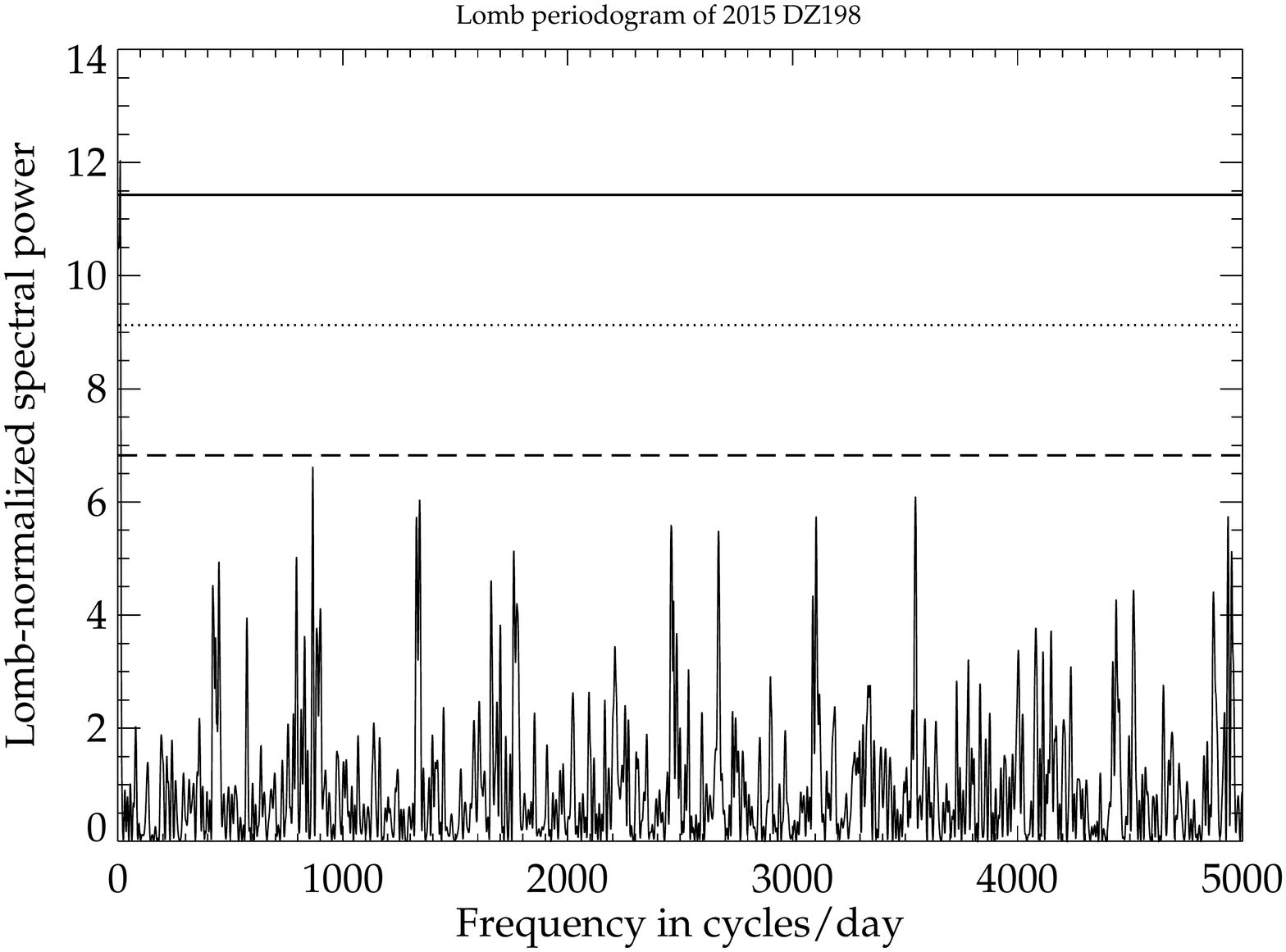}
\includegraphics[width=8cm, angle=0]{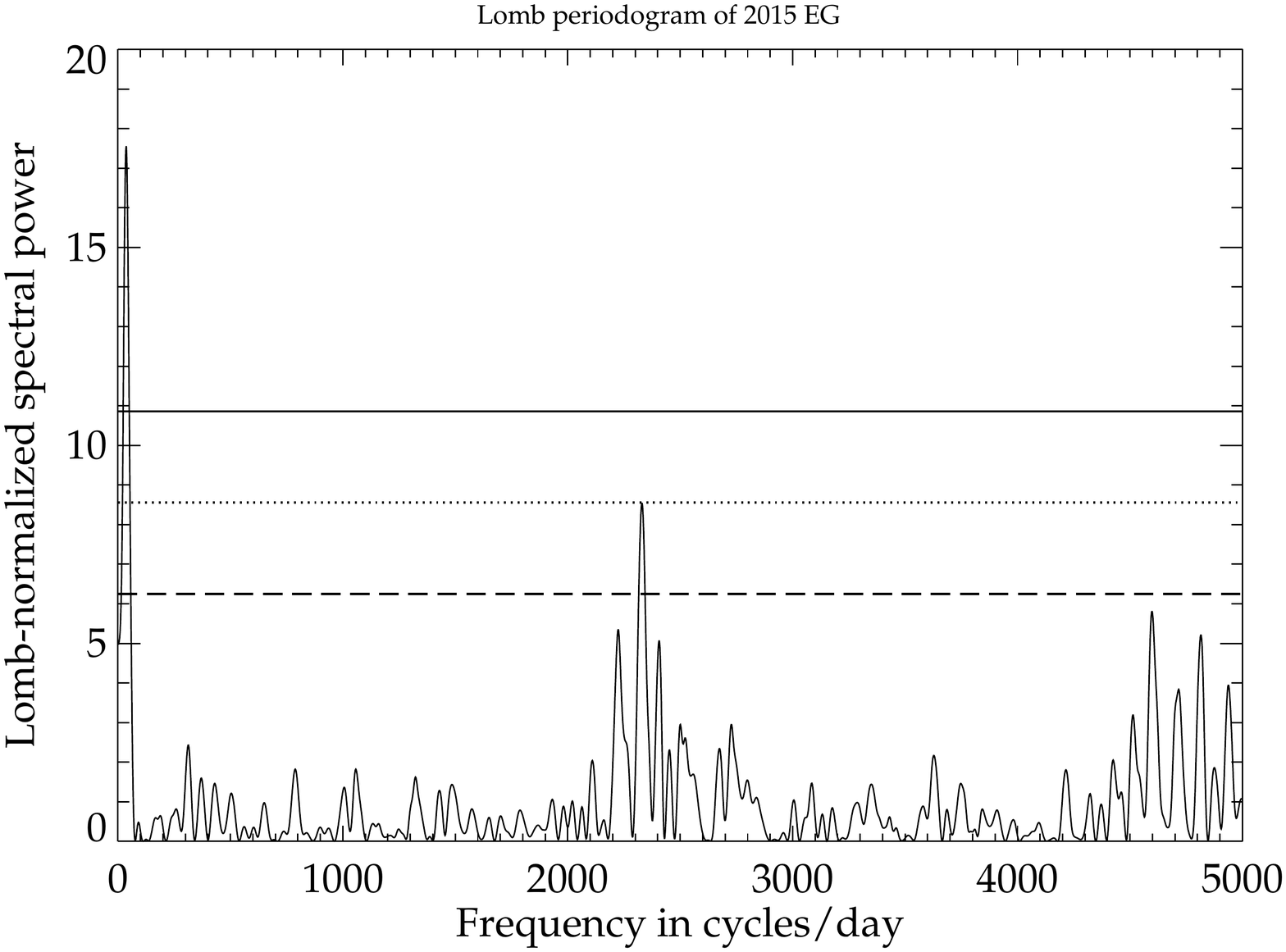}
\includegraphics[width=8cm, angle=0]{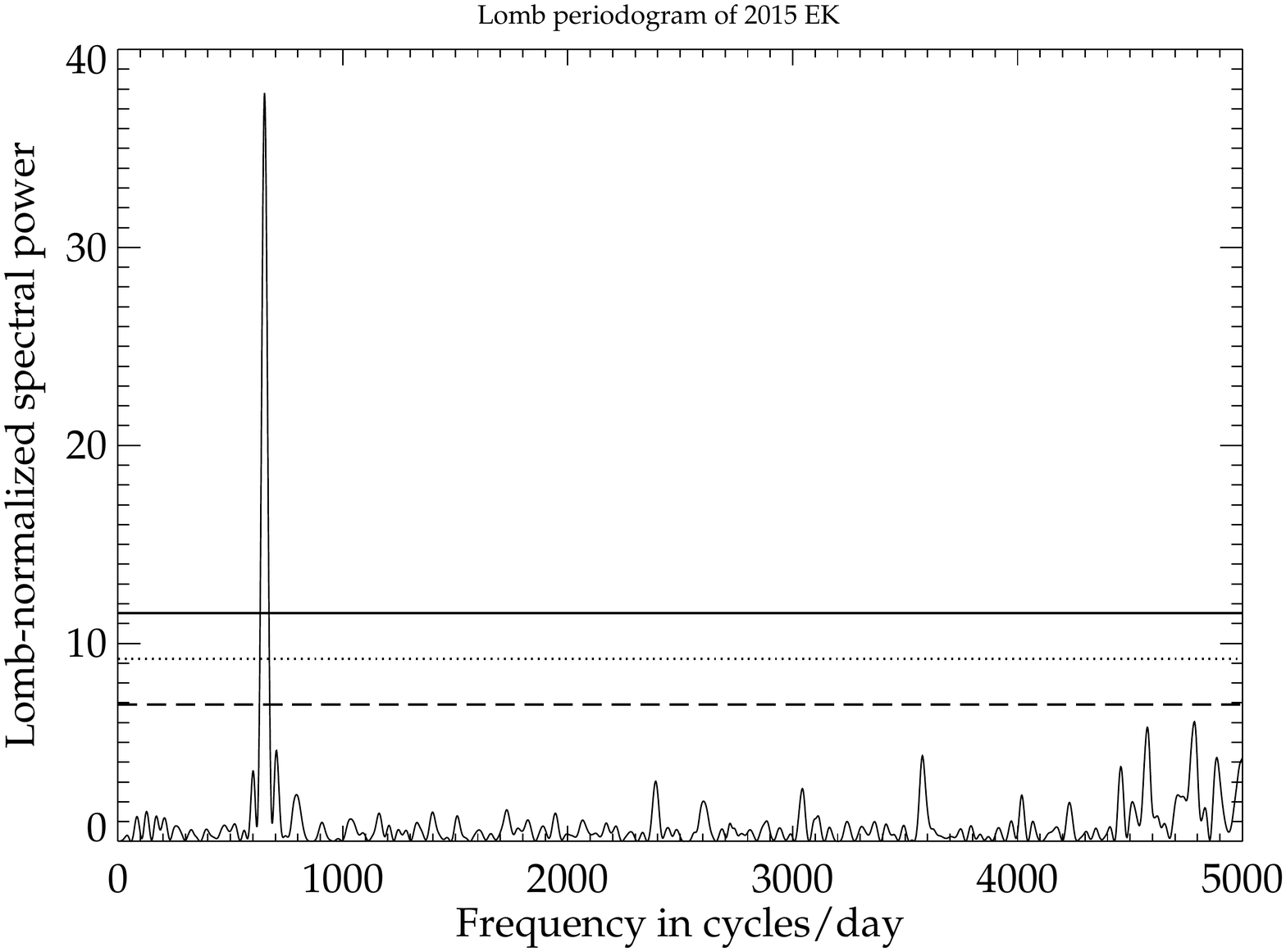}

\caption {\textit{Lomb-normalized spectral power versus Frequency}: Lomb periodograms of MANOS objects are plotted. Continuous line represents a 99.9$\%$ confidence level, dotted line a confidence level of 99$\%$, and the dashed line corresponds to a confidence level of 90$\%$.  }
\label{fig:Lomb7}
\end{figure}

\begin{figure}
\includegraphics[width=8cm, angle=0]{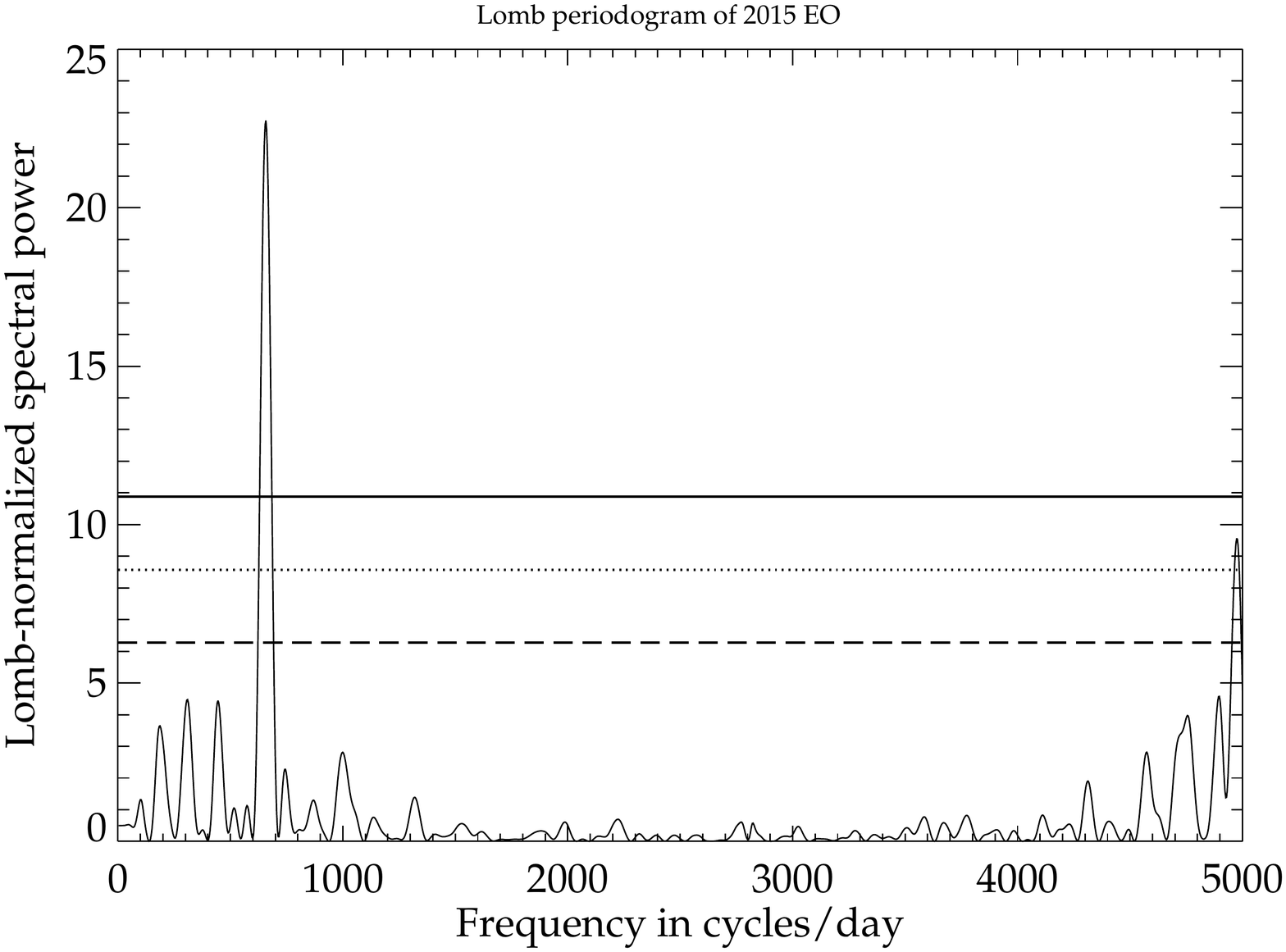}
\includegraphics[width=8cm, angle=0]{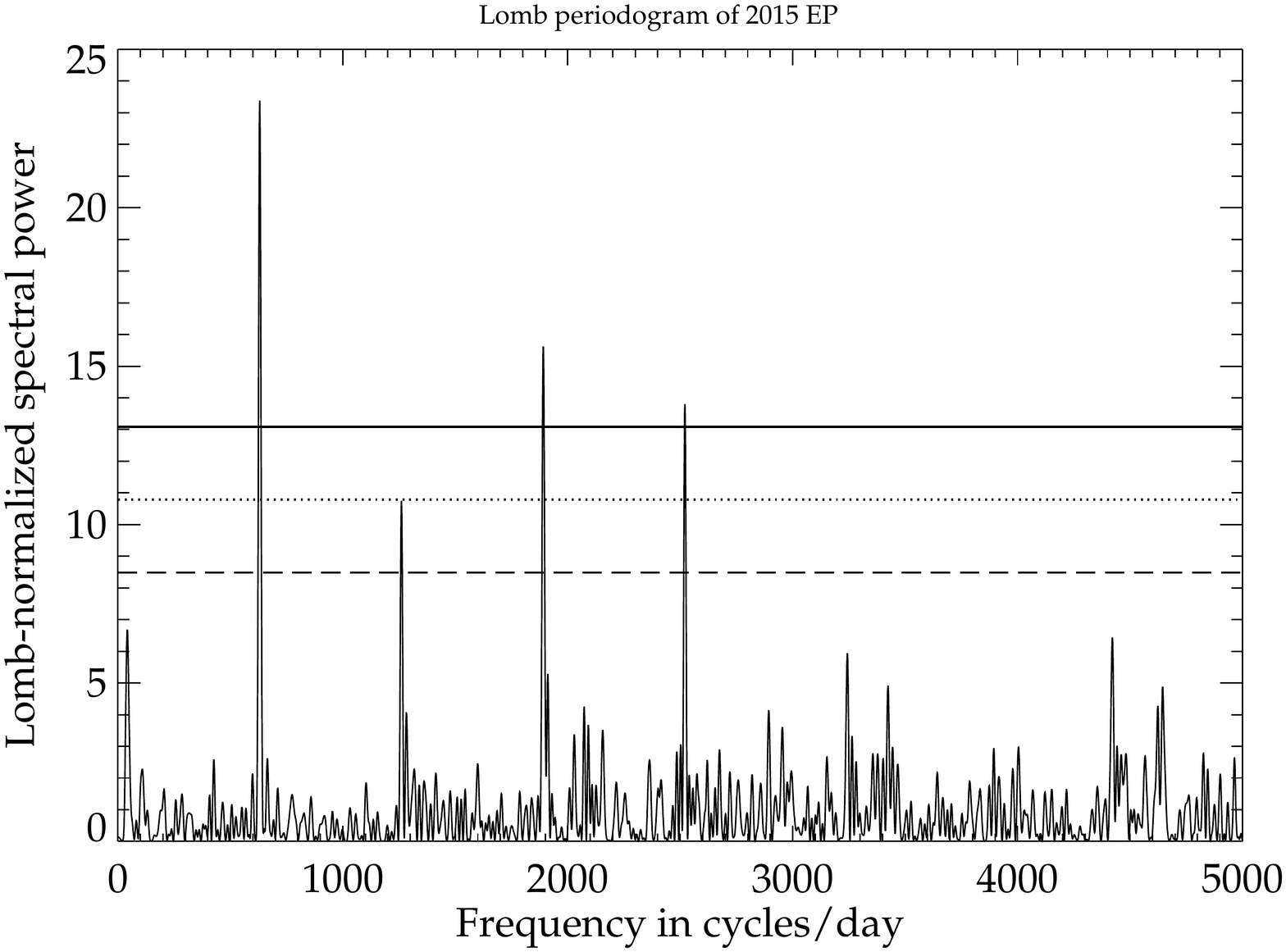}
\includegraphics[width=8cm, angle=0]{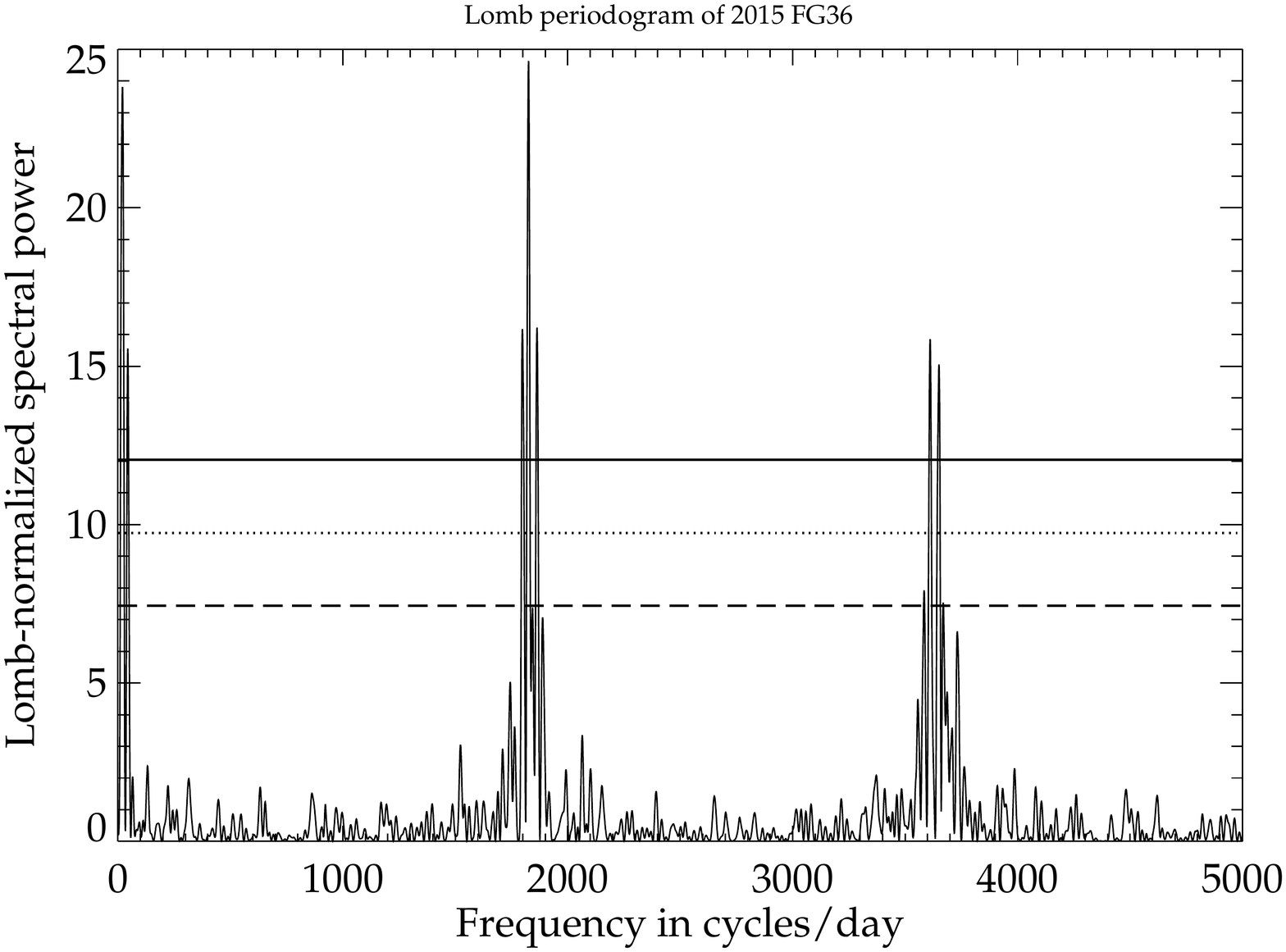}
\includegraphics[width=8cm, angle=0]{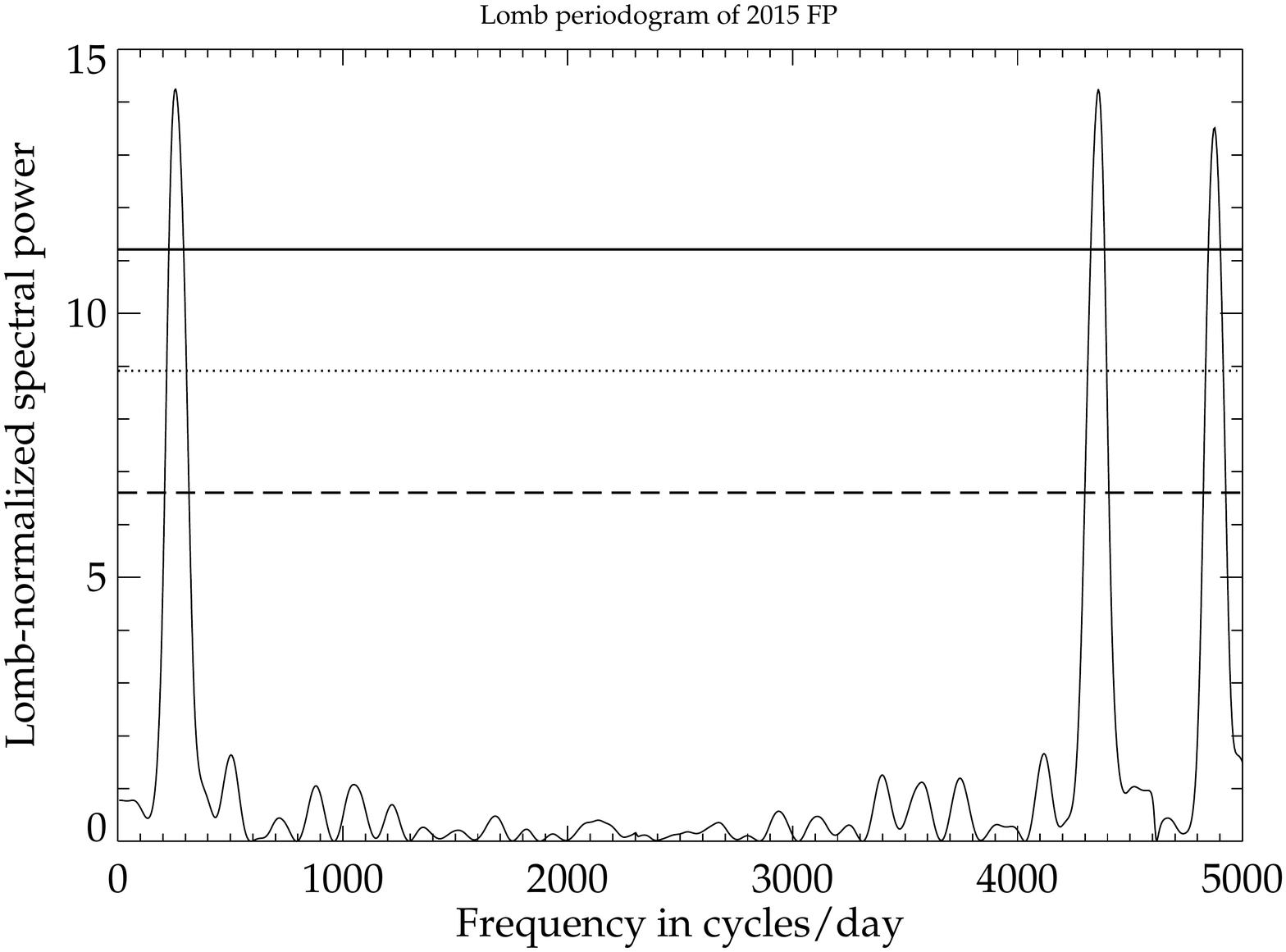}
\includegraphics[width=8cm, angle=0]{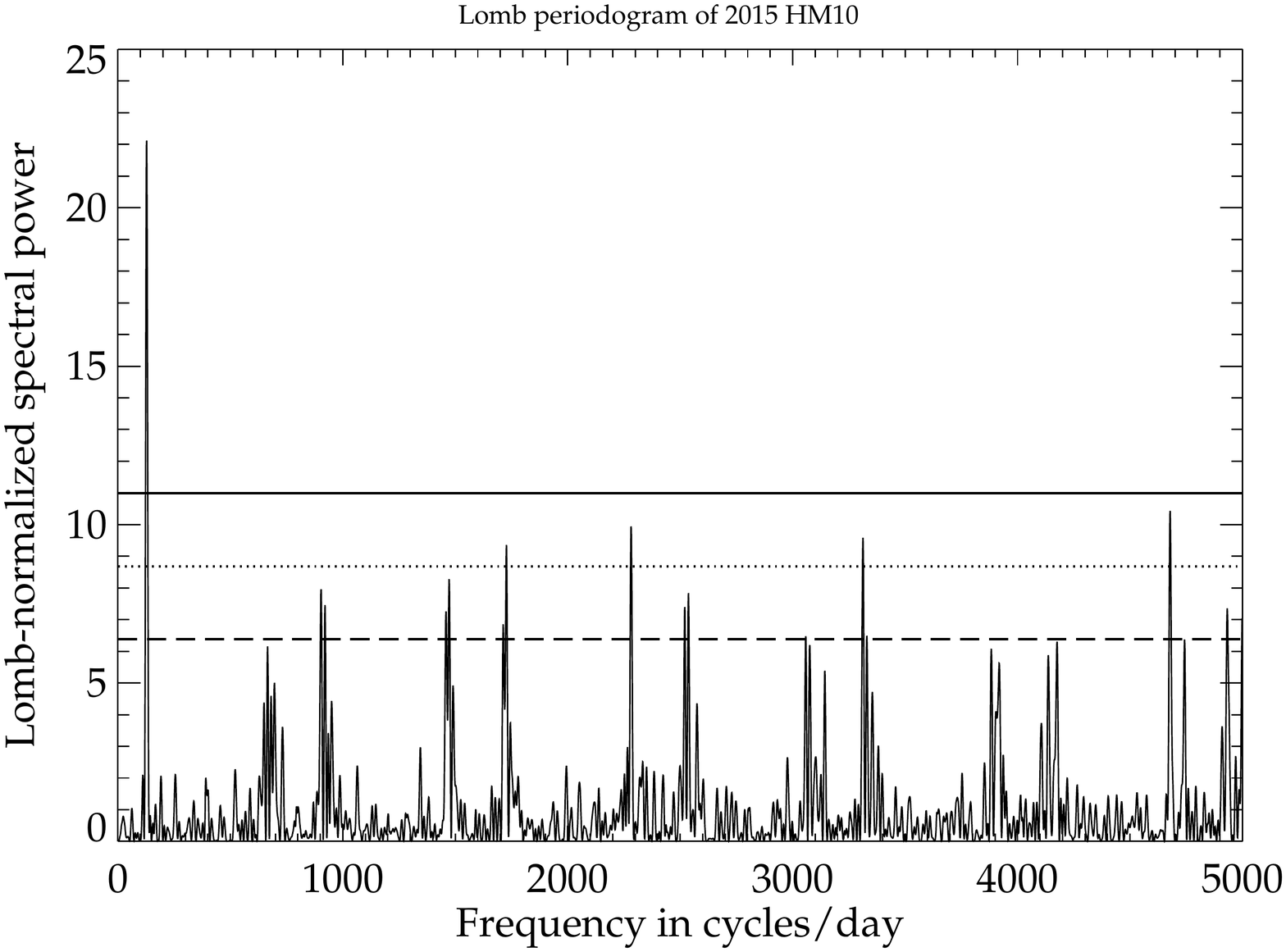}
\includegraphics[width=8cm, angle=0]{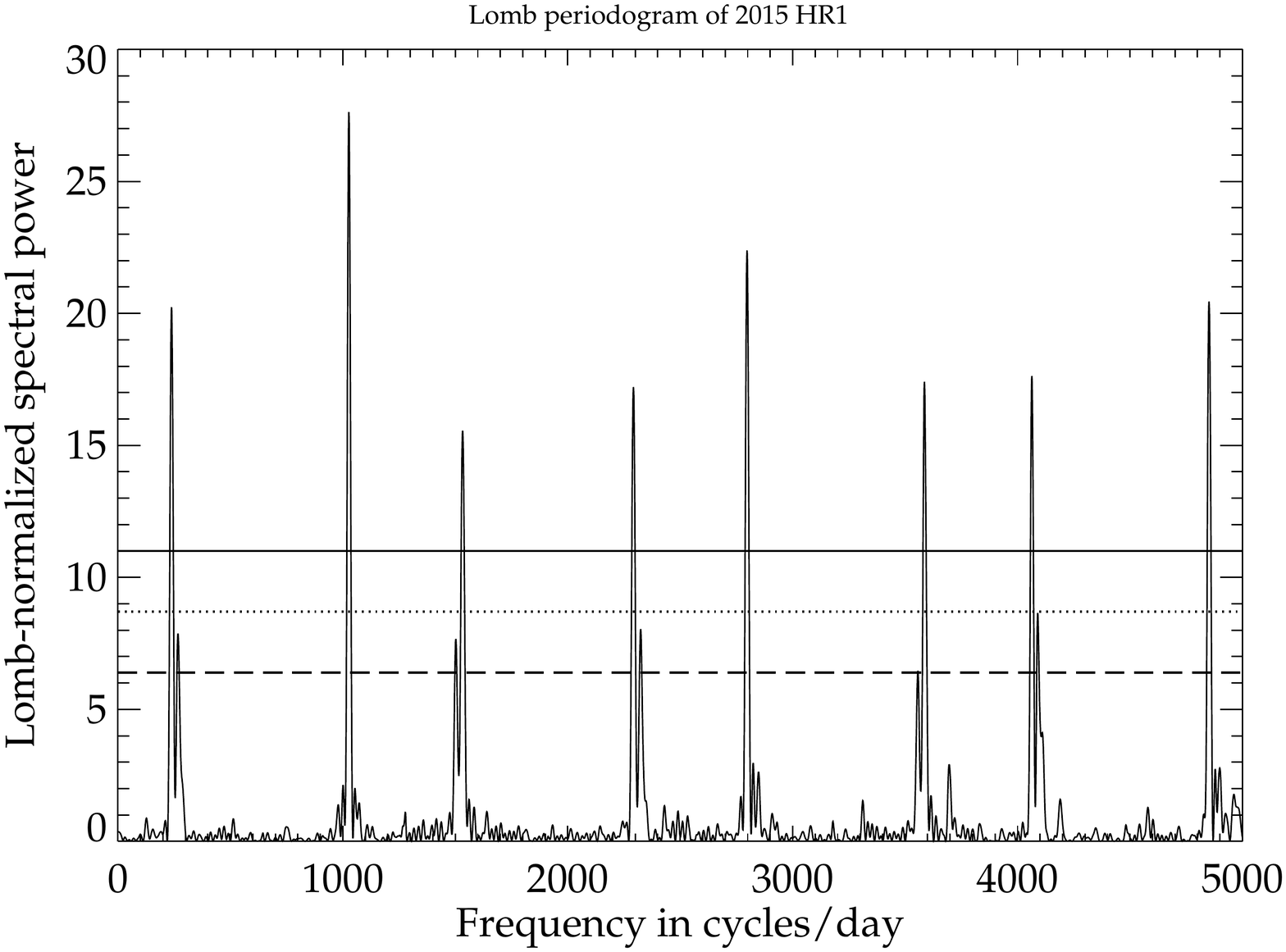}

\caption {\textit{Lomb-normalized spectral power versus Frequency}: Lomb periodograms of MANOS objects are plotted. Continuous line represents a 99.9$\%$ confidence level, dotted line a confidence level of 99$\%$, and the dashed line corresponds to a confidence level of 90$\%$.  }
\label{fig:Lomb8}
\end{figure}

\begin{figure}
\includegraphics[width=8cm, angle=0]{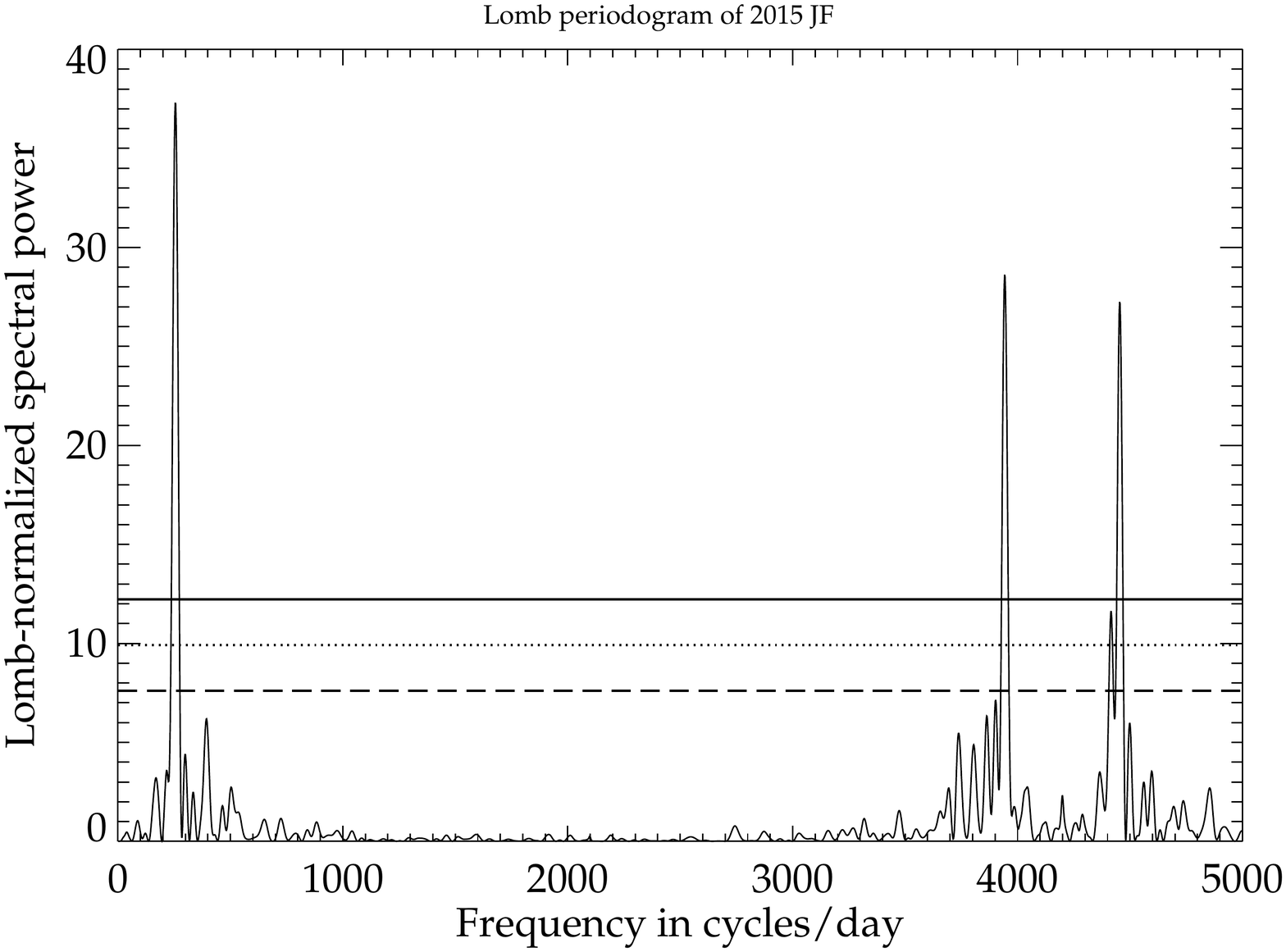}
\includegraphics[width=8cm, angle=0]{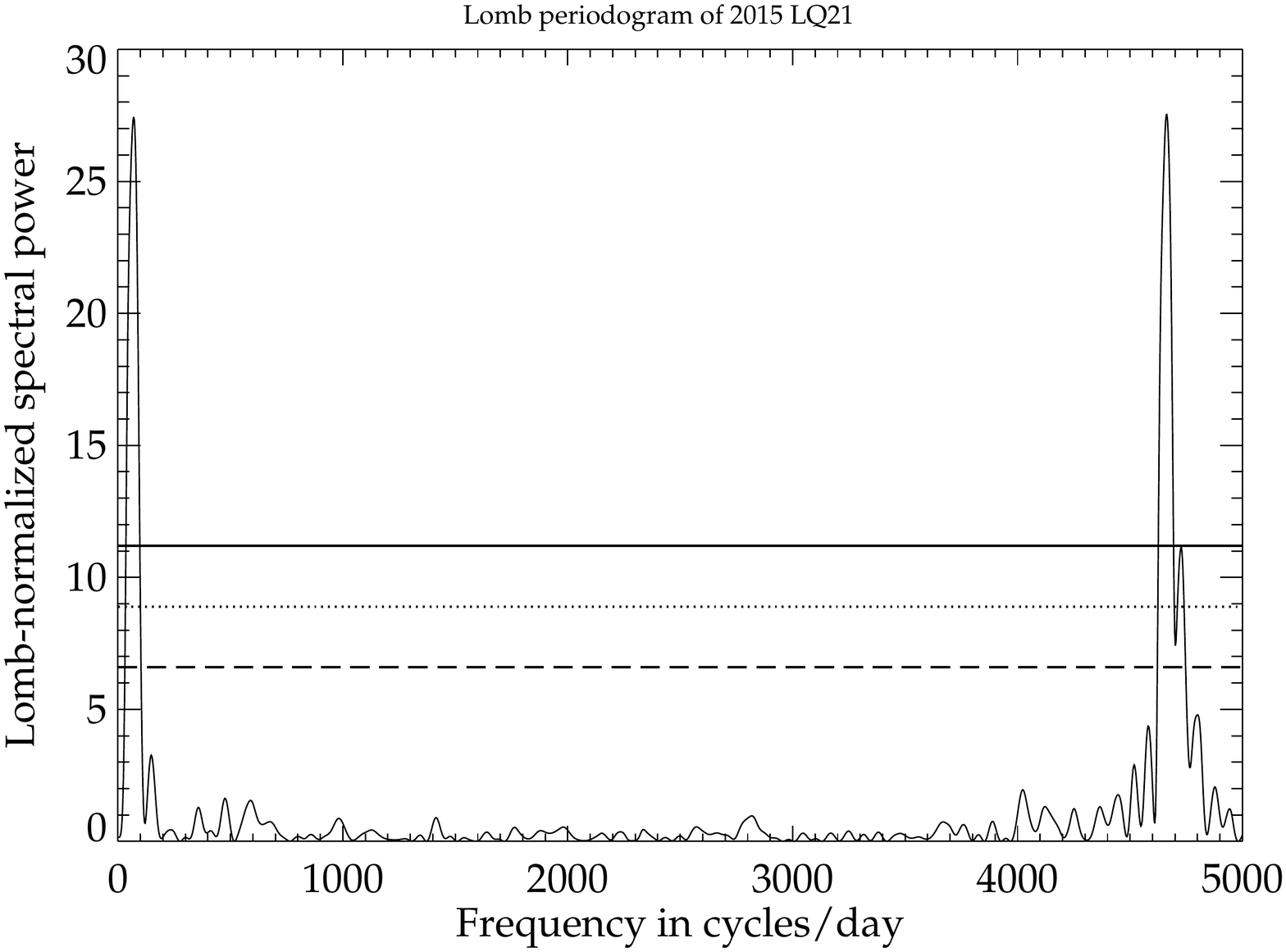}
\includegraphics[width=8cm, angle=0]{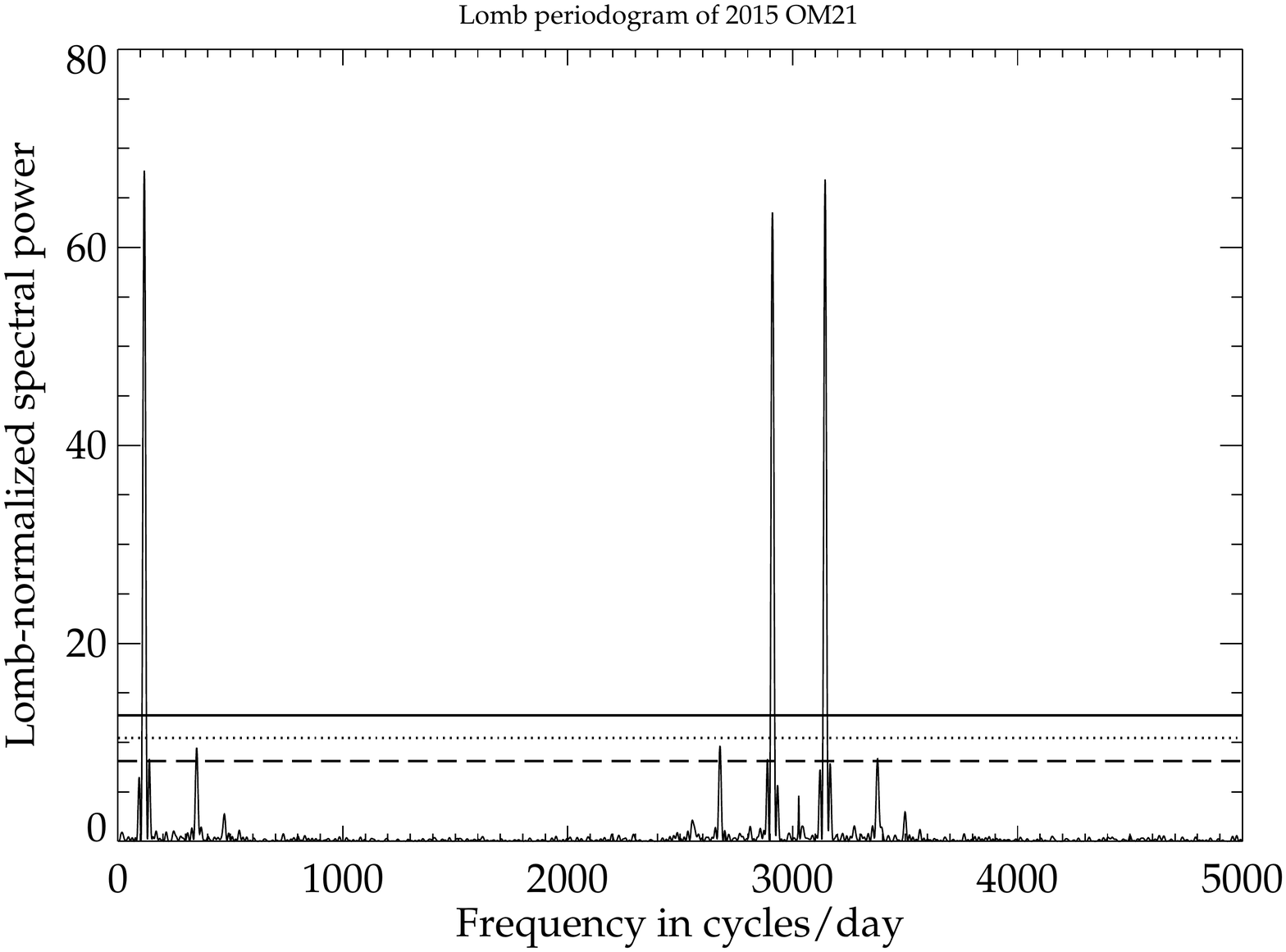}
\includegraphics[width=8cm, angle=0]{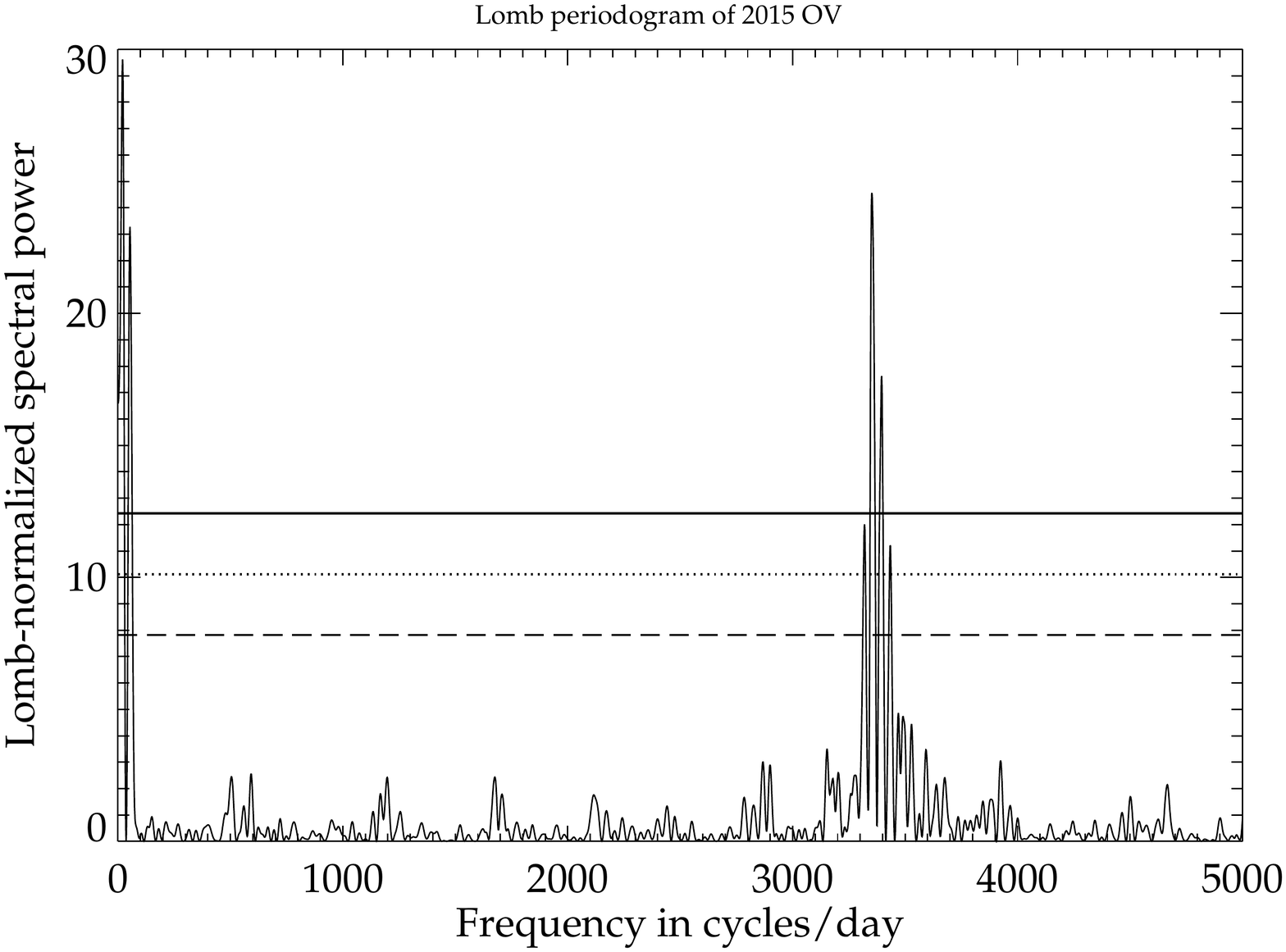}
\includegraphics[width=8cm, angle=0]{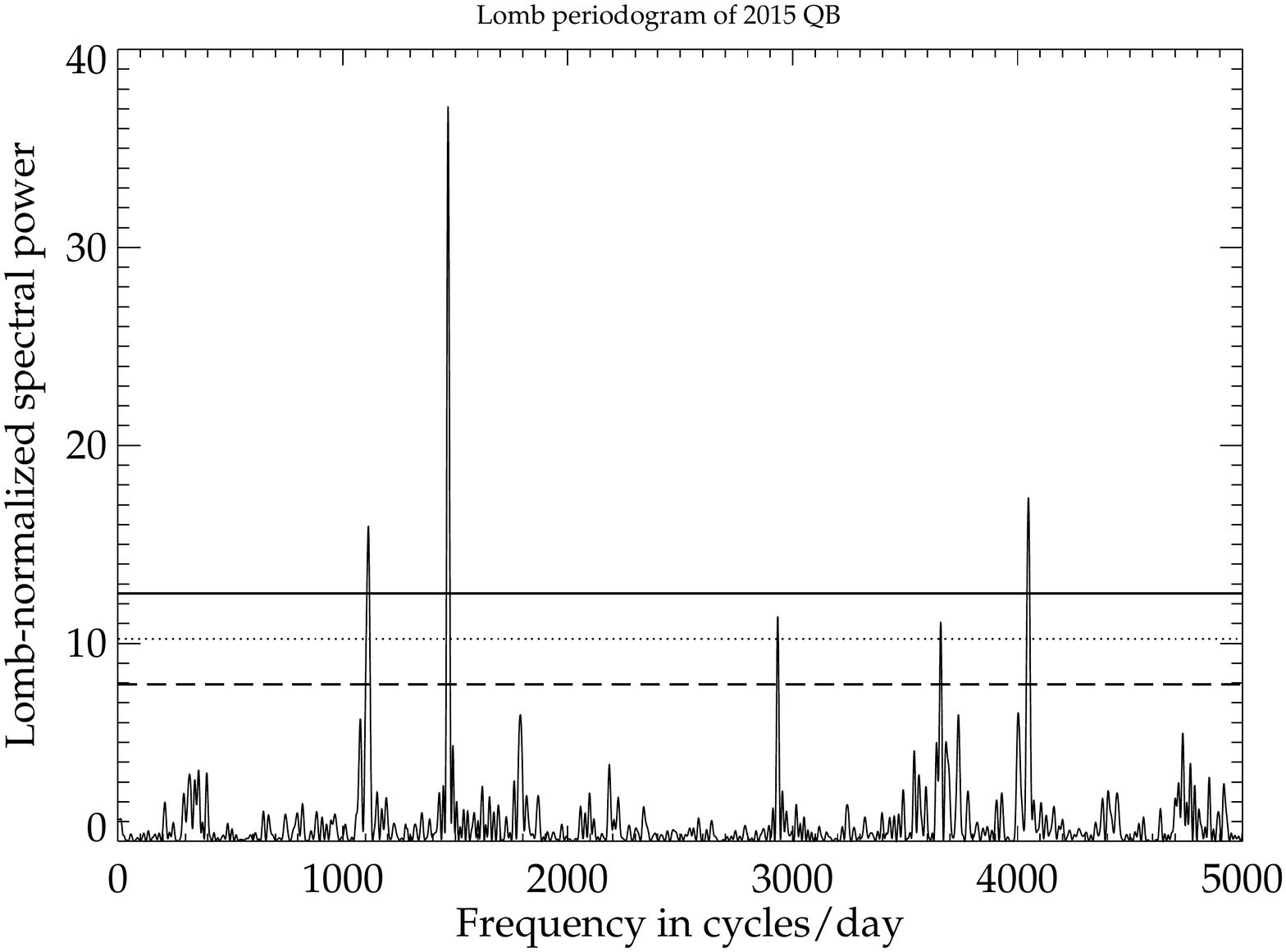}
\includegraphics[width=8cm, angle=0]{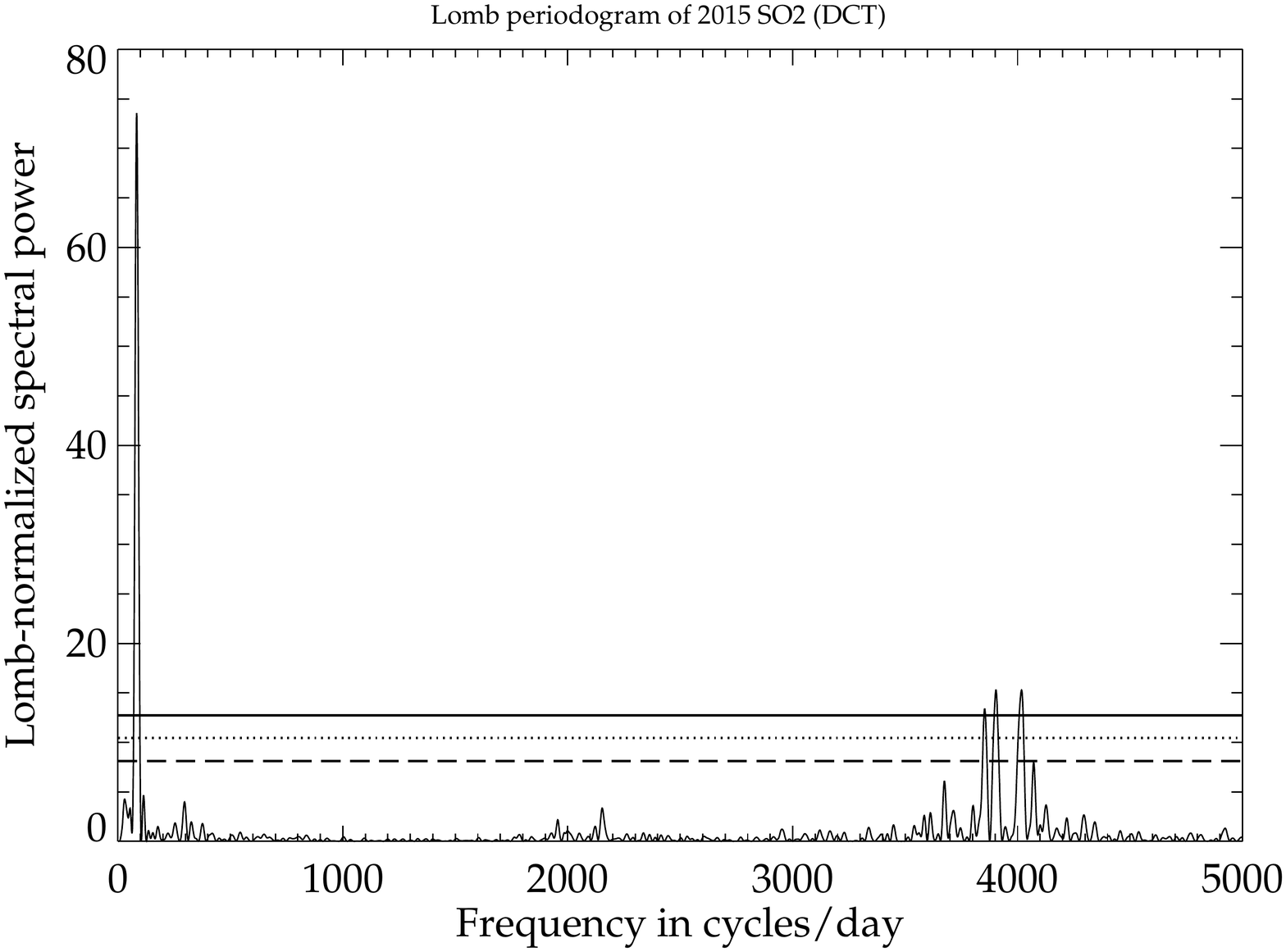}

\caption {\textit{Lomb-normalized spectral power versus Frequency}: Lomb periodograms of MANOS objects are plotted. Continuous line represents a 99.9$\%$ confidence level, dotted line a confidence level of 99$\%$, and the dashed line corresponds to a confidence level of 90$\%$.  }
\label{fig:Lomb9}
\end{figure}

\begin{figure}
\includegraphics[width=8cm, angle=0]{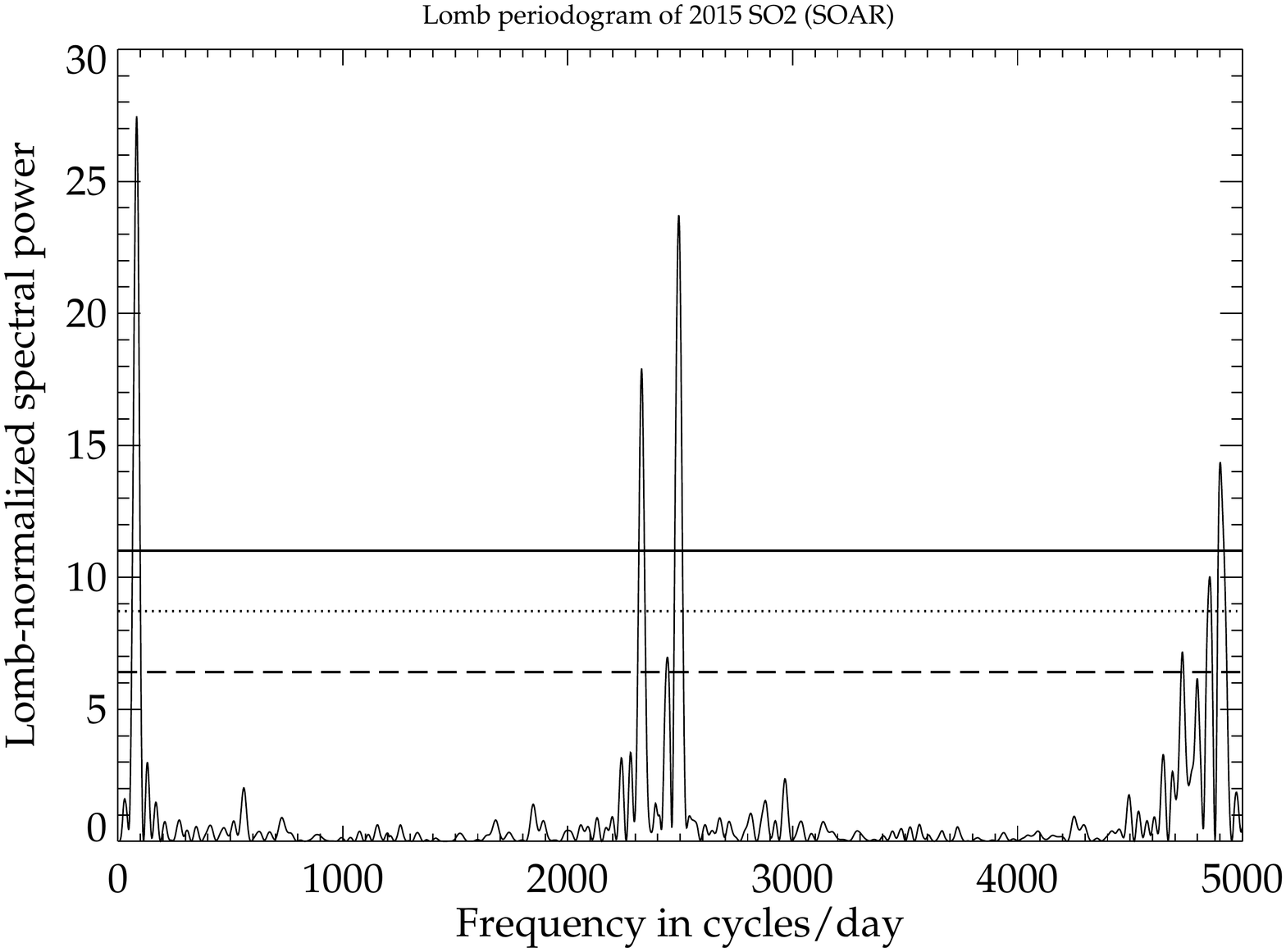}
\includegraphics[width=8cm, angle=0]{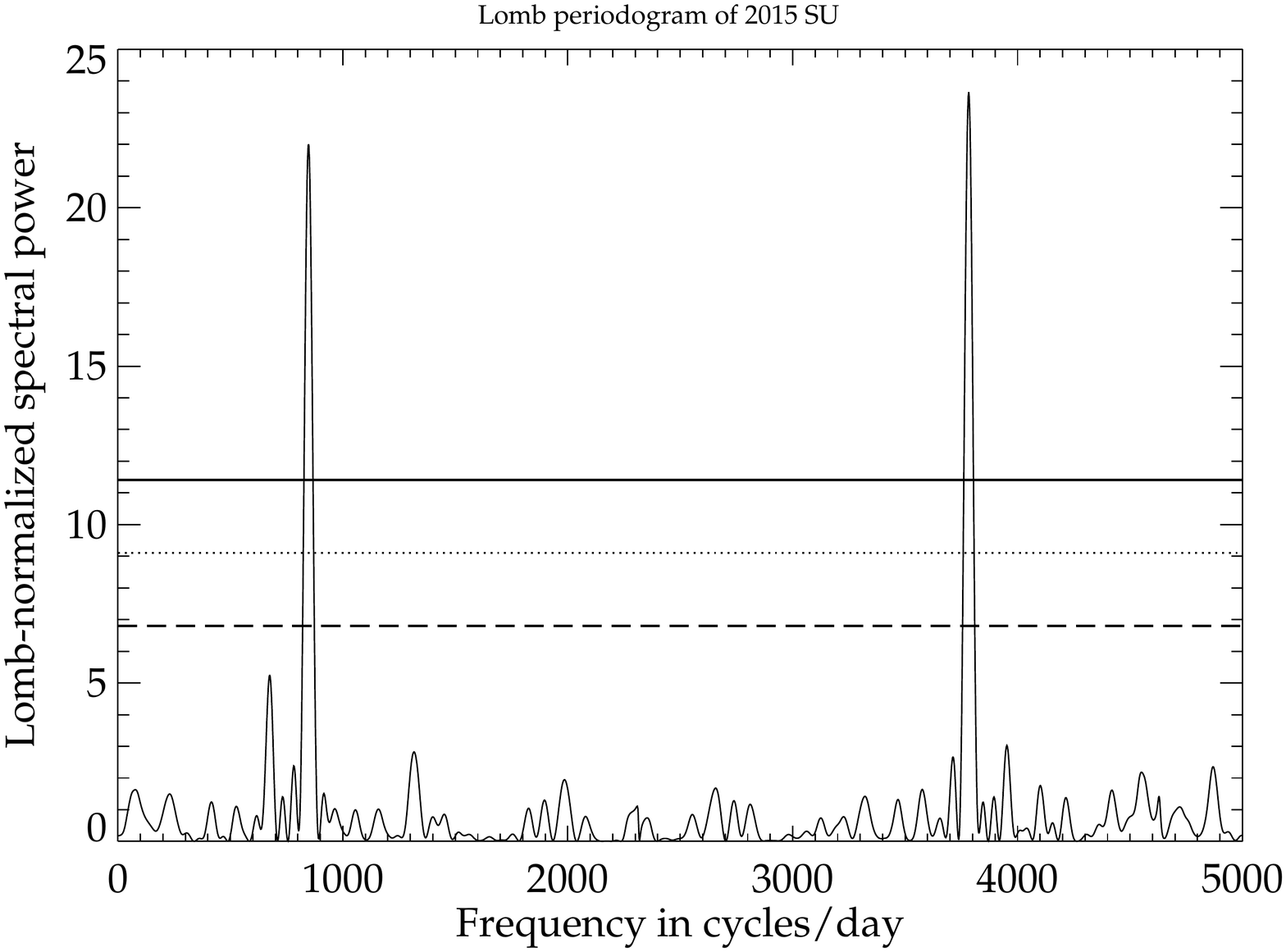}
\includegraphics[width=8cm, angle=0]{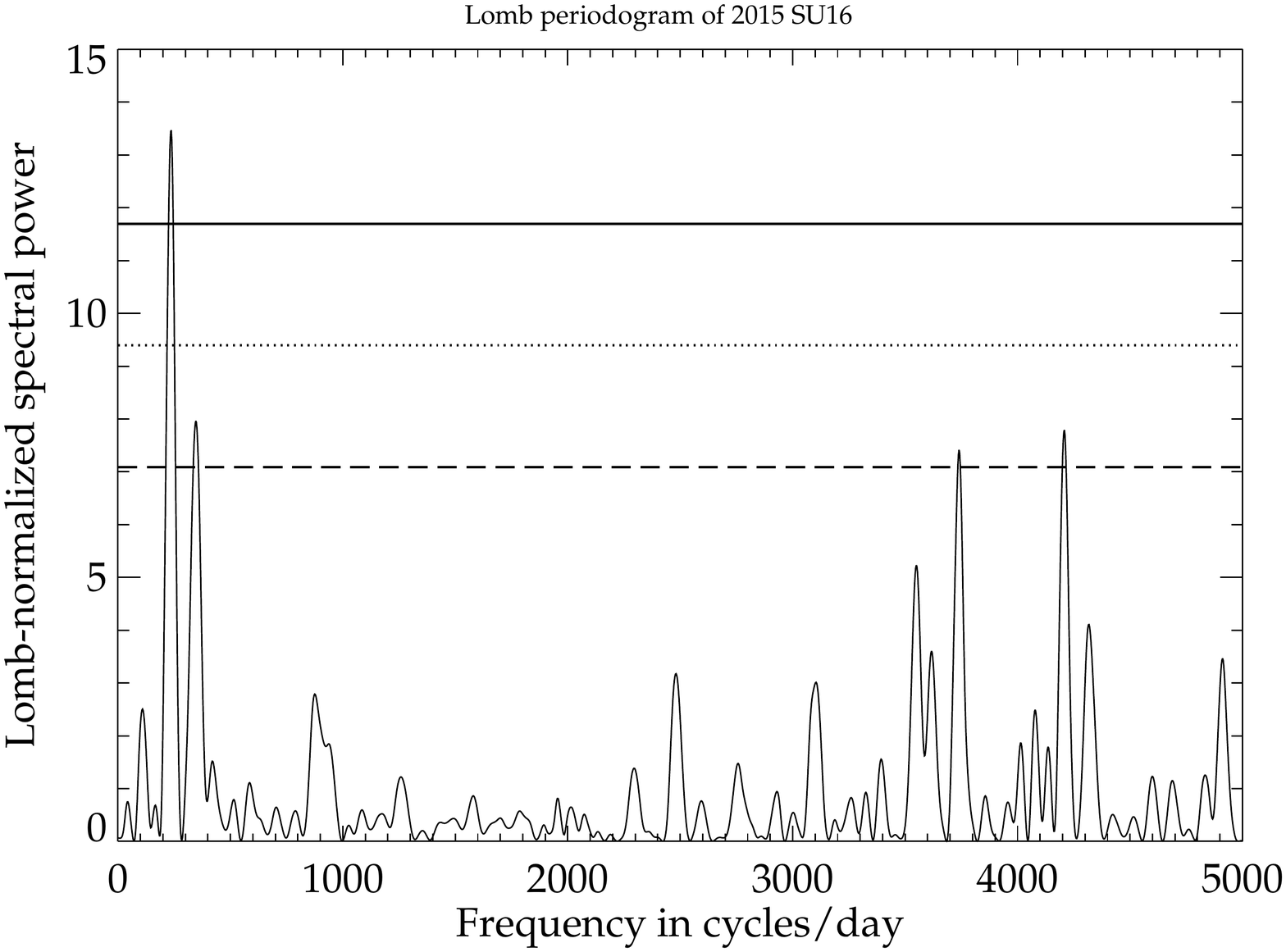}
\includegraphics[width=8cm, angle=0]{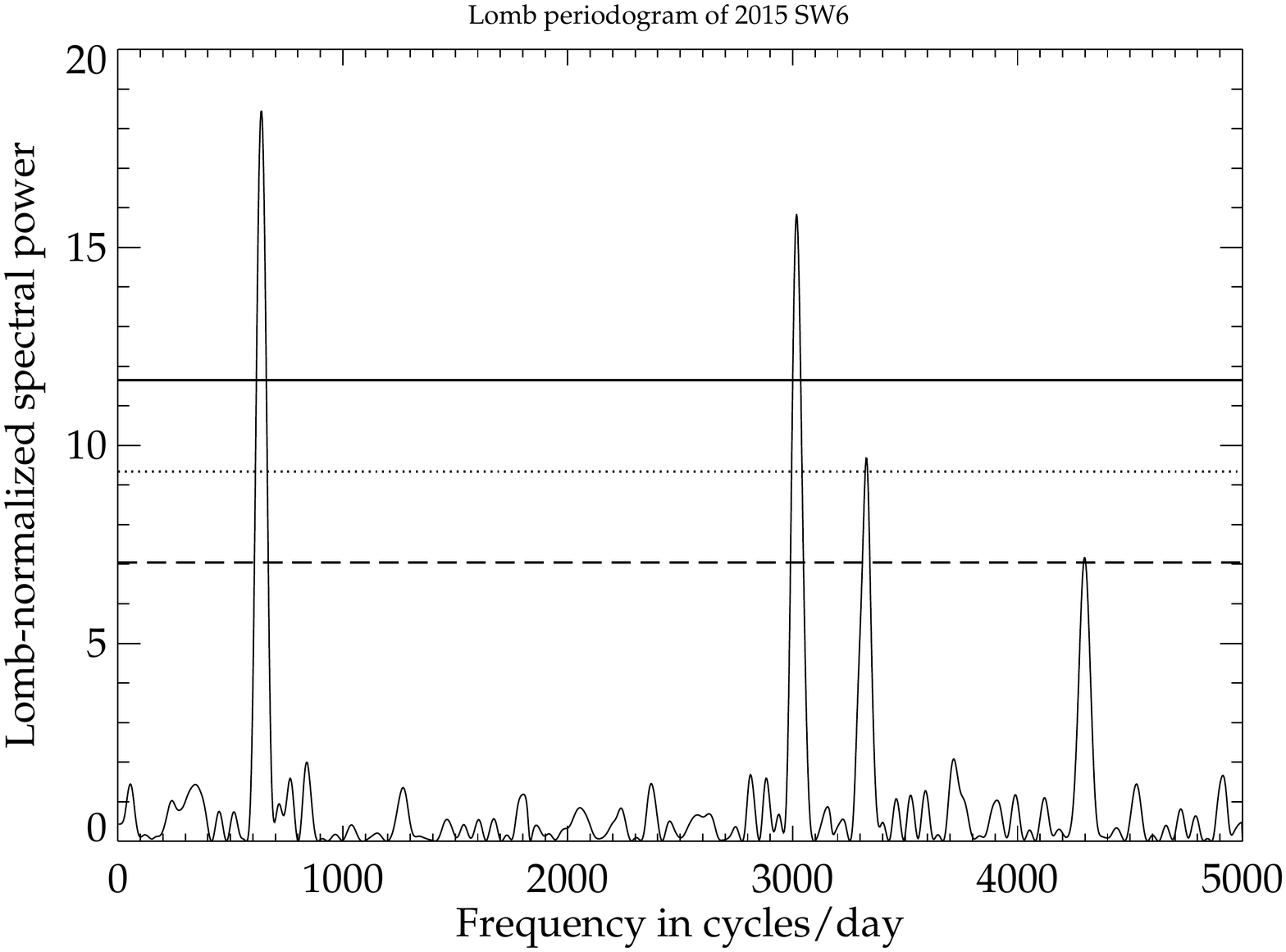}
\includegraphics[width=8cm, angle=0]{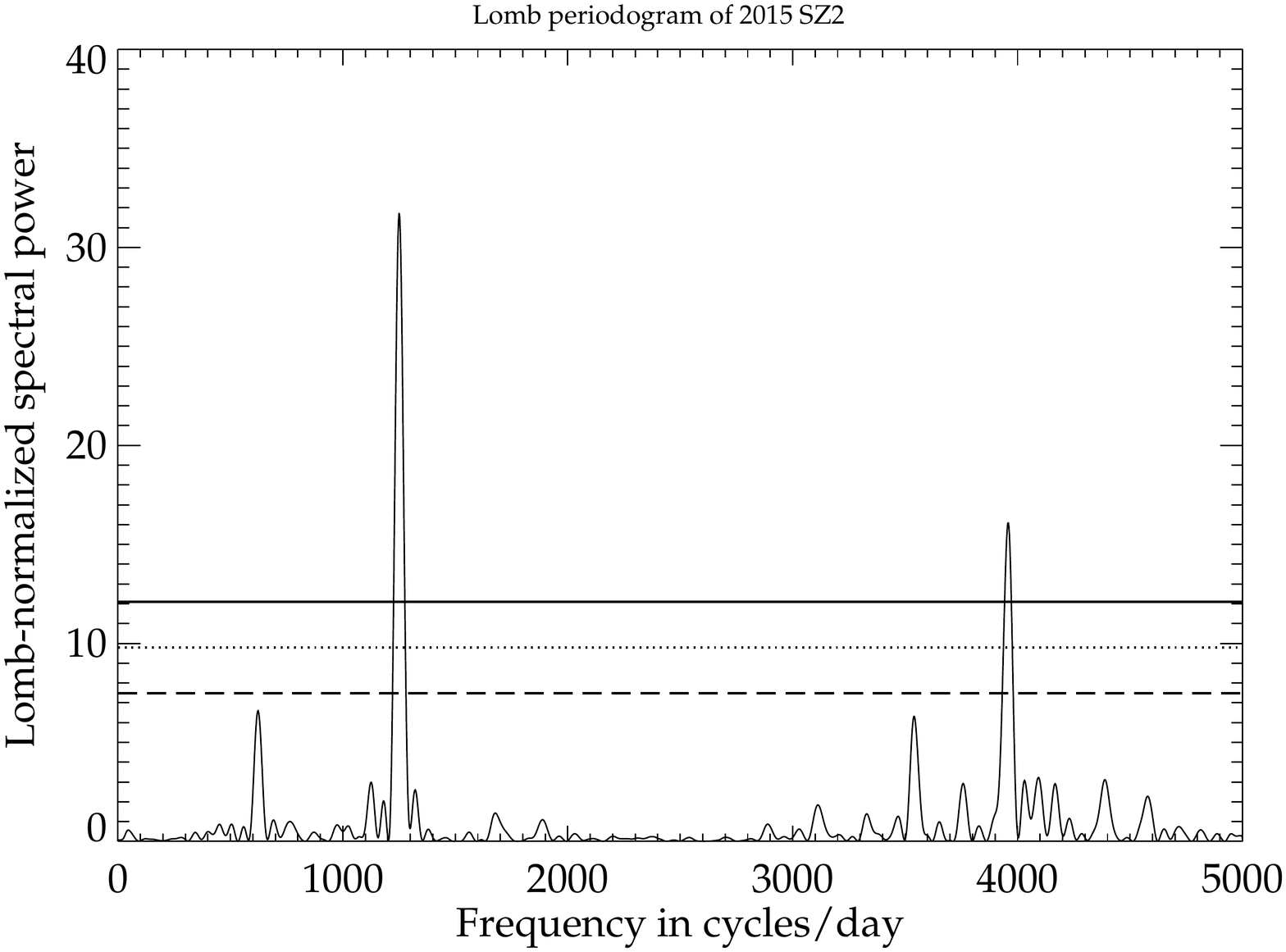}

\caption {\textit{Lomb-normalized spectral power versus Frequency}: Lomb periodograms of MANOS objects are plotted. Continuous line represents a 99.9$\%$ confidence level, dotted line a confidence level of 99$\%$, and the dashed line corresponds to a confidence level of 90$\%$.  }
\label{fig:Lomb10}
\end{figure}

When a possible rotational period is identified, it is useful to know how confident that estimation is. The confidence level is given by:  
\begin{equation}
P(> z) = 1 -(1 - e^{-z})^M
\end{equation}
where M is the number of independent frequencies, and z is the spectral power \citep{Scargle1982, Press1992}. Lomb periodograms with confidence levels of 90$\%$, 99$\%$, and 99.9$\%$ are plotted in Figure~\ref{fig:Lomb1} to Figure~\ref{fig:Lomb10}. Only periodograms for objects with an estimated rotational period are plotted (i.e. periodograms of flat and partial lightcurves are not reported). We considered a large range of frequency up to 5,000~cycles/day which allow us to test our data for short and long periodicities. Care has to be taken to interpret the peaks of the periodogram. For example, there are frequencies not randomly spaced in time such as the exposure time, the read-out time of the instrument, the duration of the observing block, the duration of the pointings, and the aliases of the main peak. These frequencies can be confunded with the main peak (the tallest peak) which corresponds to the rotation of the object. Rotational periods reported in Table~\ref{Summary_photo} correspond to the highest peak. Error bar for the rotational period is the width of the main peak.

%%%%%%%%%%%%%%%%%%%%%%%%%%%%%%%%%%%%%%%%%%%%
%%%% Photometric results
%%%%%%%%%%%%%%%%%%%%%%%%%%%%%%%%%%%%%%%%%%%%

\section{Photometric results}
\label{sec:photo}

\begin{figure}
\center
\includegraphics[width=22cm, angle=90]{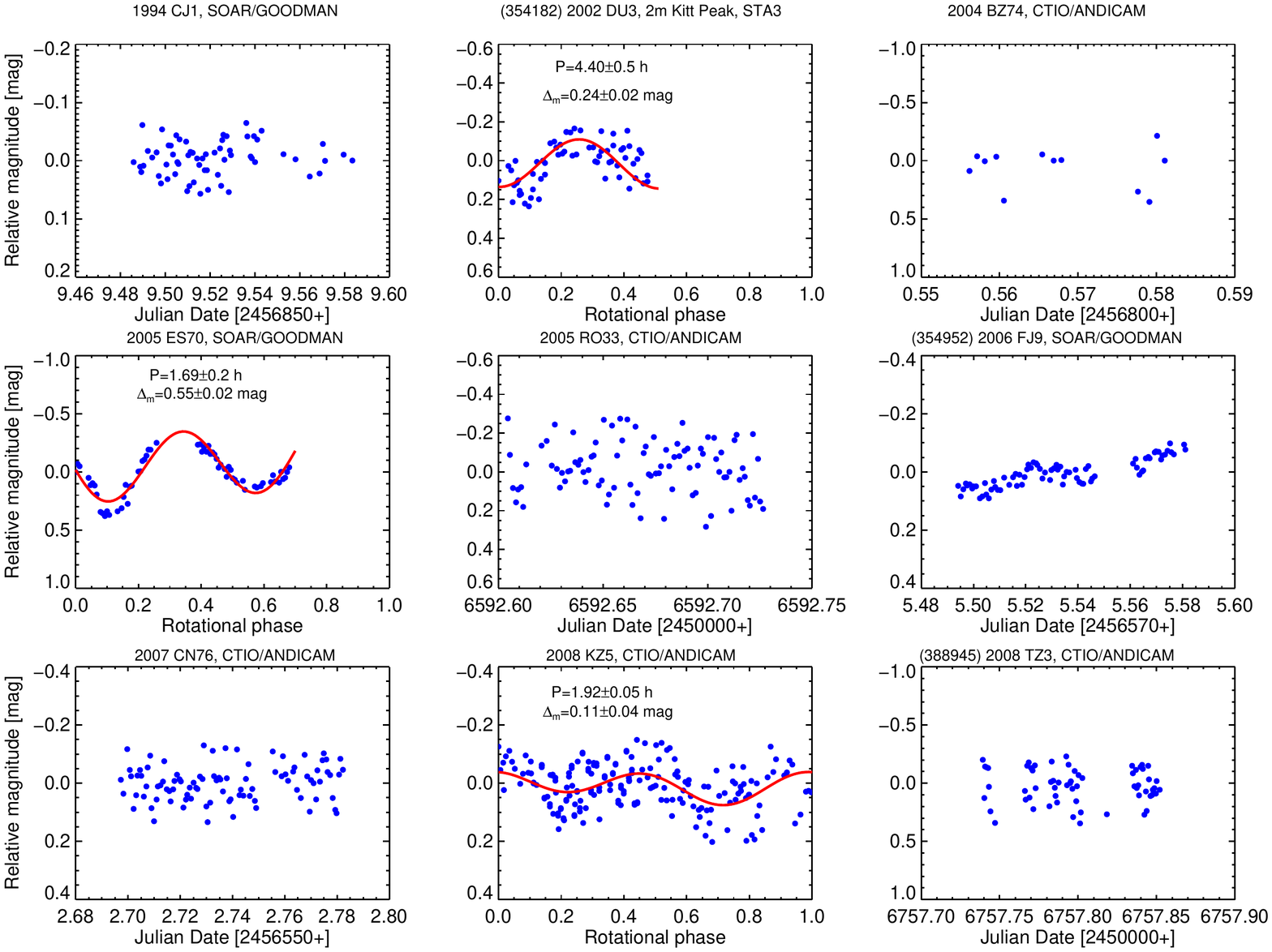}
\caption {\textit{Relative magnitude versus Rotational phase or Julian Date}: MANOS results are plotted.  }
\label{fig:LC1}
\end{figure}

\begin{figure}
\center
\includegraphics[width=22cm, angle=90]{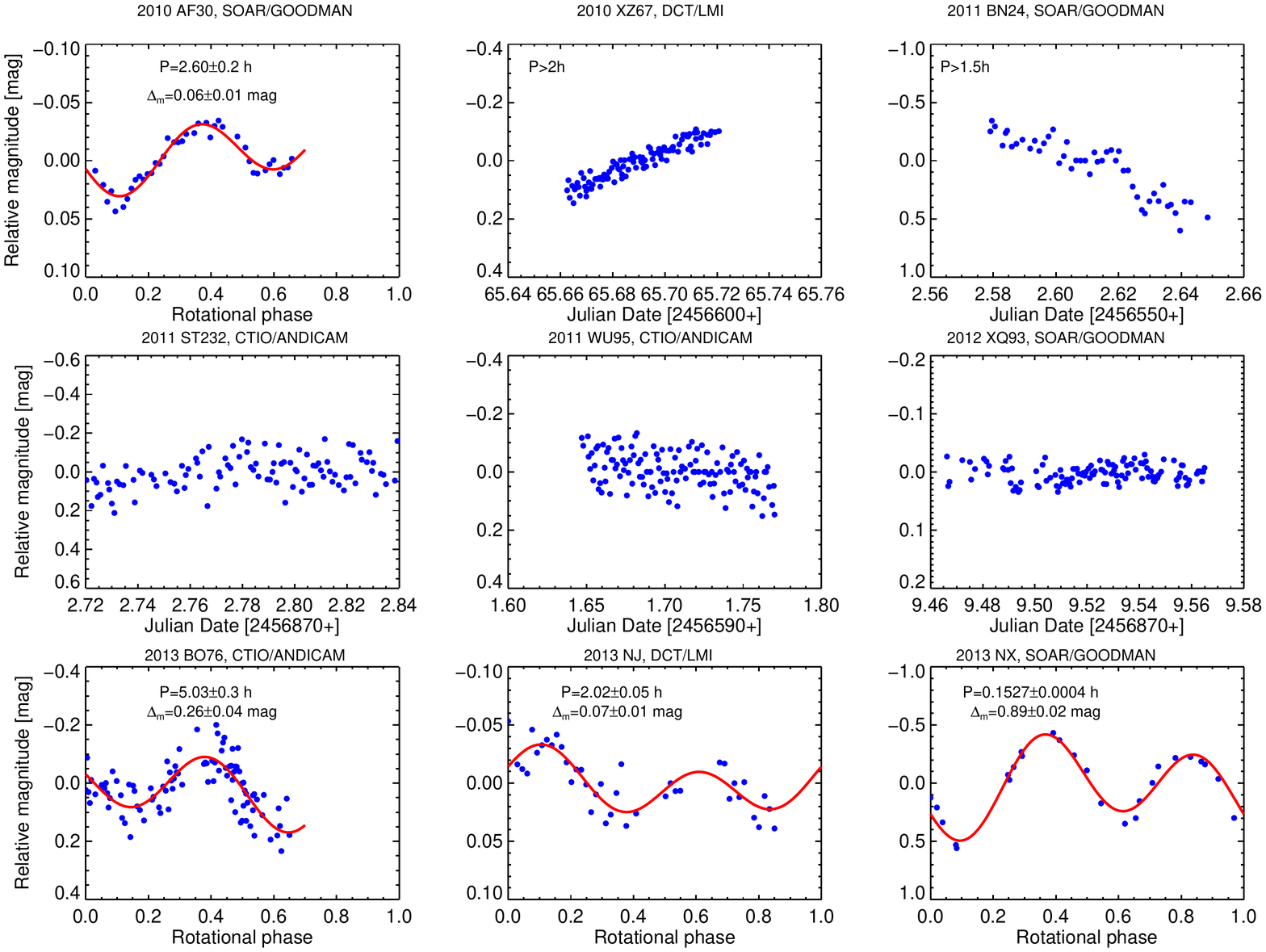}
\caption {\textit{Relative magnitude versus Rotational phase or Julian Date}: MANOS results are plotted.  }
\label{fig:LC2}
\end{figure}

\begin{figure}
\center
\includegraphics[width=22cm, angle=90]{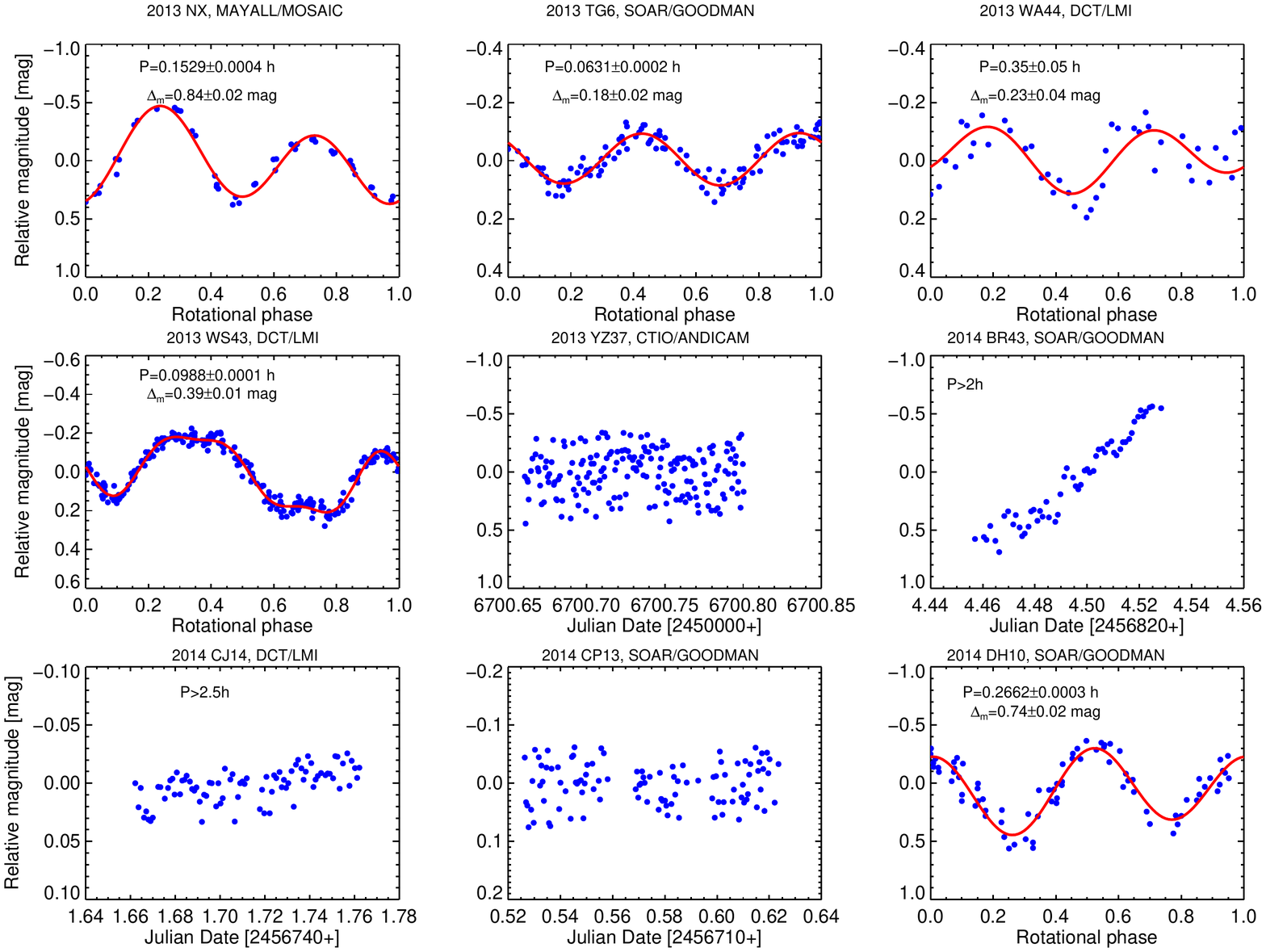}
\caption {\textit{Relative magnitude versus Rotational phase or Julian Date}: MANOS results are plotted.  }
\label{fig:LC3}
\end{figure}

\begin{figure}
\center
\includegraphics[width=22cm, angle=90]{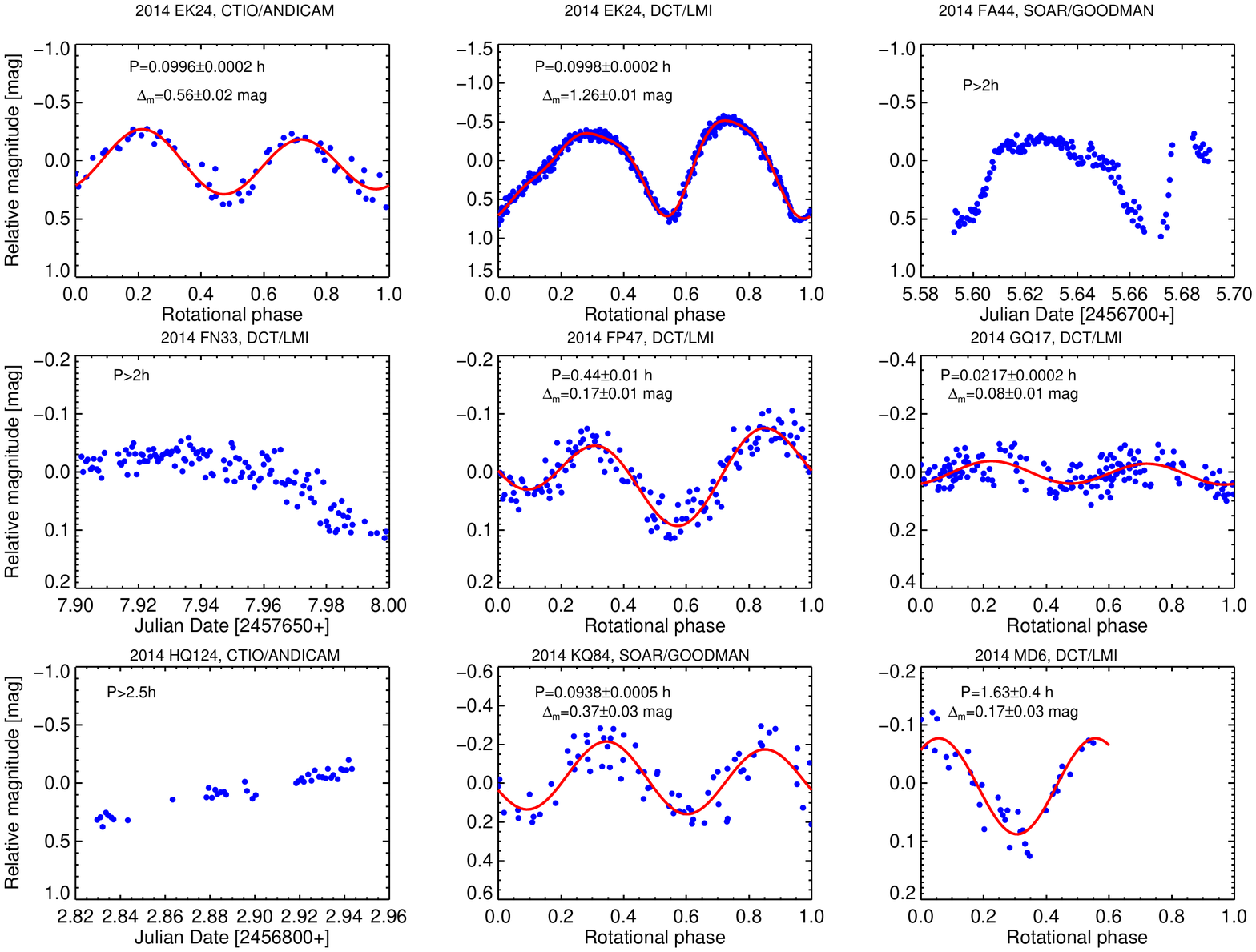}
\caption {\textit{Relative magnitude versus Rotational phase or Julian Date}: MANOS results are plotted.  }
\label{fig:LC4}
\end{figure}

\begin{figure}
\center
\includegraphics[width=22cm, angle=90]{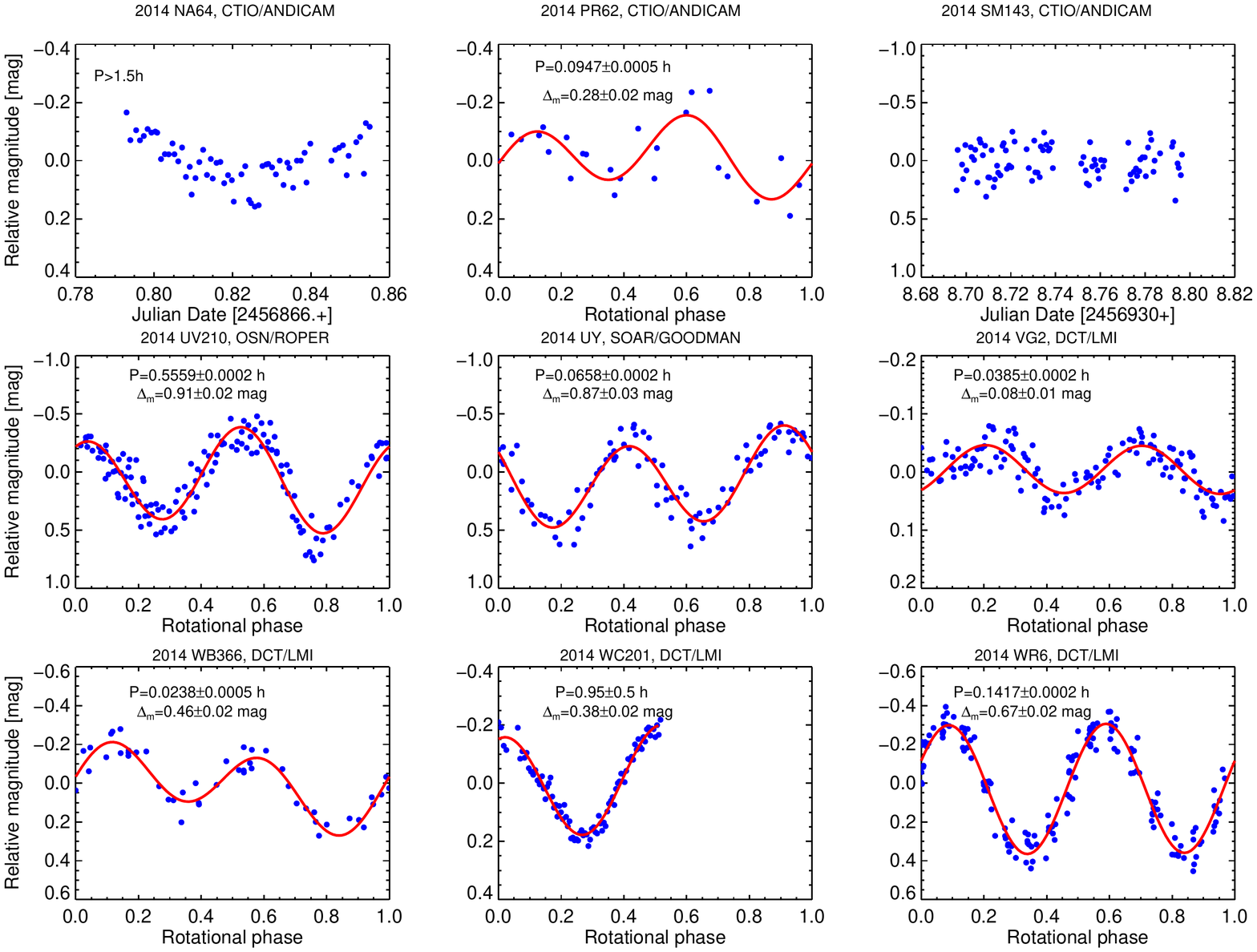}
\caption {\textit{Relative magnitude versus Rotational phase or Julian Date}: MANOS results are plotted. }
\label{fig:LC5}
\end{figure}

\begin{figure}
\center
\includegraphics[width=22cm, angle=90]{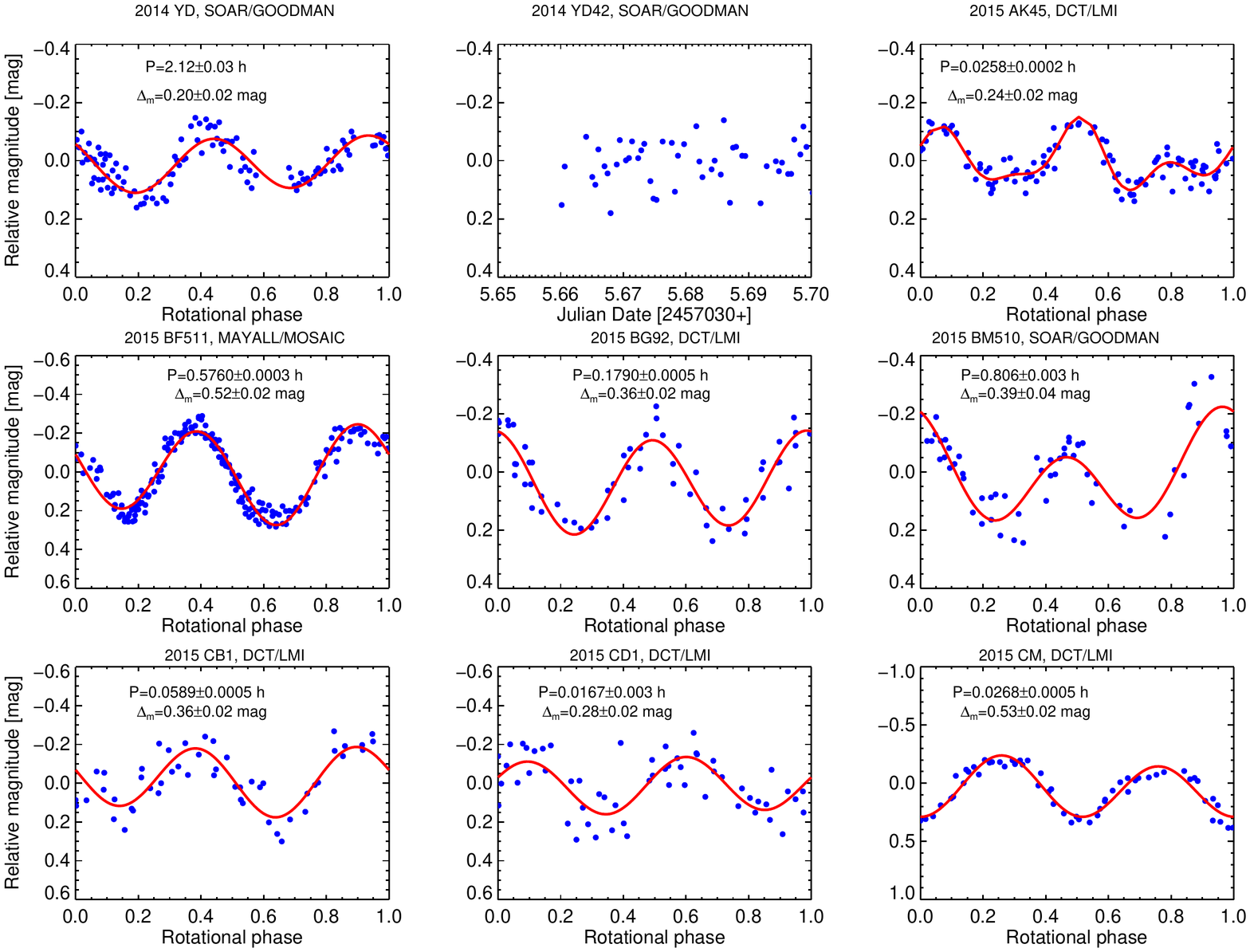}
\caption {\textit{Relative magnitude versus Rotational phase or Julian Date}: MANOS results are plotted.  }
\label{fig:LC6}
\end{figure}

\begin{figure}
\center
\includegraphics[width=22cm, angle=90]{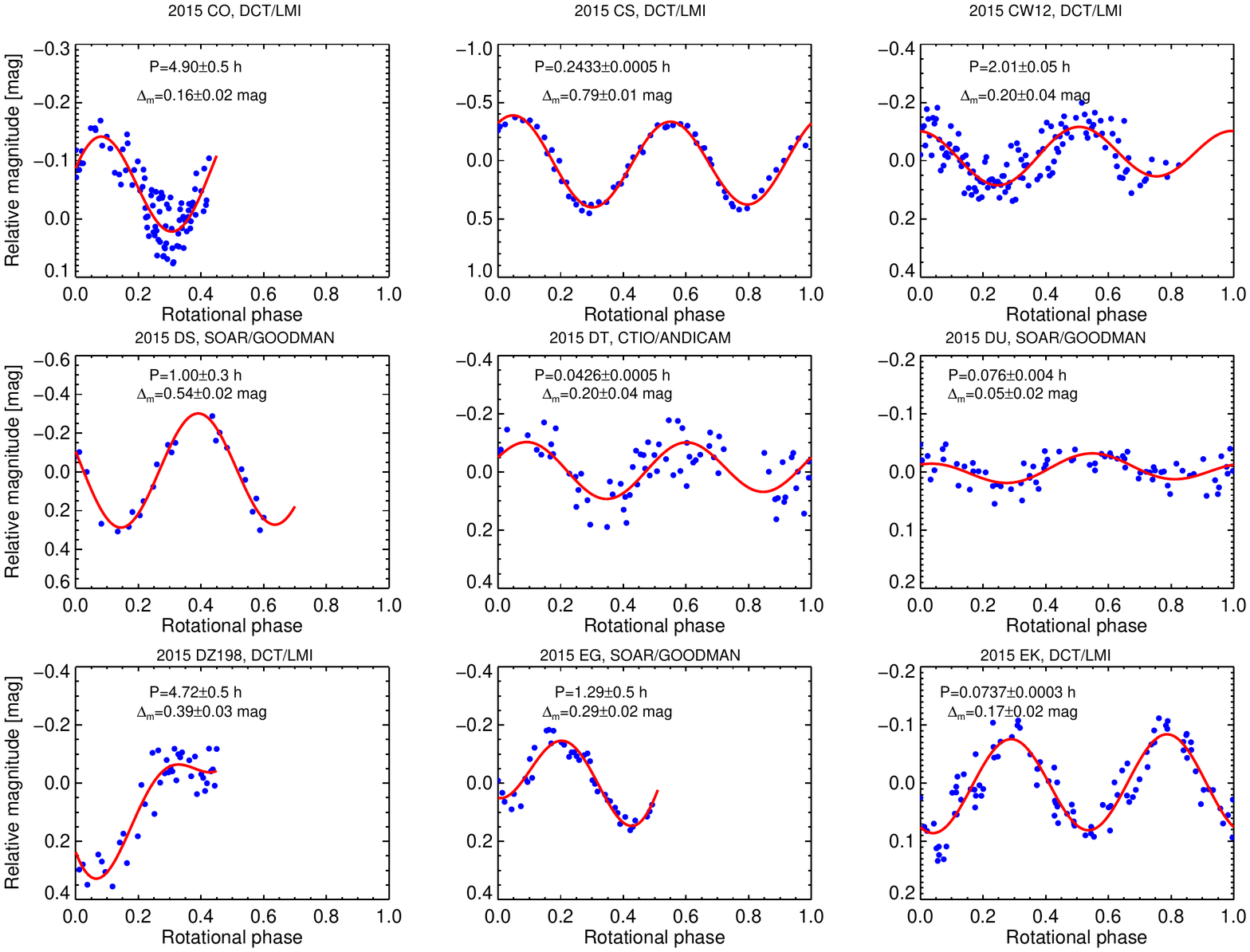}
\caption {\textit{Relative magnitude versus Rotational phase}: MANOS results are plotted.  }
\label{fig:LC7}
\end{figure}

\begin{figure}
\center
\includegraphics[width=22cm, angle=90]{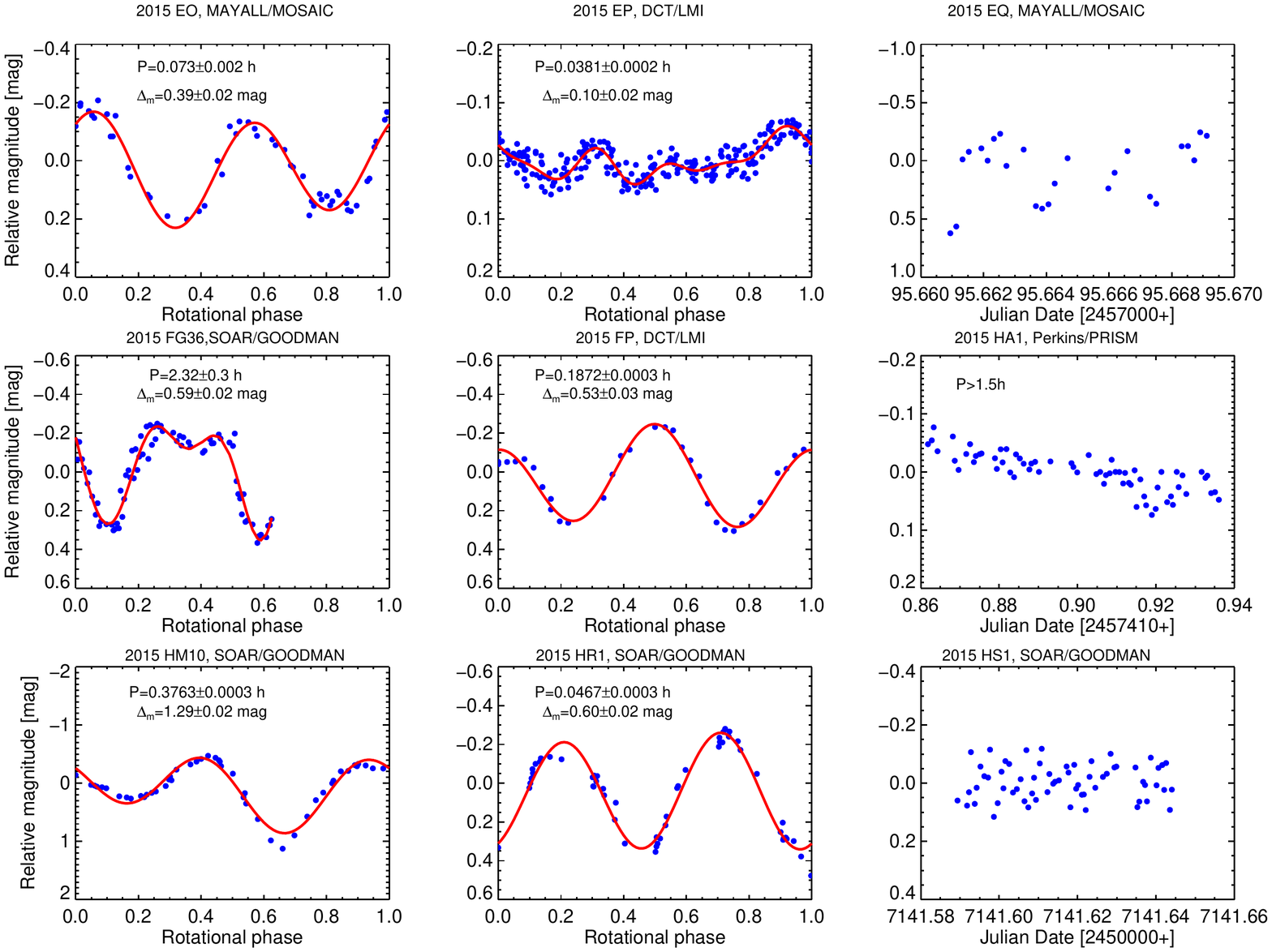}
\caption {\textit{Relative magnitude versus Rotational phase or Julian Date}: MANOS results are plotted.  }
\label{fig:LC0}
\end{figure}

\begin{figure}
\center
\includegraphics[width=22cm, angle=90]{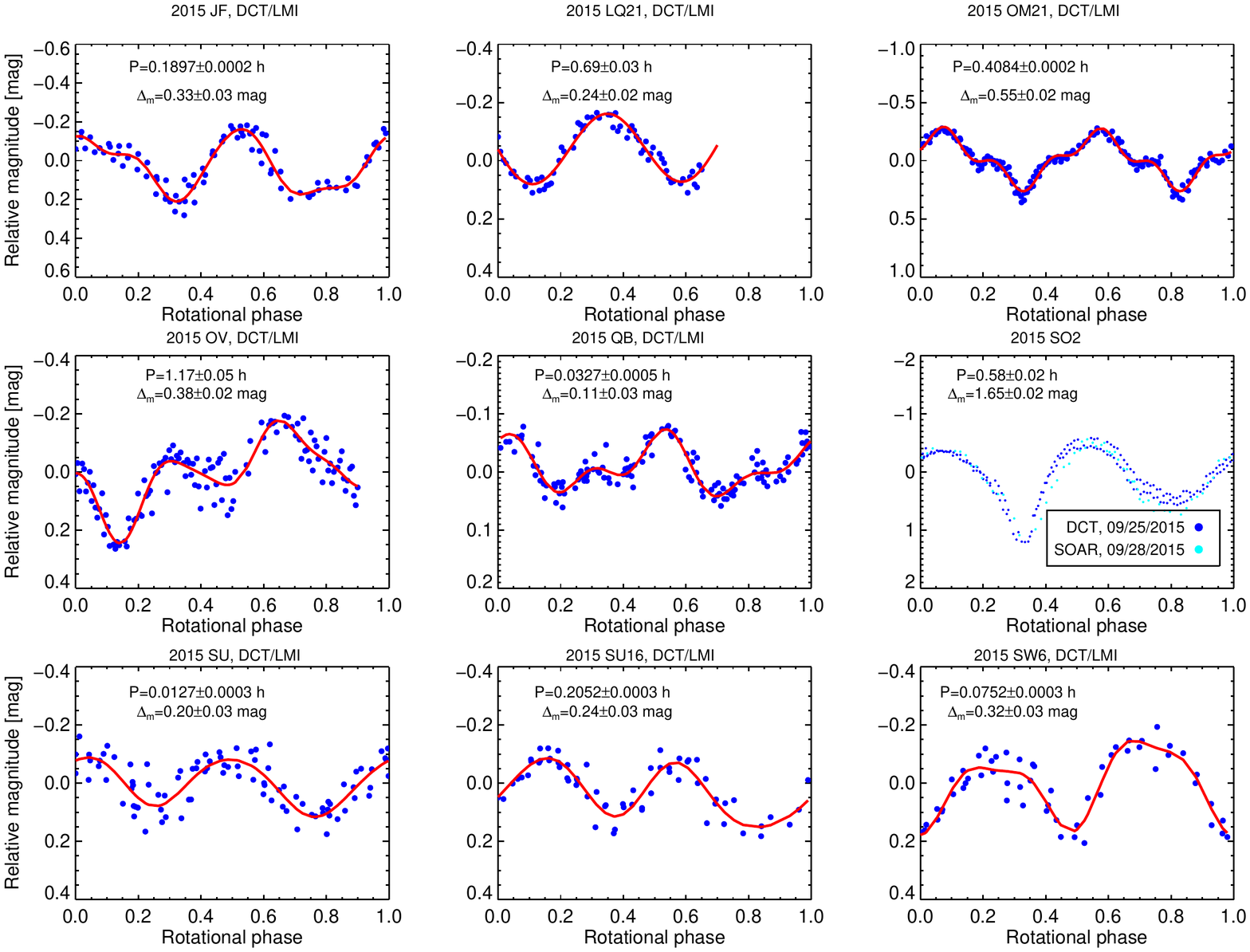}
\caption {\textit{Relative magnitude versus Rotational phase}: MANOS results are plotted.  }
\label{fig:LC8}
\end{figure}

\begin{figure}
\center
\includegraphics[width=22cm, angle=90]{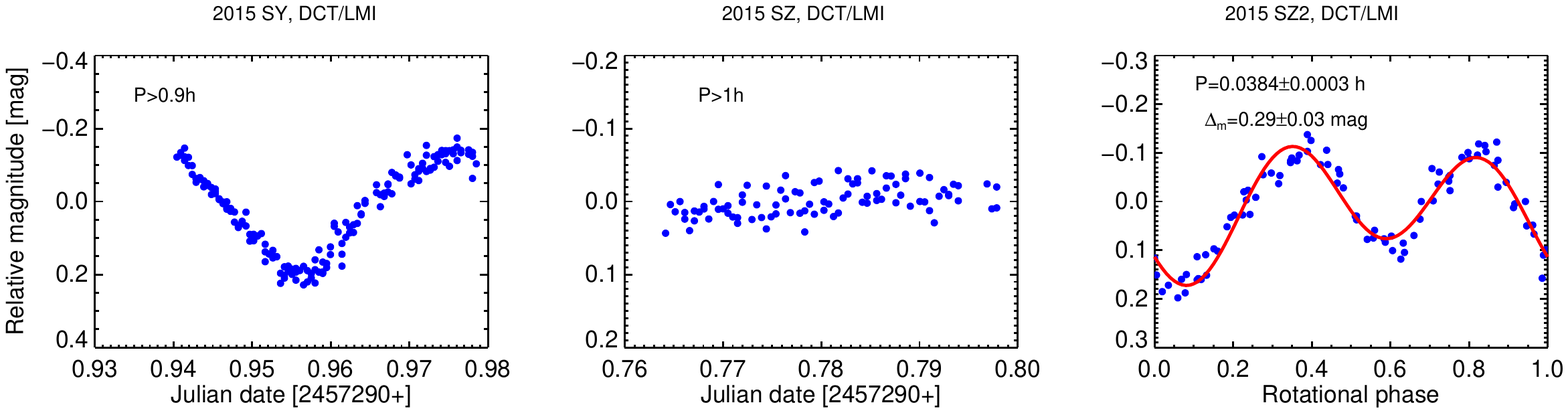}
\caption {\textit{Relative magnitude versus Rotational phase or Julian Date}: MANOS results are plotted.  }
\label{fig:LC9}
\end{figure}

Lightcurves are plotted\footnote{Alternative versions of the lightcurves are available online at \url{https://manosobs.wordpress.com/observations/neo-observing-log/}} in Figure~\ref{fig:LC1} to Figure~\ref{fig:LC9}. For each lightcurve, a Fourier series is fit to the photometric data. The order of the fit depends on the lightcurve morphology. Lightcurves with a clear rotational signature are plotted over one full cycle (rotational phase from 0 to 1). Times for zero phase, without light time correction, corresponding to the beginning of the integration are reported in Table~\ref{Summary_photo}. Error bars for the measurements are not shown on the lightcurves for clarity. Typical error is about 0.02-0.03~mag, but it can be up to 0.04-0.05~mag in case of faint and/or fast moving object. The full photometry with error bars will ultimately be made available in NASA's Planetary Data System (PDS).  
 
We report rotational periods and peak-to-peak amplitudes for 60 objects ($\sim$70$\%$ of our sample), lower limits for amplitude and periodicity for 13 objects ($\sim$15$\%$), and 13 objects ($\sim$15$\%$) that show flat lightcurves without any significant amplitude variation. All relevant geometric information about the objects at the dates of observation, the number of images and filters used are summarized in Table~\ref{Summary_photo}. We divide our sample into three groups: i) \textit{full lightcurve} with at least one full rotation or a significant fraction of one rotation to provide a clear period estimate, ii) \textit{partial lightcurve} with only an increasing or decreasing trend in apparent magnitude, resulting in no periodicity estimate, and iii) \textit{flat lightcurve} with no evident trend in magnitude and no periodicity estimate \citep{Lacerda2003}.

We highlight two ultra-fast rotators: 2014~RC, and 2015~SV$_{6}$. 2014~RC (diameter of $\sim$12~m) had a close encounter with the Earth in September 2014. We obtained several lightcurves before and close to the fly-by and derived a rotational period of 15.8~s. Such fast rotation makes this object the fastest rotator known to date \citep{Warner2009}. MANOS also discovered the second fastest rotator, 2015~SV$_{6}$ (diameter of $\sim$8~m) with a rotational period of about 18~s. A complete study of ultra-rapid rotators is in preparation (Polishook et al, In prep). We include those rotation periods and lightcurve amplitudes in our ensemble analysis.  

Several candidates for tumbling or non-principal axis rotation were identified and will be the topic of future work: 2015~LJ, 2015~CG, 2014~DJ$_{80}$, and 2015~HB$_{177}$ (not reported in Table~\ref{Summary_photo}). Detailled studies of asteroids 2015~AZ$_{43}$ and 2015~TC$_{25}$ will be presented in Kwiatkowski et al. (In prep), and Reddy et al. (In prep), respectively. The rotation periods and amplitudes of these objects are used here as part of our statistical analysis. 

\subsection{Lightcurves}

Photometric brightness variations are produced by several effects: i) albedo variations across a body's surface, ii) non-spherical shapes, and/or iii) contact or eclipsing binary systems. Based on a lack of large scale albedo heterogeneity (i.e. detectable in unresolved images) amongst those NEOs visited by spacecraft (e.g., \citet{Clark2002, Saito2006}) and the lack of binary systems at sizes below $\sim$100~m \citep{Margot2002} the general expectation is that the photometric variability of our targets will be dominated by shape effects and thus the lightcurves will display two maxima and two minima with each rotation. We highlight that the changes of phase angle, and geocentric/heliocentric distances during the observing runs will introduce variations in the lightcurve. However, because we are only observing our targets during a couple of hours, such changes are minimal and are not affecting our lightcurves (see Table~\ref{Summary_photo}).

\subsubsection{Symmetric/Asymmetric lightcurves}

A symmetric lightcurve is one where both peaks reach the same relative magnitude. Only seven MANOS objects have symmetric lightcurves; most of the reported lightcurves are asymmetric with peaks that are not of the same amplitude. In our sample, the typical asymmetry is $<$0.15~mag. However, five objects show larger variations: 2013~NX, 2015~BM$_{510}$, 2015~HM$_{10}$, 2015~SV$_{6}$, and 2015~SO$_{2}$.  
\begin{itemize}
\item 2013~NX was observed at two epochs in March and August 2015 with solar phase angles corresponding to $\sim$56$^{\circ}$ and $\sim$31$^{\circ}$ respectively. Lightcurve period and amplitude were derived from both epoch. Both lightcurves are asymmetric with peaks that differ in amplitude by about 0.2~mag (Figure~\ref{fig:LC2}, and  Figure~\ref{fig:LC3}). 
\item 2015~BM$_{510}$ was observed with the 4~m SOAR telescope in February 2015. The peaks differ in amplitude by 0.17~mag (Figure~\ref{fig:LC6}).  
\item 2015~HM$_{10}$ was observed with the 4~m SOAR telescope in June 2015. 2015~HM$_{10}$ shows an asymmetry of about 0.5~mag (Figure~\ref{fig:LC0}). 2015~HM$_{10}$ was observed with the Goldstone radar facility\footnote{\url{http://echo.jpl.nasa.gov/asteroids/}} during its close approach with Earth on the 7 July 2015 \citep{Busch2015}. A rotational period of $\sim$0.4~h and elongated shape as suggested by our lightcurve data were confirmed by the radar observations.  
\item 2015~SO$_{2}$ was observed on 25 September 2015 and 28 September 2015 with DCT and SOAR, respectively. Phase angles changed from $\sim$58$^{\circ}$ to $\sim$63$^{\circ}$ across those dates. A rotation period of 0.58~h and lightcurve amplitude of 1.65~mag are consistent in both datasets. This object has the highest variability in our sample. The morphology of the lightcurve is also noteworthy (Figure~\ref{fig:LC8}). The first peak has a V-shape characteristic of a contact binary object, and an amplitude of about 0.9~mag compared to the first minimum. In contrast, the second minimum is U-shaped and is deeper by about 0.2~mag.  
\item 2015~SV$_{6}$ has an absolute magnitude of 27.7 which corresponds to a diameter of 8~m assuming an albedo of 0.2. Observations of 2015~SV$_{6}$ were challenging because of the rapid sky motion of the object ($\sim$155$\arcsec$/min), and its apparent brightness (visual magnitude of $\sim$18.8). Due to this rapid sky motion, we used an exposure time of 1~s to avoid appreciable trailing of the object. With such short exposures we were sensitive to its very fast rotation of $\sim$18~s. The second interesting feature of this object is its lightcurve with a strong asymmetry of about 0.3~mag (Polishook et al., In prep).     
\end{itemize}

\subsubsection{Complex shape}

Ten MANOS targets display complex lightcurves that require higher order harmonics (i.e. more than two harmonics) in the fit: 2013~WS$_{43}$, 2015~AK$_{45}$, 2015~EP, 2015~FG$_{36}$, 2015~JF, 2015~OM$_{21}$, 2015~OV, 2015~QB, 2015~SW$_{6}$, and 2015~TC$_{25}$. Such curves can only be explained by complex shape and/or strong albedo variations. Such objects present an opportunity for future shape modeling pending the addition of multi-epoch observations. None of these objects have been observed by radar to provide any additional shape information.  

\subsubsection{Partial and flat lightcurves}

We also report partial lightcurves that show only a trend of increasing or decreasing magnitude. Because the data cover less than half of the object's rotation, we are not able to derive a secure rotational period. Therefore, we report lower limits for these objects' rotation periods and amplitudes. These results are only limits and are not used for our statistical study (see next section). 

Some objects show no measurable photometric variations. In such a case, lightcurves are called flat lightcurves. Several causes can explain these lightcurves. The object may have i) a long rotational period undetectable during our observing block, ii) an almost pole-on orientation, iii) a spherical shape, iv) a very rapid rotation period comparable to or much less than the integration time per exposure. Additional observations would be needed to secure a rotational period for these objects.

\subsubsection{Lightcurves with mutual eclispes}

For binary systems, the lightcurve may present mutual eclipses due to the companion passing in front or behind the primary \citep{Pravec2006}. In our sample, the lightcurve of 2014~FA$_{44}$ may present some evidence of mutual eclipse (Figure 5.). However, the lightcurve is incomplete and only more data at several epochs can confirm or not the presence of a companion. 

\section{Discussion}
\subsection{Dataset}

We report a dataset reduced and analyzed with the same methods that is well suited to statistical study. However, we have also used the lightcurve database by \citet{Warner2009} to increase the sample size (LCDB refers to the lightcurve database hereafter). Merging our sample and the NEOs in the LCDB results in a sample of 906 objects (with rotational period and lightcurve amplitude). The LCDB uses a reliability code or quality rating to categorize lightcurves. This system is based on the work of \citet{Lagerkvist1989}, who defined a reliability code from 1 (tentative result) to 4 (multiple apparition coverage, and pole position reported). \citet{Warner2009} suggested that a pole solution did not necessarily reflect the quality of the lightcurve, and thus removed the \textit{4} code. Their ranking is: \textit{1}: result based on fragmentary lightcurve(s), may be completely wrong, \textit{2}: result based on less than full coverage, so period may be wrong by 30$\%$ or so, \textit{3}: secure result. MANOS lightcurves have a reliability code of 2 or 3 based on this classification. 
 
Fifty-four objects ($\approx$5~$\%$ of the LCDB) have a code of 1, 365 objects ($\approx$37~$\%$) with a code of 2, 396 objects ($\approx$40~$\%$) with a code of 3, and 180 objects ($\approx$18~$\%$) with no code (only upper/lower limit on the rotational period and/or lightcurve amplitude). We do not make use of the \citet{Warner2009} \textit{+} and \textit{-} sub-division codes. Here, we consider nearly all objects independent of their reliability code. In fact, only 5$\%$ of the LCDB is composed of code values = 1, and thus their contribution to a statistical study is minor. Furthermore, \citet{Binzel1989} pointed out several biases inherent to  asteroid lightcurve literature that argues against removing low quality data. They stressed that excluding poor reliability objects results in overweighting objects with large amplitude and short rotational periods. The exception to our inclusion of the full LCDB are those objects with no reliability code. These objects do not have constraints on their lightcurve properties and cannot be included in our analysis.

\subsection{Rotational frequency distribution}
\label{sec:freq}

\begin{figure}
\centerline{\includegraphics[width=12cm]{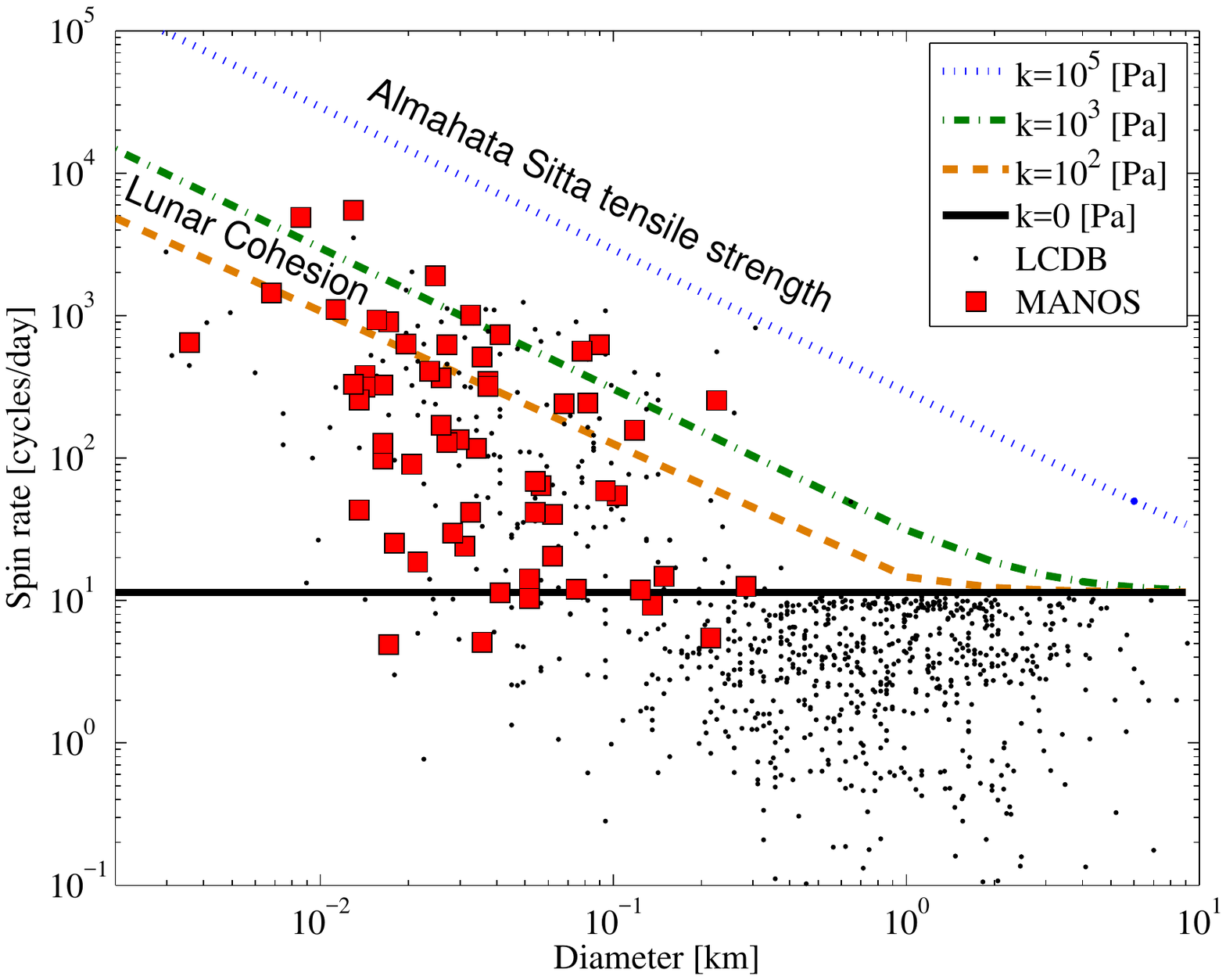}}
\caption{The diameter-spin diagram of asteroids in the LCDB (dots) and the MANOS sample (squares). The lines represent boundaries for zero cohesion (solid line), 100~Pa (lower limit for lunar regolith; dashed line), 1000~Pa (higher limit for lunar regolith; dot dash line), and 10$^{5}$~Pa (a lower limit for the tensile strength of the Almahata Sitta meteorite, dotted line). The lines were determined for bodies with $\rho$=3.3~g~cm$^{-3}$ and lightcurve amplitude of 0.5~mag, but it should be noted that the minimal cohesion is hardly sensitive to these parameters, and using $\rho$=2~g~cm$^{-3}$ and lightcurve amplitude of 0.1~mag give us similar results. Diameter was computed assuming an albedo of 0.2.
\label{fig:Spin}}
\end{figure}

All asteroids  with rotational periods reported in the literature are plotted in Figure~\ref{fig:Spin} (LCDB by \citet{Warner2009}). It has been shown that asteroids with sizes from a few hundred meters ($\sim$200~m) up to about 10~km show a spin deformation limit at $\sim$2.2~h \citep{Pravec2002}. In other words, this boundary is interpreted as a critical spin limit for rubble piles in the gravity regime. This limit disappears at diameters less than 200~m suggesting that cohesion is important for the smallest asteroids. In fact, fast rotators cannot be held together by self-gravitation only (see Section~\ref{sec:barrier} for more details). 

MANOS objects are clustered in the left upper part of Figure~\ref{fig:Spin}. As already mentioned, our survey focuses on small NEOs and is sensitive to rotational periods from about 16~s to $\sim$5~h. Approximately 50$\%$ of MANOS objects are spinning fast, in less than 5~min which is expected in this size range. However, we highlight the ``slow" rotation of 2015~CO. This object has an absolute magnitude of 26.2, so an approximate diameter of 17~m assuming an albedo of 0.2, and its rotational period is 4.9~h. 2015~CO is the slowest rotator amongst objects smaller than 20~m. In Figure~\ref{fig:Spin}, asteroid diameters are reported assuming an albedo of 0.2. The diameter (D) according to \citet{Pravec2007}, can be estimated by: 

 \begin{equation}
D = {\frac{K}{\sqrt{p} }} 10^{-0.2H}
 \end{equation}
where p is the geometric albedo, and H is the absolute magnitude. The constant K is: 
 \begin{equation}
K = 2 AU \times 10^{\frac{V_{sun}}{5}}
 \end{equation}
where V$_{sun}$ is the visual magnitude of the Sun. 
Previous formulae are wavelength-dependent \citep{Pravec2007}. The constant K is 1329 in the V-band and 1137 in the R-band. Absolute magnitudes in Table~\ref{Summary_photo} are from the Minor Planet Center\footnote{Absolute magnitudes are available at: \url{http://www.minorplanetcenter.net/iau/lists/MPLists.html}. Absolute magnitudes listed in Table~\ref{Summary_photo} are from February 2016.} (MPC) and have a typical error bar of 0.5~mag \citep{Juric2002, Pravec2012}. Assuming the R-band (V-band), we derive a diameter of 5$\pm$1~m (4$\pm$1~m) for an object with an absolute magnitude of H=29$\pm$0.5. Geometric albedo is not available for all objects reported here, so we used a default value of 0.2. Within the error bars, the values are consistent, but one must keep in mind that such values are only rough estimates. Diameters for MANOS objects have been estimated assuming the R-band and are summarized in Table~\ref{Summary_photo}.     

\begin{figure}
\center
\includegraphics[width=18cm, angle=0]{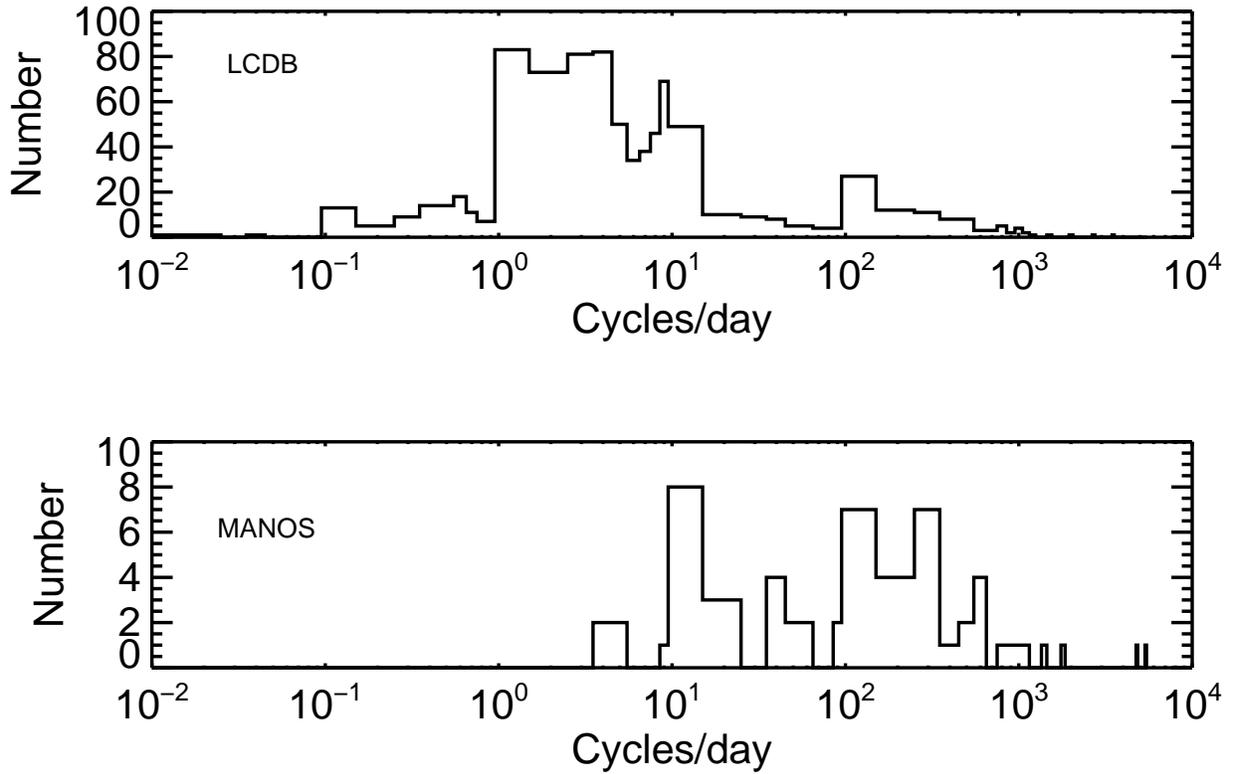}
\caption {\textit{Rotational frequency distributions}: We plotted all NEOs with a known rotational period reported in the lightcurve database (LCDB by \citet{Warner2009}, upper plot), and MANOS results (lower plot). There
are several biases in these datasets: i) lack of objects with a rotational period
longer than a 1~day, and ii) lack of ultra-rapid rotators. }
\label{fig:LCDBMANOS}
\end{figure}

Figure~\ref{fig:LCDBMANOS} shows all NEOs reported in the LCDB and in the MANOS sample. There are several biases in these datasets. First bias is the lack of objects with a rotational period longer than a single day. Long rotation periods are difficult to determine due to alias effects and a requirement for long duration observations. Furthermore, null results or failed attempts to derive lightcurves are rarely published, which exacerbates this bias in the literature. Our observing strategy is not sensitive to slow rotation, and no objects with periods greater than about 5 hours are reported in our sample (see Section~\ref{sec:obs} for more details).

The second bias is against ultra-rapid rotators. From sampling theory (e.g., \citet{Nyquist1928}) it is known that periodic signals can only be reconstructed when sampled at a rate of at least twice the signal frequency. Larger aperture telescopes can generally employ shorter exposure times and thus are more sensitive to short rotational periods. For example, 2014~RC and 2015~SV$_{6}$ were observed with exposure times of $\sim$~1-2~s. As such, MANOS is sensitive to ultra-rapid rotation due to the regular use of 4-meter-aperture facilities, but that is generally not the case for the majority of current asteroid lightcurve surveys. Our sensitivity to rotational periods $<$~1~minute is a novel benefit of the MANOS observing strategy, but we are still not able to probe periods comparable to observational cadences set by individual exposure times. In other words, we are not able to directly measure whether rotation periods $<<$~10~seconds exist among small NEOs. 

\subsubsection{Rotational period versus Absolute magnitude}  

\begin{figure}
\includegraphics[width=18cm, angle=0]{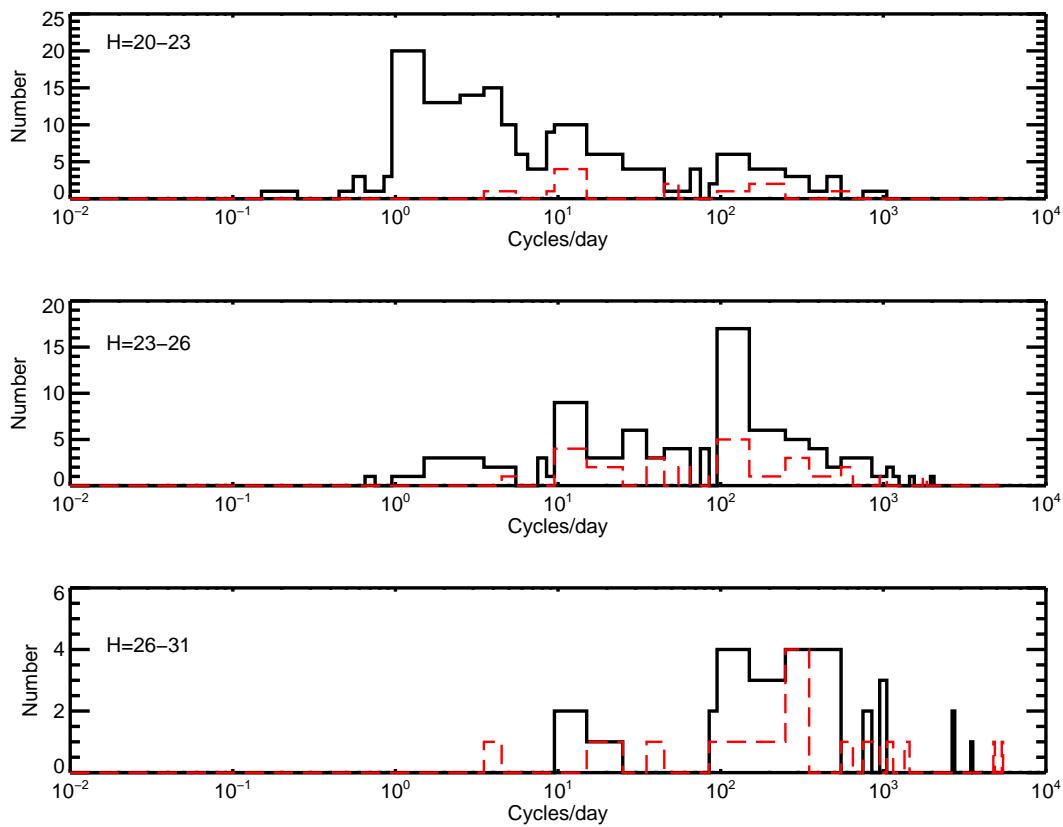}
\caption {\textit{Number of Objects versus cycles/day}: All NEOs with a known rotational period reported in the lightcurve database (black, \citet{Warner2009}), and MANOS results (red) reported in this work are plotted. Objects have been sub-divided based on their absolute magnitude. For objects with H=20-23, the mean rotational frequency is 71~cycles/day with a standard deviation of 171~cycles/day. For objects in the range
H=23-26 (H=26-31), the mean frequency is 270~cycles/day (745~cycles/day) and the
standard deviation is 380~cycles/day (1201~cycles/day). 
 }
\label{fig:HistoSize}
\end{figure}

In Figure~\ref{fig:HistoSize}, we plot NEOs with a rotational period available from the LCDB and the MANOS sample. This full sample has been divided according to the absolute magnitude (H) of the objects: H=20-23, 23-26 and a large bin for the smallest objects with H=26-31. We only consider objects with an absolute magnitude higher than 20 (i.e. diameter smaller than $\sim$300~m assuming an albedo of 0.2) due to the MANOS focus on objects in this size range. 

Despite the still limited MANOS sample, we note that our distribution is similar to the LCDB. In the size range H=26-31, the distributions are sparse and it is difficult to distinguish an underlying distribution. For objects in the size range H=20-23, the mean rotational frequency is 71~cycles/day with a standard deviation of 171~cycles/day. For objects in the range H=23-26 (H=26-31), the mean frequency is 270~cycles/day (745~cycles/day) and the standard deviation is 380~cycles/day (1201~cycles/day). Based on Figure~\ref{fig:HistoSize}, it is clear that rotational frequency distribution is size-dependent.  

\citet{Binzel1989} concluded that for asteroids with a diameter D$>$125 km, a Maxwellian distribution is
able to fit the observed rotation rate distributions implying that their rotation rates may be determined
by collisional evolution. However, for asteroids with a diameter D$<$125 km, there is an
excess of slow rotators and their non-Maxwellian distributions suggests that their rotation rates are
more strongly influenced by other process. An updated version of \citet{Binzel1989} by \citet{Pravec2002} showed that the rotational frequency distribution for large main belt asteroids (diameter larger than 40~km) can be fit by a Maxwellian distribution, but for very small NEOs, a Maxwellian fit is not able to match the observations. Based on our sample, we have an excess of slow (objects rotating in hours) and fast rotators (objects rotating in few minutes) which do not allow us to fit a Maxwellian distribution. 

Several ideas have been proposed to explain the existence of these fast and slow rotators. The main processes to consider are radiation pressure effects (YORP), and gravitational interactions with planets during close encounters \citep{Richardson1998, Scheeres2000, Rubincam2000, Pravec2000, Bottke2002}. These effects can spin up or spin down objects, thus broadening the overall distribution of rotation rates. It is also thought that small objects are fragments of larger objects that have suffered a catastrophic collision \citep{Morbidelli2002}. This kind of collision produces fragments that are ejected and may have fast rotations. Tidal evolution in a binary system can slow down rotation rates, but there are no known binary systems among objects with a diameter $<$~100~m.

\subsubsection{Rotational period versus Dynamical class} 

The NEO population is traditionally divided into four sub-categories: i) \textit{Amor} with a semi-major axis a$>$1~AU and a perihelion distance q where 1.017$<$q$<$1.3~AU, ii) \textit{Apollo} with a$>$1~AU and q$<$1.017~AU, iii) \textit{Aten} with a$<$1~AU and an aphelion distance Q$>$0.983~AU, and iv) \textit{Atira} with a$<$1~AU and an aphelion distance Q$<$0.983~AU. Atira NEOs, with orbits entirely interior to the Earth's, are difficult to detect and make up such a small fraction of the known NEO population ($<<1\%$) that we do not consider them further here. As of April 2016, the Minor Planet Center (MPC) cataloged 6080 Amor ($\sim$43$\%$ of the entire NEO population), 7038 Apollo ($\sim$50$\%$), and 1048 Aten ($\sim$7$\%$) NEOs. Based on de-biased distributions, \citet{Bottke2002} estimated that 62$\%$ of known NEOs are Apollo, 32$\%$ are Amor, and 6$\%$ are Aten. Here, we report 39 Apollo ($\sim$46$\%$ of our sample), 36 Amor ($\sim$43$\%$), and 10 Aten ($\sim$11$\%$). In spite of our focus on low $\Delta$v objects, the distribution of our targets within each of these dynamical classes is reasonably close to the de-biased relative fractions in \citet{Bottke2002}. Recent estimates by \citet{Greenstreet2012} are consistent with \citet{Bottke2002} results, but \citet{Mainzer2012} estimates differ a little with an Aten population of 8$\pm$4$\%$, 55$\pm$18$\%$ for the Apollo group and 37$\pm$16$\%$ for the Amor sample.

\begin{figure}
\center
\includegraphics[width=19cm, angle=0]{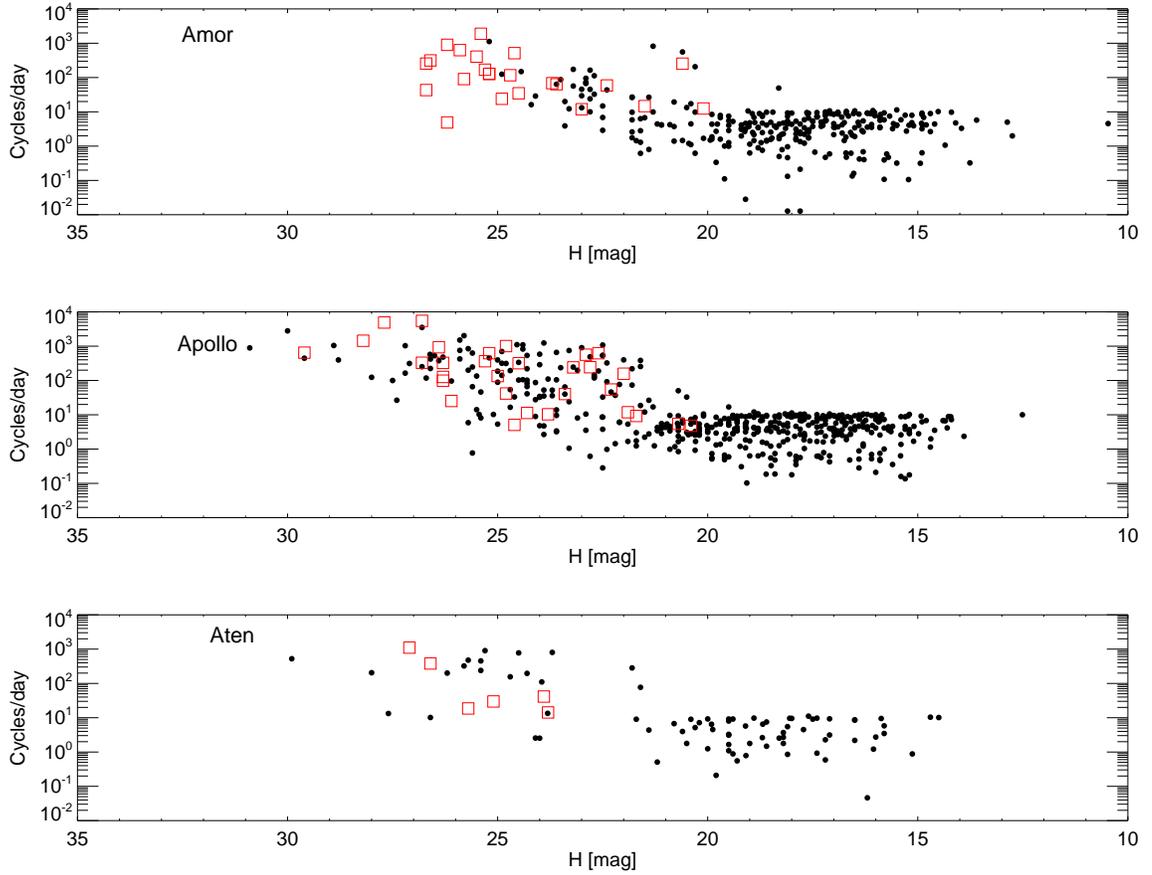}
\caption {\textit{Rotational frequency versus absolute magnitude (H)}: All NEOs with a known rotational period reported in the lightcurve database (black circles, \citet{Warner2009}), and MANOS results (red squares) reported in this work are plotted. Objects have been sub-divided according to their dynamical class (i.e. Amor, Apollo, and Aten). MANOS is finding a significant number of small, fast rotating Amors which do not appear in the LCDB.
}
\label{fig:Class}
\end{figure}

Figure~\ref{fig:Class} shows all NEOs reported in the LCDB as well as MANOS objects. Traditionally, the distribution of fast  and small rotators is more extended in the Apollo sub-population as compared to the Amors. However, MANOS is finding a significant number of small, fast rotating Amors which do not appear in the LCDB. This could be attributable to our use of large aperture facilities and the corresponding ability to probe small Amors, which often are fainter than the observational limits of smaller telescopic facilities. This trend is probably due to an observational bias. We performed a 2D Kolmogorov-Smirnov test (K-S test) to compare the three datasets and find if the samples are significantly different (or not) from one another. The KS test estimates the maximum deviation between the cumulative distribution of both datasets to test the similarity (or not) between the two distributions (Df). Significance level of the KS test is a value between 0 and 1. Small values show that the cumulative distribution of the first dataset is significantly different from the second dataset. Comparing the Amor population to the Apollo population, we obtained a value of Df=0.11, and a significance level of 0.01, indicating that the two samples are not significantly different. Whereas the Aten population compared to the Apollo group gave a Df=0.08 and a significance level of 0.72 suggesting that the populations are not significantly different. However, we must point out the limited number of Aten asteriods with a measured rotational period.

\subsection{Axis ratio and lightcurve amplitude }
\label{sec:amplitudecorrection}

Estimating the axis ratio of an object from its lightcurve amplitude is useful to constrain the object elongation. 

The observed lightcurve of a minor body depends on the geometrical circumstances during the observing run. Three angles have to be considered: i) the \textit{phase angle, $\alpha$} is the angular distance between the Sun and the observer as seen from the asteroid, ii) the \textit{viewing (or aspect) angle, $\xi$} is the angle between the rotation axis and the line of sight, and iii) the \textit{obliquity} is the angle between the spin vector and the orbital plane. Figure 1 of \citet{Barucci1982} summarizes these angles.

Based on \citet{Binzel1989}, if we assume NEOs are triaxial ellipsoids with axes a$>$b$>$c rotating along the c-axis, the lightcurve amplitude ($\Delta$${m}$) varies as a function of the viewing angle $\xi$ as: 
\begin{equation}
\Delta m = 2.5 \log \left(\frac{a}{b}\right) - 1.25 \log \left(\frac{a^2 \cos ^2 \xi + c^2 \sin ^2 \xi}{b^2 \cos ^2 \xi + c^2 \sin ^2 \xi}\right)
\end{equation}
The lower limit for the object elongation (a/b) is obtained assuming an equatorial view ($\xi$=90$^\circ$): 
\begin{equation}
\Delta m_{max}= 2.5 \log \left(\frac{a}{b}\right) \rightarrow  \frac{a}{b}  \geq   10^{0.4 \Delta m}
\end{equation}

However, this approach does not consider the phase angle effect, and $\Delta$m from the previous equation is the lightcurve amplitude obtained only during a given observing run (i.e. observation at a certain phase angle, obliquity, and viewing angle). 
The relation between lightcurve amplitude and phase angle is well known \citep{Gehrels1956} (see \citet{Muinonen2002} for a complete review). Specifically, the lightcurve amplitude increases with increasing phase angle. This relation results in an overestimation of the true axial ratio of the object. Only lightcurves obtained at very different phase angles can be used to correct the amplitude ($\Delta$m). In general, phase curves are not available for small NEOs, so an approximation has to be used. The lightcurve amplitude can be corrected as follows: 
\begin{equation}
\Delta m (\alpha = 0^\circ) = \frac{\Delta m (\alpha)}{1 + s \alpha }
\end{equation}
where s is the slope that correlates the amplitude with the phase angle, and $\Delta m (\alpha = 0^\circ)$ is the lightcurve amplitude at zero phase \citep{Zappala1990}. 
Combining Equations 4 and 5, and assuming a viewing angle of 90$^\circ$, an obliquity of 0$^\circ$, and a phase angle $\alpha$, we obtain:
\begin{equation}
\frac{a}{b}  \geq   10^{0.4 \Delta m (\alpha)/(1 + s \alpha)}
\end{equation}

Analyzing lightcurves of more than 30 asteroids from different taxonomic types (S, M, and C), \citet{Zappala1990} concluded that the slope (s) depends on the taxonomic type. They found a slope of 0.013~mag~deg$^{-1}$ for M-type asteroid, s=0.015~mag~deg$^{-1}$ for C-type and, s=0.030~mag~deg$^{-1}$  for S-type. Based on numerical models, \citet{Gutierrez2006} found different slopes depending on the object's topography, and surface scattering properties. They suggest an upper limit of 0.03~mag~deg$^{-1}$. We chose a slope of 0.03~mag~deg$^{-1}$ for the purpose of our work. 

\begin{figure}
\center
\includegraphics[width=12cm, angle=0]{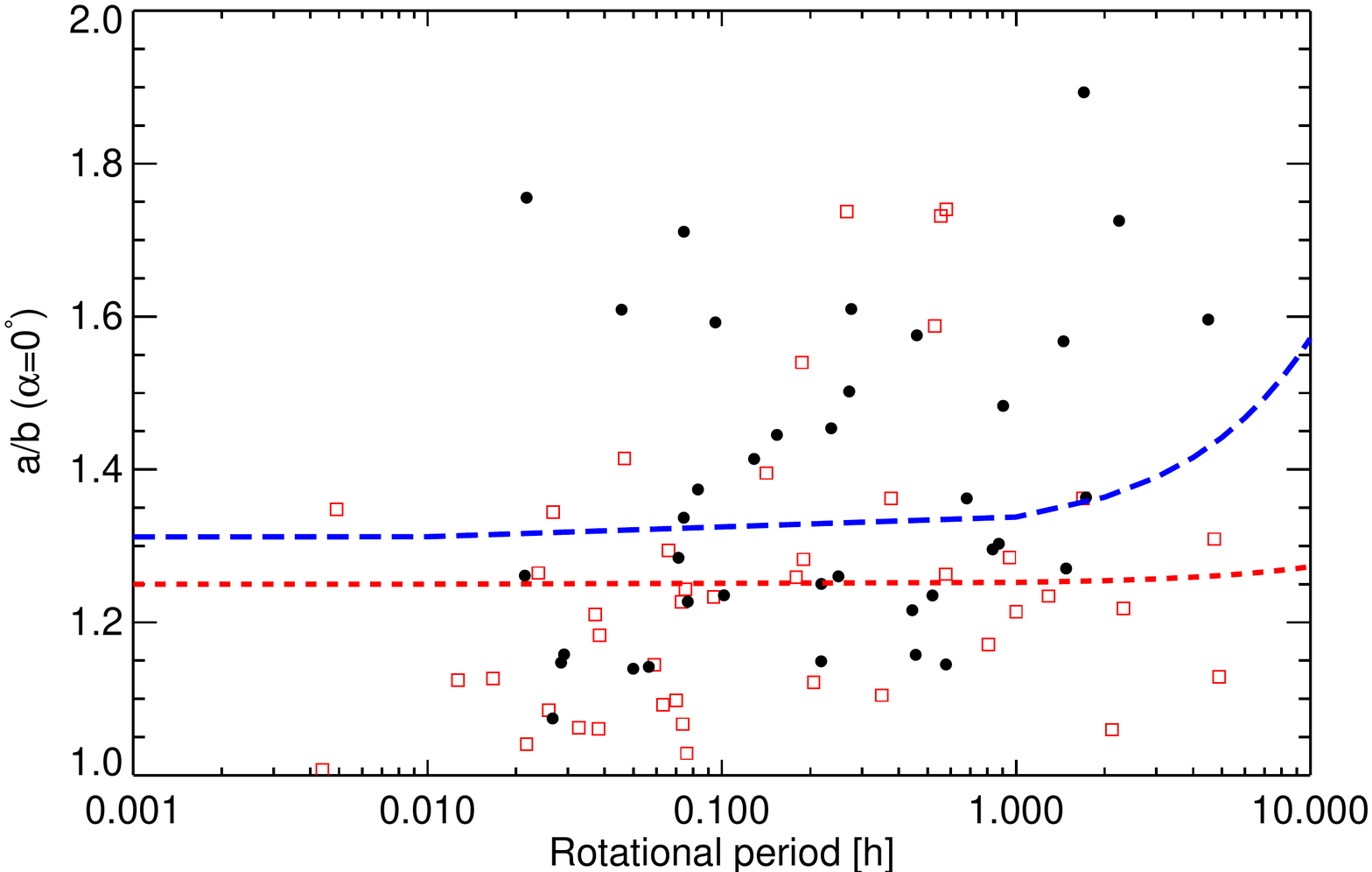}
\includegraphics[width=12cm, angle=0]{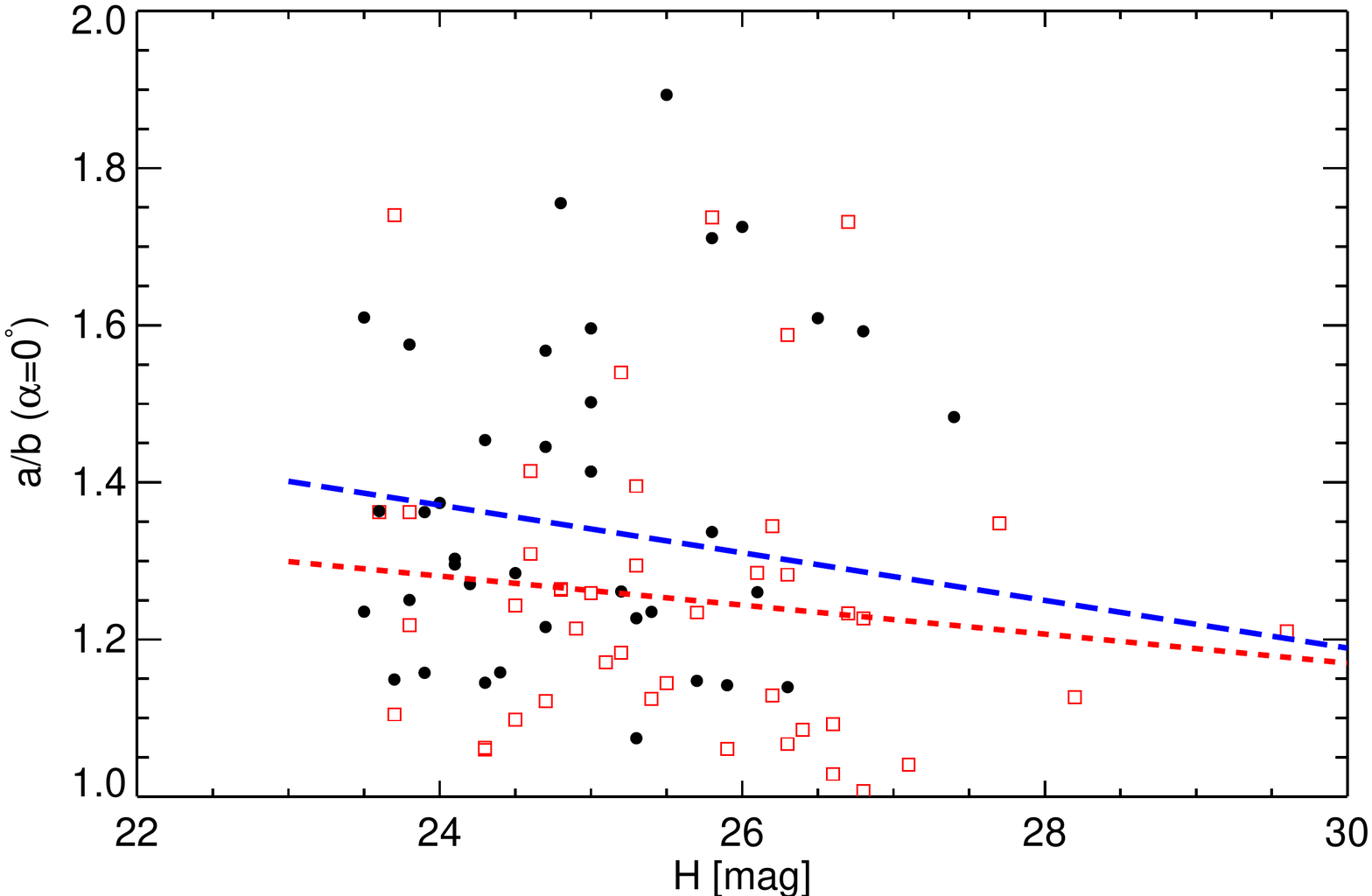}
\caption {\textit{Axis ratio (a/b) versus rotational period and absolute magnitude (H)}: MANOS (red square) and \citet{Hatch2015} data (black circle) are plotted. Axis ratio has been corrected for phase angle. Red (MANOS sample) and blue lines (MANOS+\citet{Hatch2015} datasets) are linear fits. We find an insignificant correlation between axis ratio and rotational period in our sample. }
\label{fig:ab}
\end{figure}

In Figure~\ref{fig:ab}, the axis ratio (a/b) of MANOS objects have been corrected for this phase angle effect. The mean a/b ratio without phase angle correction is 1.54, whereas the phase angle correction gives a mean ratio of 1.23. The maximum axis ratio is obtained for 2015~SO$_{2}$ with a/b=4.6 (without phase angle correction), and a/b=1.7 with correction for observed phase angle of about 60$^{\circ}$.

A weak correlation between axis ratio and rotational period has been noticed in the NEO population for objects with a diameter less than 60~m, though this result may be influenced by low number statistics \citep{Hatch2015}. Similarly, for small Main Belt Asteroids with diameters less than 1~km, fast-rotating asteroids (with period $<$2.3~h) have a tendency toward low amplitudes relative to slow rotators \citep{Nakamura2011}. We examined our sample for this weak correlation. In Figure~\ref{fig:ab}, MANOS results and \citet{Hatch2015}\footnote{\citet{Hatch2015} used data from \citet{Whiteley2002, Kwiatkowski2010a, Kwiatkowski2010b, Hergenrother2011, Statler2013}.} data are plotted. We corrected the phase angle effect for both samples and plotted the axis ratio at $\alpha$=0$^\circ$ (a/b ($\alpha$=0$^\circ$)). Only objects with a diameter less than 60~m are considered (i.e. objects with an absolute magnitude higher than 23.5, assuming an albedo of 0.2). We find an insignificant correlation between axis ratio and rotational period in our sample (linear fit\footnote{For a perfect one-to-one correlation, R$^{2}$=1.} has a R$^{2}$ of 0.0002). Merging the MANOS and \citet{Hatch2015} data, the correlation is still very weak (R$^{2}$=0.073 for the linear fit). For axis ratio versus absolute magnitude, we found that small objects seem to be more spherical. However, this tendency is also very weak. The linear fit to MANOS results has R$^{2}$=0.0159 and we find R$^{2}$=0.0083 for the MANOS+\citet{Hatch2015} data.  \\

%We looked for tendency between the lightcurve amplitude (after they were corrected from phase angle) with the asteroid sizes. Our sample has been divided into three sub-samples according to absolute magnitude: i) objects with an absolute magnitude between 20 and 23 (number of object in this sample is N=14), ii) objects with H between 23 and 26 (N=29), and iii) objects with an absolute magnitude between 26 and 31 (N=17). The largest objects (H=20-23) have low/moderate lightcurve amplitude with 93$\%$ ($\sqrt{N}$=3.7) showing $\Delta m$$\leq$0.25~mag. In the absolute magnitude range 23 to 26, only 66$\%$ ($\sqrt{N}$=5.4) of our sample have a low/moderate lightcurve amplitude, whereas the percentage increases up to 76$\%$ ($\sqrt{N}$=4.1) for the smallest objects. In conclusion, most of our objects (75$\%$ of our sample) have a $\Delta m$$\leq$0.25~mag which corresponds to a low/moderate object deformation. More data are needed to confirm (or not), the highest lightcurve amplitude in the range H=23-26, and to check if our sample is biased towards low amplitudes at the smallest sizes. 

We also looked for lightcurve amplitude versus size tendency according to the objects dynamical class. The Apollo group has a mean lightcurve amplitude of $\sim$0.2~mag that is roughly constant across all sizes. The Amor group indicates a weak correlation between amplitude and absolute magnitude. The Atens show an anti-correlation, but this sample is too small (N=6) to draw any reliable conclusions.

\section{Constraints for NEOs internal structure}
\label{sec:barrier}

A gravitationally bound strengthless rubble-pile cannot spin faster than $\sim$2.2~h without disrupting \citep{Pravec2002}. However, very small NEOs can rotate with periods shorter than 2.2~h. In fact, many have rotational periods of as little as a few minutes. Such rotations are so fast that rubble piles without cohesion could not be held together by self-gravity. A physical interpretation of these fast rotators is that they are objects bound through some combination of cohesive and/or tensile strength rather than gravity. 
The clear distinction between fast spinning and small-sized asteroids to larger bodies with slower spins, is thought to be related to internal structure. Objects larger than $\sim$200~m are interpreted as collections of rocks, boulders, and dust loosely consolidated by gravity alone and are therefore often referred to as ``rubble piles'' \citep{Chapman1978}. The fact that asteroids smaller than $\sim$200~m can rotate much faster suggests they are monolithic in nature and might constitute the blocks from which rubble piles are made. Alternatively, these small-sized asteroids with extremely fast rotation might be ``rubble piles'', held together by strong cohesion controlled by van der Waals forces and friction between constituent regolith grains (e.g., \citet{Holsapple2007, Goldreich2009, Scheeres2010, Sanchez2014}). Cohesion forces of 100-1000~Pa were measured within Lunar samples returned by the Apollo astronauts \citep{Mitchell1974}. \citet{Sanchez2014} suggested that the cohesive strength of sampled asteroid (25143) Itokawa is $\sim$25~Pa based on its measured grain-size. These values are lower than the tensile stress of meteoritic material by at least two orders of magnitude (see Table 4 at \citet{Kwiatkowski2010b}). Since meteorites are monolithic and not to be considered as rubble piles, constraining the cohesion values of fast rotating NEOs can allow us to determine whether an asteroid has a monolithic nature, or can survive as a rubble pile. 

This test can be performed by applying the Drucker-Prager yield criterion on NEOs parameters. This criterion calculates the minimal shear stress in a rotating ellipsoidal body at breakup taking into account its size, density and spin. Here we use the formalism published by Holsapple (2004, 2007) and later used in other studies that constrained the cohesion values of a few fast rotating asteroids: 64$_{-20}^{+12}$~Pa for (29075) 1950~DA (Rozitis et al. 2014), 40-210~Pa for the precursor body of the active asteroid P/2013 R3 \citep{Hirabayashi2014}, about 100~Pa for the fast rotator (335433) 2005~UW$_{163}$ \citep{Polishook2016} and 150-450~Pa for (60716) 2000~GD$_{65}$ \citep{Polishook2016}.

We applied the Drucker-Prager yield criterion on our MANOS targets and on the LCDB data (only NEOs considered) and compared the resulting minimal cohesion. From the lightcurves we use the measured rotation periods and the amplitude that we translate to the physical ratio a/b (using Equation 4). We assume b=c in order to derive a conservative lower value for the cohesion. The diameters were derived from the absolute magnitude H and an assumed fixed albedo of 0.2 (Equation 1). Density was set to 2.0-3.3~g~cm$^{-3}$ which is the density of the rubble pile (25143) Itokawa \citep{Fujiwara2006} and the mean density of ordinary chondrite \citep{Carry2012}, respectively, though the derived cohesion value is hardly sensitive to this density range. We find that $\sim$70$\%$ of the MANOS asteroids have minimal cohesion values of less than 100~Pa which is smaller than the cohesion of the lunar regolith; $\sim$20$\%$ have cohesion similar to that of the Moon (100-1000~Pa), and $\sim$10$\%$ have a minimal cohesion larger than lunar cohesion. Still, the largest minimal cohesion we derive is $\sim$3000~Pa, two order of magnitude smaller than the cohesion measured in meteorites (we used Almahata Sitta meteorite as a reference, \citet{Kwiatkowski2010b}). For comparison, $\sim94\%$ of the asteroids in the LCDB list have minimal cohesion values of less than 100~Pa. This difference is most probably due to the size difference between the two samples (the mean diameter of LCDB is 1000~m and is 50~m for the MANOS data). When considering LCDB asteroids that are similar in size to the MANOS objects, $\sim$80$\%$ have minimal cohesion values of less than 100~Pa. It is important to note that none of the LCDB asteroids reach minimal cohesion values that are comparable to the tensile stress measured in the monolithic meteorites, meaning that we cannot reject the notion that even a single one of them is a rubble pile held by shear stress (i.e., cohesion) against a fast spin. However, the dearth of asteroids with diameters larger than $\sim$200~m that rotate faster than $\sim$2~h and are limited by the cohesion lines in Figure~\ref{fig:Spin}, makes this notion less likely.

%________________________________________________________________

\section{Summary and Conclusions}

We present a homogeneous dataset composed of 86 objects (data reduced and analyzed the same way). We report rotational periods and lightcuve amplitude for most of them, but in some cases, we only report constraints on these properties.  

We report that 70~$\%$ of our sample shows at least one full rotation or a partial lightcurve with a period estimate. We report partial lightcurves for 14~$\%$ of our sample, i.e. objects that do not show a clear rotation period but do show a clear increase or decrease in magnitude. Finally, 16~$\%$ of the lightcurves are flat. Most of the observed obejcts are small and fast rotators, with $\sim$50$\%$ of objects spinning in less than 5~min. 

MANOS found two ultra-fast rotators: 2014~RC, and 2015~SV$_{6}$ with rotational periods of 15.8~s and 17.6~s, respectively. Discovery of these objects confirmed that MANOS is highly sensitive to the detection of fast spinning objects thanks to the use of large facilities allowing us to use short exposure time. We also highlight fast rotators in the Amor population, confirming again that our survey is sensitive to fast rotating small NEOs. 

We studied rotational frequency distribution according to size, and dynamical class. We noted an excess of both slow and fast rotators that does not allow us to fit a Maxwellian distribution to the observable distribution. Rotational periods are not significantly different in the Amor, Aten or Apollo groups. 

Axis ratio corrected from phase angle has been derived for our MANOS sample. No strong correlation between axis ratio and size or axis ratio and period has been found. 

Among the 30 mission accessible MANOS targets with complete lightcurves, six objects have rotational periods higher than 2~h, whereas three have periods between 1 and 2~h. The rest (i.e. 21 objects) have periods ranging from few seconds up to 1~h: 10 objects are rotating in less than 5~min, 3 objects have period between 5 and 10~min, and 8 with period longer than 10~min. Their sizes range from 3~m to 215~m, i.e. an absolute magnitude of 29.6 to 20.7.  

In conclusion, 33 of our 86 MANOS targets are mission accessible according to NHATS (i.e. $\Delta v$$^{NHATS}$$\leq$12~km~s$^{-1}$, launch window 2015-2040), 26 of these 33 are fully characterized with lightcurve (partial and flat lightcurves also considered) and visible spectrum. Only 7 objects of the objects presented here meet the NHATS dynamical criteria and the 1~h rotation limit \citep{Abell2009}: 2002~DU$_{3}$, 2010~AF$_{30}$, 2013~NJ, 2014~YD, 2015~CO, 2015~FG$_{36}$, and 2015~OV. Assuming a similar rate of object characterization as reported in this paper, $\sim$1,230 objects (i.e. approximately 10$\%$ of the known NEO population) need to be characterize in order to find 100 viable mission targets. Approximately 400,000 NEOs with diameter between 10~m and 1~km are estimated \citep{Tricarico2016}. To find 100 viable mission targets, $\sim$0.3$\%$ of the estimated population need to be characterized. \citet{Harris2015} estimated a population of $\sim$8$\times$10$^{7}$ objects in the 10~m-1~km size range, and so $\sim$0.002$\%$ of this population has to be characterized in order to find 100 viable targets. This means that $\sim$33,000 NEOs are expected to be mission accesible targets using the \citet{Tricarico2016} estimate, whereas the \citet{Harris2015} value gives us a total of $\sim$6,000,000 objects.  

As our main goal is to get a large set of fully characterized objects (lightcurve, and visible spectra), it is important to complete the study of some objects that have been partially characterized (incomplete/unknown lightcurve and/or no spectra, Table~\ref{Tab:candidates}). For most of these objects, their next optical windows are within the next ten years. NHATS generates the next optical windows through to the year 2040 (see NHATS webpage for more details). Unfortunately, because of their highly uncertain orbits most of these objects will be lost by then and current/future surveys will have to re-discover them. It is also important to point out that because of their uncertain orbits, their next windows of visibility can be off as well as their visual magnitude.

\acknowledgments
 
We thank an anonymous referee for her/his careful reading of the paper, and useful comments. We are grateful to the staffs of Discovery Channel Telescope (DCT), the Southern Astrophysical Research (SOAR) telescope, the Kitt Peak telescope, the SMARTS telescope, the Observatory of Sierra Nevada (OSN), and the Observatory of Anderson Mesa. This research was based in part on data obtained at the Lowell Observatory's Discovery Channel Telescope (DCT). Lowell operates the DCT in partnership with Boston University, Northern Arizona University, the University of Maryland, and the University of Toledo. Partial support of the DCT was provided by Discovery Communications. LMI was built by Lowell Observatory using funds from the National Science Foundation (AST-1005313).
This work is also based on observations obtained at the Southern Astrophysical Research (SOAR) telescope, which is a joint project of the Minist\'{e}rio da Ci\^{e}ncia, Tecnologia, e Inova\c{c}\~{a}o (MCTI) da Rep\'{u}blica Federativa do Brasil, the U.S. National Optical Astronomy Observatory (NOAO), the University of North Carolina at Chapel Hill (UNC), and Michigan State University (MSU). We also used the 1.3~m SMARTS telescope operated by the SMARTS Consortium. Based in part on observations at Kitt Peak National Observatory, National Optical Astronomy Observatory (NOAO Prop. ID:0270; PI:N. Moskovitz), which is operated by the Association of Universities for Research in Astronomy (AURA) under cooperative agreement with the National Science Foundation. 
This research was based on data obtained at the Observatorio de Sierra Nevada which is operated by the Instituto de Astrof\'{i}sica de Andaluc\'{i}a, CSIC. This research has made use of data and/or services provided by the International Astronomical Union's Minor Planet Center. Authors acknowledge support from NASA NEOO grant number NNX14AN82G, awarded to the Mission Accessible Near-Earth Object Survey (MANOS). A. Thirouin acknowledges Lowell Observatory funding. D. Polishook is grateful to the Ministry of Science, Technology and Space of the Israeli government for their Ramon fellowship for post-docs, the AXA Research Fund for their generous post-doc fellowship, and to Prof. Oded Aharonson of Weizmann Institute for his advisory work.

\clearpage

%% the file, uncomment the \plotone, \plottwo, and
%% If you need a layout that cannot be achieved with \plotone or
%% \plottwo, you can invoke the graphicx package directly with the
%% \includegraphics command or use \plotfiddle. For more information,
%% please see the tutorial on "Using Electronic Art with AASTeX" in the
%% documentation section at the AASTeX Web site, http://aastex.aas.org/
%%
%% The examples below also include sample markup for submission of
%% supplemental electronic materials. As always, be sure to check
%% the instructions to authors for the journal you are submitting to
%% for specific submissions guidelines as they vary from

%% This example uses \plotone to include an EPS file scaled to
%% 80% of its natural size with \epsscale. Its caption
%% has been written to indicate that additional figure parts will be
%% available in the electronic journal.

\clearpage

\clearpage

 \clearpage

 \clearpage

\clearpage

\begin{deluxetable}{lccccccc|ccc|cccc}
\tabletypesize{\tiny}
\rotate
\tablecaption{\label{Summary_photo} Summary of observations$^{a}$. }
\tablewidth{0pt}
\tablehead{
Object  & Date & $N_{im}$ & r$_{h}$& $\Delta$ & $\alpha$ & Filter &Tel. & Rot. per. & $\Delta$m &  $\varphi_{0}$ [JD] & H &  Diameter  & Dyn. &$\Delta$v$^{SH}$\\
  &   &  &[AU]& [AU]& [$^\circ$]& & & [h] &  [mag] & [2450000+] & [mag]  & [m] &   class & [km s$^{-1}$]\\
}
\startdata \textit{Full lightcurve:} & & & & & & & &		&			&	&		&   &	&    \\
\textit{Symmetric} & & & & & & & &		&			&	&		&   &	&    \\
\textit{lightcurve:} & & & & & & & &		&			&	&		&   &	&    \\
2014~GQ$_{17}$ 	& 04/10/14& 233& 1.019&  0.020& 28.12-28.49 & VR &DCT	& 	0.0217 	& 	0.08$\pm$0.01 	& 6757.64339& 	27.1 	& 11 &   Aten	&6.86\\
2014~MD$_{6}$ 	& 08/20/14& 79&1.032 &0.108 &76.20-76.16 & VR& DCT&	1.63 	& 	0.17$\pm$0.03 	& 6889.91573 & 	21.5 	& 148	 	& Amor & 6.12 \\
2014~VG$_{2}$ 	& 11/26/14& 155 & 1.055&0.105 & 47.10-47.18& VR& DCT&	0.0385 	& 	0.08$\pm$0.01 	& 6987.66738& 	22.6 	& 89	 &  Apollo&5.15	\\
2014~WR$_{6}$ 	& 11/26/14& 207& 1.027 &0.046 & 27.70-27.97&VR &DCT &	0.1416 	& 	0.67$\pm$0.02 	& 6987.57843& 	25.3 	& 25	 &   Amor	& 5.78\\
2015~DS 	& 03/10/15&158 &1.019 & 0.044 &51.92-51.95 & r'&	SOAR	& 1.00	 	& 	0.54$\pm$0.02 	& 7091.53468& 	24.9 	& 31 &   Amor	& 4.51	\\
2015~EK 	& 03/11/15 & 130& 1.002-1.004&0.013-0.015 & 46.03-42.09 & VR&DCT	& 	0.0737 	& 	0.17$\pm$0.02 	& 7092.63387& 	26.3 	& 16  &   Apollo &5.57	\\
2015~LQ$_{21}$ 	& 06/21/15& 85& 1.042& 0.035& 42.29-42.23&VR &	DCT& 	0.69 	& 	0.24$\pm$0.02 	& 7194.94385& 	24.5 	& 37  	& Amor	& 6.07\\
\textit{Asymmetric} & & & & & & & &		&			&	&		&   &	&    \\
\textit{ightcurve:} & & & & & & & &		&			&	&		&   &	&    \\
2005~ES$_{70}$ &03/10/15 &115 &1.054 &0.066 & 21.34-21.07 & r' & SOAR	& 	1.69 	& 	0.55$\pm$0.02 	& 7091.82669	& 	23.8 	& 51 &	  Aten & 12.22\\
2008~KZ$_{5}$ 	& 09/27/13& 200& 1.042 &0.109 & 65.75-65.65 & V&CTIO & 	1.92 	& 	0.11$\pm$0.04 	& 6563.68999& 	20.1 	& 283 &   Amor	& 5.74\\
2010~AF$_{30}$  &08/02/13 & 68& 1.114 &  0.113 &27.13-27.03 & V &	SOAR& 	2.6 	& 	0.06$\pm$0.01 	& 6507.67354	& 	21.7 	& 135	&  Apollo & 5.63\\
2013~BO$_{76}$ 	& 08/28,08/31,09/01/13& 135& 1.232-1.265 &0.225-0.259 &0.84-0.10 & V &	CTIO& 	5.03 	& 	0.26$\pm$0.04 	& 6533.71073& 	20.4 	&  247	&  Apollo & 6.29\\
2013~NJ & 01/08/14&55 &1.159 & 0.192-0.193 &22.02-22.04 & VR &	DCT& 	2.02 	& 	0.06$\pm$0.02 	& 6665.87196& 	21.9 	& 123 &   Apollo	& 4.87\\
2013~NX$^{*}$ & 03/14/15   &53 &1.076 &0.166 & 56.44-56.38& wh&  KP4& 	0.1529 	& 	0.84$\pm$0.02 	& 7095.96172& 	22.0 	& 118 &  Apollo& 5.49	\\
2013~NX$^{*}$ & 08/02/13   & 55 &1.085 &0.084 &  31.66-31.88   & r' &  SOAR  & 	0.1527 	& 	0.89$\pm$0.02 	& 6507.49100	& 	... 	& ... & ... &...	\\
2013~TG$_{6}$ 	& 10/09/13&127 & 1.023  &0.247 &73.76-73.71 & V &SOAR &	0.0631 	& 	0.18$\pm$0.02 	& 6575.63856	& 	26.6 	& 14	&  Aten & 5.1 \\
2013~WA$_{44}$ 	& 01/08/14 &63 & 1.143& 0.213&37.59-37.67 & VR& DCT&	0.35 	& 	0.23$\pm$0.04 	& 6665.75749& 	23.7 	& 54 &   Amor	& 4.17\\
2014~DH$_{10}$ 	& 03/26/14& 130 &  1.089 &  0.092- 0.093 & 7.83-7.85 &  R &	SOAR& 	0.2662 	& 	0.74$\pm$0.02 	& 6742.63103& 	25.8 	& 20   	& Amor & 5.37\\
2014~EK$_{24}$$^{*}$ & 03/25/14  & 76&1.057 &  0.068 &27.73-27.78 &V  &CTIO  	& 	0.0998 	& 	 1.26$\pm$0.01 	& 6741.74154	& 	23.2 	& 68 &   Apollo	&4.88	\\
2014~EK$_{24}$$^{*}$  & 02/11/15  &432  &1.011 & 0.046&57.38-57.19 &  VR &  DCT	& 	0.0996 	& 	 0.56$\pm$0.02 	& 7064.93072& 	... 	&...   & ...&...		\\
2014~FP$_{47}$ 	& 04/10/14& 183 &1.153-1.152 & 0.157& 14.90-15.07 &VR &	DCT & 	0.44 	& 	0.17$\pm$0.01 	& 6757.75387& 	22.3 	& 103  &   Apollo& 5.2\\
2014~KQ$_{84}$ 	&05/31/14 & 108& 1.036&0.024 &   20.53-20.28& r' &SOAR	& 	0.0938 	& 	0.37$\pm$0.03 	& 6809.76403	& 	26.7 	& 13	 &  Amor	 & 4.89\\
2014~PR$_{62}$	&09/10/14 & 49&1.1800 & 0.189 & 21.80-21.83& V & CTIO	& 	0.0947 	& 	0.28$\pm$0.02 	& 6910.66708& 	20.6 	& 	225 & Amor & 6.78\\
2014~RC$^{*}$  	& & & & & & &		& 	0.00439 	& 	0.01$\pm$0.01 	& 	& 	26.8 	& 12 &  Apollo&5.75	\\
2014~UV$_{210}$ &12/08/14 & 99 &2.158 &1.184 &  5.27-5.32 & Cle &OSN	& 	0.5559 	& 	0.91$\pm$0.02 	& 7000.32630& 	26.7 	& 	13 &   Amor &3.93\\
2014~UY &12/03/14 &116 &0.995 & 0.036 &  73.11-73.24& r' &SOAR& 	0.0658 	& 	0.87$\pm$0.03 	& 6995.79968& 	25.3 	& 	25 &   Apollo & 4.43\\
2014~WB$_{366}$ & 12/17/14&78 &1.024 &0.045 &26.77-26.78 & VR& DCT & 	0.0238 	& 	0.46$\pm$0.02 	& 7008.70556 & 	24.8 	& 32 & Apollo& 5.19\\
2014~WC$_{201}$ & 11/28/14 & 89&1.010 &0.024 & 13.48-13.51& VR& DCT& 	0.95 	& 	0.38$\pm$0.02 	& 6990.01152& 	26.1 	& 17 &  Apollo &6.28		\\
2014~YD & 01/13/15& 151 &0.996 &0.044 &72.66-72.76 &r' &SOAR	& 	2.12 	& 	0.20$\pm$0.02 	& 7035.77312& 	24.3 	& 41 & Apollo &3.98	\\
2015~AZ$_{43}$$^{*}$  	& 02/10/15& 333& 1.005-1.004& 0.0219& 51.36-51.44&wh &KP4	& 	0.5969 	& 	0.39$\pm$0.02 	& 7063.93078& 	23.4 	& 62  & Apollo	&5.65	\\
2015~BF$_{511}$ 	&02/10/15 & 263& 1.020& 0.041& 34.98-35.13& wh&KP4	& 	0.576	& 	0.52$\pm$0.02	& 7063.83662	& 24.8	  	&  32 &   Apollo	& 4.92\\
2015~BG$_{92}$ 	& 01/29/15& 175& 1.013 & 0.029 &14.21-13.98 &VR & DCT& 	0.179 	& 	0.36$\pm$0.02	& 7051.89156& 	25.0 	& 29 & Apollo& 5.24	\\
2015~BM$_{510}$ &02/05/15 &193 &1.011 &0.036-0.037 &45.24-45.56 &r' &SOAR& 	0.806 	& 	0.39$\pm$0.04 	& 7059.52984& 	25.1 	& 28   & Aten	& 4.79	\\
2015~CB$_{1}$ 	& 02/11/15	& 67 &1.005 &0.027 &48.65-48.76 & VR& DCT&	0.0589 	& 	0.36$\pm$0.02 	& 7064.85546& 	25.5 	& 23	 &  Amor& 5.69	\\
2015~CD$_{1}$ 	& 02/11/15	&68 & 0.996 & 0.012& 38.7&VR & DCT&	0.0167 	& 	0.28$\pm$0.02 	& 7064.79444& 	28.2 	& 6 & Apollo&4.92	\\
2015~CM 	& 02/11/15&62 &  1.009 & 0.023 &16.77-16.85 &VR &	DCT	& 	0.02678 	& 	0.53$\pm$0.02 	& 7064.81885& 	26.2 	& 	17  &  Amor & 6.27\\
2015~CS 	& 02/11/15& 72& 1.008&0.022 &19.22-19.17 & VR&	DCT	& 	0.2433 	& 	0.79$\pm$0.01 	& 7064.83603& 	26.3 	& 	16  &  Apollo & 5.81\\
2015~CW$_{12}$ 	& 03/11/15& 183& 1.067& 0.104 & 42.48-42.56& VR &DCT	& 	2.01 	& 	0.20$\pm$0.04 	& 7092.94971& 	23.0 	& 74 &  Amor	& 5.2\\
2015~DT 	& 02/22/15& 120& 1.080-1.079& 0.093-0.092& 11.33-11.28& V&	CTIO& 	0.0426 	& 	0.20$\pm$0.04 	& 7075.66190& 	22.9 	& 78  &   Apollo& 6.68		\\
2015~DU 	& 03/10/15& 110&1.017 &0.026 &  20.64-20.68& r'&	SOAR	& 	0.076 	& 	0.05$\pm$0.02 	& 7091.63277	& 	26.6 	& 14 &   Amor	& 3.97	\\
2015~EG 	& 03/10/15&62 &1.022 & 0.030&9.45-9.49 & r' &		SOAR& 	1.29 	& 	0.29$\pm$0.02 	& 7091.79723& 	25.7 	& 21  &   Aten	& 7.96\\
2015~EO      	& 03/14/15& 110 &1.007-1.006 &0.014-0.013 &23.01-23.31 & wh&  KP4  & 	0.073 	& 	0.39$\pm$0.02 	& 7095.67617 & 	26.8   	&  	12  & Apollo & 6.76\\
2015~FP 	&03/25/15 & 45&   1.025 &  0.028 & 4.17-4.18 & VR&	DCT	& 	0.1872 	& 	0.53$\pm$0.03 	& 7106.86738& 	25.2 	& 27  &   Amor	&5.51	\\
2015~HM$_{10}$ & 06/29/15&79 & 1.011& 0.042& 96.10-96.25& r'& SOAR&	0.3763 	& 	1.29$\pm$0.02 	& 7202.48903& 	 23.6	& 56 &  Amor	& 6.3\\
2015~HR$_{1}$ 	& 04/29/15& 60&1.138 &0.132 &5.47-5.48 &r' &SOAR &	0.0467 	& 	0.60$\pm$0.20 	& 7141.64653	& 	24.6 	& 	35  &  Amor & 6.53\\ 
2015~SO$_{2}$      	& 09/25/15 & 188 & 1.024  & 0.041 & 58.07-58.12 &  VR &DCT & 0.58  & 1.65$\pm$0.02	 &7290.85686	& 23.9	&  49	 	& Aten & 6.01 \\
2015~SO$_{2}$   &  09/28/15 & 64 & 1.019 & 0.038 &62.7 & r'  & SOAR & ... & 	... & 	7293.80409  	&  ...	& ... 	    	& ... & ... \\
2015~SU      	& 09/25/15 & 94& 1.040 & 0.039 & 18.98-19.11  & VR& DCT  & 	0.0127 & 0.20$\pm$0.03  	& 7290.74336& 25.4  & 24 	&  Amor & 6.47 \\
2015~SU$_{16}$      	& 09/25/15 & 79 & 1.048 & 0.053 & 30.81-30.74  &VR & DCT  & 0.2052 & 	0.24$\pm$0.03  	& 	7290.91196  &  24.7 &34  &  Amor   & 6.73\\
2015~SV$_{6}$$^{*}$   & 09/25/15 & 40& 1.009-1.008 & 0.007 & 42.47-43.06  & VR& DCT  & 0.00490	 & 	 0.74$\pm$0.03 	& 7290.61019	&  27.7	&  8	    	&  Apollo & 7.56\\
2015~SZ$_{2}$      	& 09/25/15 & 99& 1.020 & 0.019-0.018  & 19.63  &VR & DCT  & 	0.0384 & 	0.29$\pm$0.03  	& 	7290.62516 &  25.2	&  	27   	& Apollo & 4.91\\
\textit{Complex} & & & & & & & &		&			&	&		&   &	&    \\
\textit{lightcurve:} & & & & & & & &		&			&	&		&   &	&    \\
2013~WS$_{43}$ 	& 01/08/14&281 & 0.999&0.0606 &72.97-73.20 & VR&	DCT &	0.0988 	& 	0.39$\pm$0.01 	& 6665.94480	& 	22.8 	& 81 &   Apollo &5.51	\\
2015~AK$_{45}$ 	& 01/28/15& 130 &0.992 &0.015 &58.23-58.92 & VR&	DCT& 	0.0258 	& 	0.24$\pm$0.02 	& 7050.99769& 	26.4 	& 15	 &  Apollo&6.59	\\
2015~EP 	& 03/11/15&305 &1.012 &0.020 & 18.78-18.46 & VR &	DCT	& 	0.0381 	& 	0.10$\pm$0.02 	& 7092.66571& 	25.9 	& 19  &   Amor	& 5.55	\\
2015~FG$_{36}$ 	& 04/04/15 & 109&  1.032 & 0.064 & 58.36-58.46 & r' &SOAR	& 	2.32 	& 	0.59$\pm$0.02 	& 7116.60147& 	23.8 	& 51  &   Apollo& 4.67	\\
2015~JF & 05/12/15& 124 &1.035 &  0.026 &11.38 & VR&DCT& 	0.1897 	& 	0.33$\pm$0.03 	& 7154.74071& 	26.3 	& 16  &   Apollo	&5.21	\\
2015~OM$_{21}$    &07/29/15 &209 & 1.092& 0.110& 43.90-43.93&VR &DCT     & 	0.4084 	& 	0.55$\pm$0.02 	& 7232.90854& 	22.4   	&  98   	& Amor & 6.51\\
2015~OV      	& 08/18/15&143 &1.055 & 0.079& 55.69-55.76& VR&  DCT  & 	 1.17	& 	0.38$\pm$0.02 	&7252.92100&     23.4	&  62	  	& Apollo& 5.18
 \\
2015~QB      	& 08/18/15&174 &1.068-1.069 & 0.061& 22.60-22.61&VR & DCT   & 	 0.0327	& 	0.11$\pm$0.03 	& 7252.79872	&     24.3	&  41	 	& Amor&5.4\\
2015~SW$_{6}$      	& 09/25/15 & 64 & 1.073 & 0.071 & 11.77-11.79  & VR&  DCT & 	0.0752 & 	0.32$\pm$0.03  	& 7290.71933&  24.5	&  37	   &   Apollo & 6.50\\
2015~TC$_{25}$$^{*}$   & 10/12/15  & &  &  &   &  &    & 	0.03715  & 	 0.40$\pm$0.03 	& 	&  29.6&  3  &    Apollo & 4.73\\
\textit{Partially} & & & & & & & &		&			&	&		&   &	&    \\
\textit{constrained:} & & & & & & & &		&			&	&		&   &	&    \\
2002~DU$_{3}$ 	& 10/27/13 &70 &1.321-1.322 &0.369-0.372 & 23.6-23.9& r'& KP2&	4.4 	& 	0.24$\pm$0.02 	& 6593.84555& 	20.7 	& 215 &	  Apollo &5.63\\
2015~CO 	& 02/10/15&220 & 1.039 & 0.053& 7.15-7.24 & VR&	DCT	& 	4.9 	& 	0.16$\pm$0.02 	& 7063.80413& 	26.2 	& 17	 &  Amor& 5.09	\\
2015~DZ$_{198}$ &03/11/15 & 51&1.099-1.100 &0.108 & 11.19-11.12 &VR &DCT	& 	4.72 	& 	0.39$\pm$0.03 	& 7092.82531& 	24.6 	& 35  &  Apollo &5.99	\\
\hline
\hline
\textit{Partial lightcurve:} &		& & & & & & &	&			&	&		 	& &\\
2006~FJ$_{9}$   & 	10/09/13	& 83 & 1.469&  0.730-0.731& 37.89 & V &SOAR &$>$2 	& 	$>$0.2 	& 6575.49351& 	 19.3	&  410	 	& Amor&5.85\\
2010~XZ$_{67}$   & 01/08/14	&148 &  1.059-1.060 & 0.081& 18.72-18.84& VR &DCT &	$>$2 	& 	$>$0.3 	& 6665.66272 & 	 19.7 	&  341	  & Amor &6.81\\
2011~BN$_{24}$   & 09/16/13& 48 &1.344 &0.340 &4.35-4.34 &V & SOAR&	$>$1.5 	& 	$>$0.7 	& 6552.57891	& 	20.9  	&  196 &   Apollo &5.49\\
2014~BR$_{43}$   & 04/18/14& 112&1.285 &0.302 &23.57-23.59& Cle & SOAR&	$>$2 	& 	$>$1.2	& 6824.45356	& 	 21.5 	&  148	 &   Amor	& 6.09 \\
2014~CJ$_{14}$   & 03/25/14 & 92&1.173 &  0.229-0.230&36.20-36.29 & VR& DCT&	$>$2.5 	& 	$>$0.04	& 6741.66115& 	21.0  	&  187	 &  Amor	&6.17\\
2014~FA$_{44}$   & 04/18/14&179 &1.069 &  0.069-0.068&17.85-17.92 & Cle &SOAR &	$>$2 	& 	$>$1 	& 6765.59264	& 	  24.8	&  32 &  Amor& 4.49		\\
2014~FN$_{33}$   & 04/10/14 & 177&1.200 & 0.231-0.230 &27.83-27.88  & VR &DCT &	$>$2 	& 	$>$0.15	& 6757.90190& 	  21.0	&  187	 & Amor& 5.52\\
2014~HQ$_{124}$$^{**}$   &05/25/14& 47& 1.042-1.041 &0.112-0.111 &  72.21-72.23 & V&CTIO& 	$>$2.5 	& 	$>$0.5 	& 6802.82845& 	 18.9 	&  493	  & Aten& 12.29	\\
2014~NA$_{64}$   & 07/28/14	& 219&1.204 & 0.193 & 10.44-10.46 &V &CTIO &	$>$1.5 	& 	$>$0.3 	& 6870.72426& 	 19.8 	&  325	&    Amor& 6.85 \\
2015~HA$_{1}$  & 04/28/15 & 89& 1.119 &0.166 &44.36-44.29 & VR & Perkins &	$>$1.5 	& 	$>$0.1 	& 7140.86188& 	 21.2 	&  171	 &   Aten&8.96\\
2015~SY       	&09/25/15  & 171 & 1.079-1.080 & 0.077  & 4.98-4.82  & VR & DCT& 	$>$0.9 & 	$>$0.3  	& 7290.94053	&  23.1	&  71	  &    Aten & 10.59 \\
2015~SZ       	& 09/25/15 & 102& 1.082 & 0.085-0.084 & 19.98-19.96  & VR & DCT  & $>$1	 & 	 $>$0.1 	& 7290.76432 & 23.3	&  65	   &  Amor & 5.82 \\
\hline\hline
\textit{Flat lightcurve:} &		& & & & & & &	&			&	 		&&&	\\
1994~CJ$_{1}$   & 07/20/14 & 159 &1.131 & 0.140-0.141&33.08-33.02  &r' &SOAR &	 - 	& 	 -	& 6859.48559	& 	21.4 	&  155  &  Amor &4.7\\
2004~BZ$_{74}$   & 05/23/14&45 &1.486 &0.593 & 29.40-29.44&V &CTIO &	-  	& 	- 	& 6800.55622	& 	 18.1	& 712  &  Apollo& 10.77	\\
2005~RO$_{33}$   & 10/26/13& 100 &1.270 &0.314 &24.93-24.85 & V & CTIO&	-  	& 	- 	& 6592.60443	& 	20.1 	&  283 &   Amor &6.30\\
2007~CN$_{26}$   & 09/17/13	& 160&1.064 &0.083-0.084 & 43.43-43.30 &V &CTIO &	-  	& 	- 	& 6552.69735	& 	21.1 	& 179 &  Apollo	&5.09 \\
2008~TZ$_{3}$   & 04/10/14& 75& 1.122& 0.158& 37.82-37.86& V & CTIO&	-  	& 	- 	& 6757.73932	& 	 20.4	& 247  &  Apollo&5.53	 \\
2011~ST$_{232}$   & 08/03/14&141 &1.156 &0.145 &  11.04-11.03& V& CTIO&	-  	& 	- 	& 6872.72034	& 	 21.3	&  163 &  Amor & 6.38\\
2011~WU$_{95}$   & 10/25/13& 140&1.317-1.318 & 0.335 & 13.17-13.05& V & CTIO&	-  	& 	 -	& 6591.64622	& 	19.4    & 391  &   Apollo &5.54\\
2012~XQ$_{93}$   & 08/10/14&175 &1.109 & 0.153& 48.33-48.54& r' & SOAR&	-  	& 	- 	& 6879.46564	& 	21.7 	& 135  &  Apollo	&5.9 \\
2013~YZ$_{37}$   & 02/11/14& 200& 1.203 & 0.225& 14.7-14.6 & V& CTIO&	-  	& 	- 	& 6700.66054& 	19.8 	& 325  & Amor&6.56\\
2014~CP$_{13}$   & 02/22/14&175 & 1.054 & 0.098 &46.40-46.55 &r' & SOAR&	-  	& 	- 	& 	  6710.52583 &18.5 	& 592  &  Amor&5.7	 \\
2014~SM$_{143}$   & 10/08/14& 114& 1.080 &0.134-0.133 & 50.19-50.24 & V&CTIO &	-  	& 	- 	& 6938.69803& 	20.3 	& 258  &   Apollo& 9.9	\\
2014~YD$_{42}$   & 01/13/15	&137 &1.267 & 0.284-0.285& 3.36-3.46& r'&SOAR &	-  	& 	- 	& 7035.66012& 	22.2 	&  107  &  Apollo& 6.83	\\
2015~EQ          & 03/14/15	&52 & 0.999& 0.013& 65.63-65.77& wh &KP4 &	- 	& 	- 	& 7095.66094& 	  26.1 	&  17	 &  Aten& 12.87 \\
2015~HS$_{1}$   & 04/29/15 & 77& 1.104 &0.098 & 4.50-4.51& r'&SOAR &-  	& 	- 	& 7141.58920& 	24.7 	&  34  &  Amor&6.15	\\

\enddata
%% Text for table notes should follow after the \enddata but before
%% the \end{deluxetable}. Make sure there is at least one \tablenotemark
%% in the table for each \tablenotetext.
\tablenotetext{a}{Dates (UT-dates, format MM/DD/YYYY), heliocentric (r$_{h}$), and geocentric ($\Delta$) distances and phase angle ($\alpha$) of the observations are reported. The number of images ($N_{im}$), photometric filter, and the telescope (Tel.) are indicated for each entry. See text for facility details. We present the preferred rotational period (Rot. per. in hour), the peak-to-peak lightcurve amplitude ($\Delta$m in magnitude, without phase angle correction), and the Julian Date ($\varphi_{0}$) corresponding to phase zero. The Julian Date is not light time corrected. We also indicate the absolute magnitude (H) from the Minor Planet Center (MPC, July 2015) database, and a crude estimate of the object diameter assuming an albedo of 0.20. Our dataset is classified into three main categories: i) object with a full lightcurve, ii) object with only a partial lightcurve and iii) object with a flat lightcurve (i.e. no significant variability during our observing). }
\tablenotetext{*}{Objects whose study will be presented in details in future works. Observing circumstancies are not reported here.  }
\tablenotetext{**}{Contact binary \citep{Benner2014}.}

\end{deluxetable}

\begin{deluxetable}{lccccccc}
 \tablecaption{\label{Tab:candidates} All 33 MANOS targets with a $\Delta$$_v$$^{NHATS}$ lower than 12 km s$^{-1}$. Parameter $\Delta$$_{v}$ using \citet{Shoemaker1978} ($\Delta$$_v$$^{SH}$), and according NHATS are indicated in the last two columns ($\Delta$$_v$$^{NHATS}$). The best candidates for future missions are indicated in bold/italic. The best candidates are objects with a long rotational period, a $\Delta$$_v$$^{NHATS}$ lower than 12 km s$^{-1}$, and are objects fully characterized (lightcurves and spectra). Next opportunity to observe the best candidates for future missions is mentioned in the latest column (dates from NHATS webpage). We also included the next window for objects with P$>$1~h, but without visible spectrum, as well as object with a potentially slow rotation or unknown rotation.  } 
\tablewidth{0pt}
\tablehead{
Object  & H  & Diameter & Rot. Period   & Vis. Spectrum &   $\Delta$$_{v}$$^{SH}$ & $\Delta$$_{v}$$^{NHATS}$   & Next     \\
        &   &  &     & yes/no?  &     &    & Optical \\
        & [mag]  & [m]  & [h] &           & [km s$^{-1}$]  & [km s$^{-1}$]  &  Window \\
}
 \startdata 
1994~CJ$_{1}$	&	21.4 & 155 &	...  &   yes	&	4.7	&	11.928	 & 05/2016 \\ 
2007~CN$_{26}$	&	21.1 &  179  &	 .... &   yes	&	5.09	&	11.112	  &  05/2016\\
2014~RC       	&	26.8 &12&	0.00439 &   yes	&	5.75	&	11.610&	  \\ 
2015~CD$_{1}$	&	28.2&	6 & 0.0167 & no	&	4.92	&	7.651	 & \\ 
2014~GQ$_{17}$	&27.1	 &11 &	0.0217 &   no	&	6.86	&	7.728	 & \\ 
2015~CM	&26.2	 &17&	0.02678 &   no	&	6.27	&	7.760 & \\ 
2015~TC$_{25}$	&29.6	 &3 &	 0.03715 &   yes	&	4.73	&	4.261	&  \\ 
2015~SZ$_{2}$	&	25.2 &27&0.0384	 &    no &	4.91	&	7.401&	  \\ 
2014~VG$_{2}$	&	22.6 &89&0.0385	 &   yes	&	5.15	&	9.847	&  \\  
2013~TG$_{6}$	&	 26.6&	14 &0.0631 &  yes	&	5.1	&	5.577	&  \\ 
2014~UY	&	25.3 &25&0.0658	 &   no	&	4.43	&	7.035	 & \\
2015~DU	&	26.6 &14 &	0.076 &   yes	&	3.97	&	5.278	&  \\  
2014~KQ$_{84}$	&	26.7 &13&	0.0938 &   yes	&	4.89	&	9.734	&  \\ 
2014~EK$_{24}$	& 23.2	&68 &	0.0998 &   yes	&	4.88	&	5.099	&  \\ 
2013~NX 	&	22.0 &118&	0.1529 &   yes	&	5.49	&	6.648	&  \\ 
2015~BG$_{92}$	&25.0	& 29&	0.179 &   yes	&	5.24	&	5.218	 & \\
2013~WA$_{43}$	&23.7	& 54&	0.35 &   yes	&	4.17	&	5.442	 & \\ 
2014~FP$_{47}$	&22.3	 & 103 &	0.44 &   yes	&5.2	&	9.054	 & \\ 
2014~UV$_{210}$	&26.7	 &13&	0.5559 &   yes	&	3.93	&	5.902	 & \\ 
2015~BF$_{511}$	&24.8	 &32&	0.576 &   yes	&	4.92	&	9.752	 & \\ 
2015~SO$_{2}$	&	23.9 &49&	0.58 &   yes	&	6.01	&	6.034	&  \\ 
2015~BM$_{510}$	&	25.1 &28&	0.806 &   yes	&	4.79	&	5.638	&  \\  
2014~WC$_{201}$	&26.1	 &17&	0.95 &   yes	&	6.28	&	11.835	&  \\ 
2015~DS	&24.9	 &31&	1.00  &   no	&	4.51	&	9.648&	01/2029  \\
\textit{\textbf{2015~OV}}	&	\textit{\textbf{23.4}} &\textit{\textbf{62}}&	\textit{\textbf{1.17}} &   \textit{\textbf{yes}}	&		\textit{\textbf{5.18}} &	\textit{\textbf{7.788}}	&  03/2022\\ 
2015~EG	&25.7	 &21 &	1.29 &  no	&	7.96	&	10.586	 & 03/2019 \\ 
2014~FA$_{44}$	&	24.8 &  32&	$>$2 &   yes	&	4.49	&	8.584	 & 08/2017 \\ 
\textit{\textbf{2013~NJ}}	&\textit{\textbf{21.9}}	 &  \textit{\textbf{123}} &	\textit{\textbf{2.02}} &   \textit{\textbf{yes}}	&	\textit{\textbf{4.87}}	&	\textit{\textbf{9.934}}	  & 05/2016\\ 
\textit{\textbf{2014~YD} }  	&\textit{\textbf{24.3}}	& \textit{\textbf{41}}&	\textit{\textbf{2.12}} &   \textit{\textbf{yes}}	&	\textit{\textbf{3.98}}	&	\textit{\textbf{5.496}}	  & 10/2024\\ 
\textit{\textbf{2015~FG$_{36}$}	}&\textit{\textbf{23.8}}	 &\textit{\textbf{51 }}&\textit{\textbf{2.32}}	 &  \textit{\textbf{yes}}	&	\textit{\textbf{4.67}}	&	\textit{\textbf{6.974}}	 & 11/2022\\ 
\textit{\textbf{2010~AF$_{30}$}}	&	\textit{\textbf{21.7 }} & \textit{\textbf{ 135 }} &	\textit{\textbf{2.6}} &   \textit{\textbf{yes}} 	&	\textit{\textbf{5.63}}	&	\textit{\textbf{11.816}}	& 07/2016 \\  
\textit{\textbf{2002~DU$_{3}$}}	&\textit{\textbf{20.7}}	 &   \textit{\textbf{215 }} &	\textit{\textbf{4.4}} &   \textit{\textbf{yes}}	&	\textit{\textbf{5.63	}}&	\textit{\textbf{9.422	}} & 09/2017\\ 
\textit{\textbf{2015~CO}}	&	\textit{\textbf{26.2 }}&\textit{\textbf{17}}&	\textit{\textbf{4.9}} &   \textit{\textbf{yes	}}&	\textit{\textbf{5.09}}	&	\textit{\textbf{11.784}}	&  none \\ 
\hline
\hline
\enddata
\end{deluxetable}

%% If you use the table environment, please indicate horizontal rules using
%% Do not put multiple tabular environments within a single table.
%% The optional \label should appear inside the \caption command.

%% If the table is more than one page long, the width of the table can vary
%% from page to page when the default \tablewidth is used, as below.  The
%% individual table widths for each page will be written to the log file; a
%% maximum tablewidth for the table can be computed from these values.
%% The \tablewidth argument can then be reset and the file reprocessed, so
%% that the table is of uniform width throughout. Try getting the widths
%% from the log file and changing the \tablewidth parameter to see how
%% adjusting this value affects table formatting.

%% The \dataset{} macro has also been applied to a few of the objects to
%% show how many observations can be tagged in a table.

%% Tables may also be prepared as separate files. See the accompanying
%% sample file table.tex for an example of an external table file.
%% To include an external file in your main document, use the \input
%% command. Uncomment the line below to include table.tex in this
%% sample file. (Note that you will need to comment out the \documentclass,
%% \begin{document}, and \end{document} commands from table.tex if you want
%% to include it in this document.)


%% The following command ends your manuscript. LaTeX will ignore any text


\begin{thebibliography}{} 
\bibitem[Abell et 
al.(2009)]{Abell2009} Abell, P.~A., Korsmeyer, D.~J., Landis, R.~R., et al.\ 2009, Meteoritics and Planetary Science, 44, 1825 
\bibitem[Barucci 
\& Fulchignoni(1982)]{Barucci1982} Barucci, M.~A., \& Fulchignoni, M.\ 1982, Moon and Planets, 27, 47 
\bibitem[Benner et al.(2015)]{Benner2015} Benner, L.~A., Busch, 
M.~W., Giogini, J.~D., Taylor, P.~A., 
\& Margot, J.~J.\ 2015, Asteroids IV, 165 
\bibitem[Benner et al.(2014)]{Benner2014} Benner, L.~A., Brozovic, 
M., Giorgini, J.~D., et al.\ 2014, AAS/Division for Planetary Sciences 
Meeting Abstracts, 46, \#409.01 
\bibitem[Binzel et al.(2002)]{Binzel2002} Binzel, R.~P., Lupishko, 
D., di Martino, M., Whiteley, R.~J., 
\& Hahn, G.~J.\ 2002, Asteroids III, 255 
\bibitem[Binzel et al.(2010)]{Binzel2010} Binzel, R.~P., 
Morbidelli, A., Merouane, S., et al.\ 2010, \nat, 463, 331 
\bibitem[Binzel et al.(1989)]{Binzel1989} Binzel, R.~P., 
Farinella, P., Zappala, V., \& Cellino, A.\ 1989, Asteroids II, 416 
\bibitem[Binzel et al.(2002)]{Binzel2002} Binzel, R.~P., Lupishko, 
D., di Martino, M., Whiteley, R.~J., 
\& Hahn, G.~J.\ 2002, Asteroids III, 255 
\bibitem[Bottke et al.(2002)]{Bottke2002} Bottke, W.~F., 
Morbidelli, A., Jedicke, R., et al.\ 2002, \icarus, 156, 399 
\bibitem[Bottke et al.(2006)]{Bottke2006} Bottke, W.~F., Jr., 
Vokrouhlick{\'y}, D., Rubincam, D.~P., 
\& Nesvorn{\'y}, D.\ 2006, Annual Review of Earth and Planetary Sciences, 34, 157 
\bibitem[Busch et al.(2015)]{Busch2015} Busch, M.~W., Benner, L.~A.~M., Naidu, S.~P., et al.\ 2015, AAS/Division for Planetary Sciences Meeting Abstracts, 47, 402.05 
\bibitem[Carry(2012)]{Carry2012} Carry, B.\ 2012, \planss, 73, 98 
\bibitem[Chapman(1978)]{Chapman1978} Chapman, C.~R.\ 1978, NASA Conference Publication, 2053,  
\bibitem[Clark et al.(2002)]{Clark2002} Clark, B.~E., 
Helfenstein, P., Bell, J.~F., et al.\ 2002, \icarus, 155, 189 
\bibitem[DeMeo 
\& Binzel(2008)]{DeMeo2008} DeMeo, F., \& Binzel, R.~P.\ 2008, \icarus, 194, 436 
\bibitem[Foster(1995)]{Foster1995} Foster, G.\ 1995, \aj, 109, 
1889 
\bibitem[Fujiwara et al.(2006)]{Fujiwara2006} Fujiwara, A., Kawaguchi, J., Yeomans, D.~K., et al.\ 2006, Science, 312, 1330 
\bibitem[Galache et al.(2015)]{Galache2015} Galache, J.~L., Beeson, C.~L., McLeod, K.~K., \& Elvis, M.\ 2015, \planss, 111, 155 
\bibitem[Gehrels(1956)]{Gehrels1956} Gehrels, T.\ 1956, \apj, 123, 
331 
\bibitem[Goldreich \& Sari(2009)]{Goldreich2009} Goldreich, P., \& Sari, R.\ 2009, \apj, 691, 54 
\bibitem[Greenstreet et al.(2012)]{Greenstreet2012} Greenstreet, S., Ngo, H., \& Gladman, B.\ 2012, \icarus, 217, 355 
\bibitem[Guti{\'e}rrez et 
al.(2006)]{Gutierrez2006} Guti{\'e}rrez, P.~J., Davidsson, B.~J.~R., Ortiz, J.~L., Rodrigo, R., \& Vidal-Nu{\~n}ez, M.~J.\ 2006, \aap, 454, 367 
\bibitem[Harris 
\& D'Abramo(2015)]{Harris2015} Harris, A.~W., \& D'Abramo, G.\ 2015, \icarus, 257, 302 
\bibitem[Hatch 
\& Wiegert(2015)]{Hatch2015} Hatch, P., \& Wiegert, P.~A.\ 2015, \planss, 111, 100 
\bibitem[Hergenrother 
\& Whiteley(2011)]{Hergenrother2011} Hergenrother, C.~W., \& Whiteley, R.~J.\ 2011, \icarus, 214, 194 
\bibitem[Hirabayashi et al.(2014)]{Hirabayashi2014} Hirabayashi, M., Scheeres, D.~J., S{\'a}nchez, D.~P., Gabriel, T.\ 2014.\ Constraints on the Physical Properties of Main Belt Comet P/2013 R3 from its Breakup Event.\ The Astrophysical Journal 789, L12. 
\bibitem[Holsapple(2001)]{Holsapple2001} Holsapple, K.~A.\ 2001, 
\icarus, 154, 432 
\bibitem[Holsapple(2004)]{Holsapple2004} Holsapple, K.~A.\ 2004, 
\icarus, 172, 272 
\bibitem[Holsapple(2007)]{Holsapple2007} Holsapple, K.~A.\ 2007, 
\icarus, 187, 500 
\bibitem[Jewitt(2002)]{Jewitt2002} Jewitt, D.~C.\ 2002, \aj, 123, 
1039 
\bibitem[Juri{\'c} et al.(2002)]{Juric2002} Juri{\'c}, M., 
Ivezi{\'c}, {\v Z}., Lupton, R.~H., et al.\ 2002, \aj, 124, 1776
\bibitem[Kwiatkowski et 
al.(2010a)]{Kwiatkowski2010a} Kwiatkowski, T., Buckley, D.~A.~H., O'Donoghue, D., et al.\ 2010, \aap, 509, A94 
\bibitem[Kwiatkowski et 
al.(2010b)]{Kwiatkowski2010b} Kwiatkowski, T., Polinska, M., Loaring, N., et al.\ 2010, \aap, 511, A49 
\bibitem[Lacerda \& Luu(2003)]{Lacerda2003} Lacerda, P., \& Luu, J.\ 2003, \icarus, 161, 174 
\bibitem[Lagerkvist et al.(1989)]{Lagerkvist1989} Lagerkvist, C.-I., 
Harris, A.~W., \& Zappala, V.\ 1989, Asteroids II, 1162 
\bibitem[Levine et al.(2012)]{Levine2012} Levine, S.~E., Bida, T.~A., Chylek, T., et al.\ 2012, \procspie, 8444, 844419 
\bibitem[Lomb(1976)]{Lomb1976} Lomb, N.~R.\ 1976, \apss, 39, 447 
\bibitem[Mainzer et al.(2012)]{Mainzer2012} Mainzer, A., Grav, T., Masiero, J., et al.\ 2012, \apj, 752, 110 
\bibitem[Margot et al.(2002)]{Margot2002} Margot, J.~L., Nolan, 
M.~C., Benner, L.~A.~M., et al.\ 2002, Science, 296, 1445 
\bibitem[Mazanek et al.(2015)]{Mazanek2015} Mazanek, D., Naasz, B., Cichy, B., Reeves, D., \& Abell, P.\ 2015, European Planetary Science Congress 2015, held 27 September - 2 October, 2015 in Nantes, France, id.EPSC2015-279, 10, EPSC2015-279 
\bibitem[Mitchell et al.(1974)]{Mitchell1974}Mitchell, J. K., Houston, W. N., Carrier, W. D. \& Costes, N. C. Apollo soil mechanics experiment S-200 final report. Space Sciences Laboratory Series 15, 72Ð85 (Univ. California, Berkeley, 1974); $http://www.lpi.usra.edu/lunar/documents/NASA\%20CR-134306.pdf$
\bibitem[Morbidelli et al.(2002)]{Morbidelli2002} Morbidelli, A., Bottke, W.~F., Jr., Froeschl{\'e}, C., \& Michel, P.\ 2002, Asteroids III, W.~F.~Bottke Jr., A.~Cellino, P.~Paolicchi, and R.~P.~Binzel (eds), University of Arizona Press, Tucson, p.409-422, 409 
\bibitem[Muinonen et al.(2002)]{Muinonen2002} Muinonen, K., 
Piironen, J., Shkuratov, Y.~G., Ovcharenko, A., 
\& Clark, B.~E.\ 2002, Asteroids III, 123 
\bibitem[Nakamura et al.(2011)]{Nakamura2011} Nakamura, T., 
Dermawan, B., \& Yoshida, F.\ 2011, \pasj, 63, 577 
\bibitem[Nyquist(1928)]{Nyquist1928} Nyquist, H.\ 1928, 
Transactions of the American Institute of Electrical Engineers, Volume 47, 
Issue 2, pp.~617-624, 47, 617 
\bibitem[Polishook \& Brosch(2009)]{Polishook2009} Polishook, D., \& Brosch, N.\ 2009, \icarus, 199, 319 
\bibitem[Polishook et al.(2016)]{Polishook2016} Polishook, D., and 13 colleagues 2016.\ A 2 km-size asteroid challenging the rubble-pile spin barrier - A case for cohesion.\ Icarus 267, 243-254. 
\bibitem[Popova et al.(2013)]{Popova2013} Popova, O.~P., 
Jenniskens, P., Emel'yanenko, V., et al.\ 2013, Science, 342, 1069 
\bibitem[Pravec 
\& Harris(2000)]{Pravec2000} Pravec, P., \& Harris, A.~W.\ 2000, \icarus, 148, 12 
\bibitem[Pravec et al.(2002)]{Pravec2002} Pravec, P., Harris, 
A.~W., \& Michalowski, T.\ 2002, Asteroids III, 113 
\bibitem[Pravec et al.(2005)]{Pravec2005} Pravec, P., Harris, A.~W., Scheirich, P., et al.\ 2005, \icarus, 173, 108 
\bibitem[Pravec et al.(2006)]{Pravec2006} Pravec, P., Scheirich, P., Ku{\v s}nir{\'a}k, P., et al.\ 2006, \icarus, 181, 63 
\bibitem[Pravec 
\& Harris(2007)]{Pravec2007} Pravec, P., \& Harris, A.~W.\ 2007, \icarus, 190, 250 
\bibitem[Pravec et al.(2008)]{Pravec2008} Pravec, P., Harris, A.~W., Vokrouhlick{\'y}, D., et al.\ 2008, \icarus, 197, 497 
\bibitem[Pravec et al.(2012)]{Pravec2012} Pravec, P., Harris, A.~W., Ku{\v s}nir{\'a}k, P., Gal{\'a}d, A., \& Hornoch, K.\ 2012, \icarus, 221, 365 
\bibitem[Press et al.(1992)]{Press1992} Press, W.~H., Teukolsky, 
S.~A., Vetterling, W.~T., 
\& Flannery, B.~P.\ 1992, Cambridge: University Press, c1992, 2nd ed.,  
\bibitem[Richardson et al.(1998)]{Richardson1998} Richardson, D.~C., 
Bottke, W.~F., \& Love, S.~G.\ 1998, \icarus, 134, 47 
\bibitem[Rozitis et al.(2014)]{Rozitis2014} Rozitis, B., Maclennan, E., Emery, J.~P.\ 2014.\ Cohesive forces prevent the rotational breakup of rubble-pile asteroid (29075) 1950 DA.\ Nature 512, 174-176. 
\bibitem[Rubincam(2000)]{Rubincam2000} Rubincam, D.~P.\ 2000, 
\icarus, 148, 2 
\bibitem[Saito et al.(2006)]{Saito2006} Saito, J., Miyamoto, H., 
Nakamura, R., et al.\ 2006, Science, 312, 1341 
\bibitem[S{\'a}nchez 
\& Scheeres(2014)]{Sanchez2014} S{\'a}nchez, P., \& Scheeres, D.~J.\ 2014, Meteoritics and Planetary Science, 49, 788 
\bibitem[Scargle(1982)]{Scargle1982} Scargle, J.~D.\ 1982, \apj, 263, 835 
\bibitem[Scheeres et al.(2000)]{Scheeres2000} Scheeres, D.~J., 
Ostro, S.~J., Werner, R.~A., Asphaug, E., 
\& Hudson, R.~S.\ 2000, \icarus, 147, 106 
\bibitem[Scheeres et al.(2005)]{Scheeres2005} Scheeres, D.~J., 
Benner, L.~A.~M., Ostro, S.~J., et al.\ 2005, \icarus, 178, 281 
\bibitem[Scheeres et al.(2010)]{Scheeres2010} Scheeres, D.~J., Hartzell, C.~M., S{\'a}nchez, P., \& Swift, M.\ 2010, \icarus, 210, 968 
\bibitem[Scheeres et al. (2015)]{Scheeres2015} Scheeres, D.~J., Britt, 
D., Carry, B., \& Holsapple, K.~A.\ 2015, Asteroids IV 
\bibitem[Shoemaker 
\& Helin(1978)]{Shoemaker1978} Shoemaker, E.~M., \& Helin, E.~F.\ 1978, Reports of Planetary Geology Program, 20 
\bibitem[Spearman (1904)]{Spearman1904} Spearman C. \ 1904, The american journal of psychology, 15, 72. 
\bibitem[Statler et al.(2013)]{Statler2013} Statler, T.~S., 
Cotto-Figueroa, D., Riethmiller, D.~A., 
\& Sweeney, K.~M.\ 2013, \icarus, 225, 141 
\bibitem[Stellingwerf(1978)]{Stellingwerf1978} Stellingwerf, R.~F.\ 
1978, \apj, 224, 953 
\bibitem[Stetson(1987)]{Stetson1987} Stetson, P.~B.\ 1987, \pasp, 
99, 191 
\bibitem[Thomas et al.(2011)]{Thomas2011} Thomas, C.~A., Trilling, D.~E., Emery, J.~P., et al.\ 2011, \aj, 142, 85
\bibitem[Thomas et al.(2014)]{Thomas2014} Thomas, C.~A., Emery, J.~P., Trilling, D.~E., et al.\ 2014, \icarus, 228, 217  
\bibitem[Tricarico(2016)]{Tricarico2016} Tricarico, P.\ 2016, arXiv:1604.06328 
\bibitem[Warner et al.(2009)]{Warner2009} Warner, B.~D., Harris, 
A.~W., \& Pravec, P.\ 2009, \icarus, 202, 134 
\bibitem[Whiteley et al.(2002)]{Whiteley2002} Whiteley, R.~J., 
Tholen, D.~J., \& Hergenrother, C.~W.\ 2002, \icarus, 157, 139 
\bibitem[Zappala et 
al.(1990)]{Zappala1990} Zappala, V., Cellino, A., Barucci, A.~M., Fulchignoni, M., \& Lupishko, D.~F.\ 1990, \aap, 231, 548 



\end{thebibliography}
\end{document}